# Prediction of novel final phases in aged uranium-niobium alloys

Xiao L. Pan[1, 2], Hao Wang[1, 2], Lei L. Zhang[2], Yu F. Wang[2], Xiang R. Chen[1]*, Hua Y. Geng[2, 3]*, Ying Chen[4]

[1] *College of Physics, Sichuan University, Chengdu 610065, P. R. China;*

[2] *National Key Laboratory of Shock Wave and Detonation Physics, Institute of Fluid Physics, China Academy of Engineering Physics, Mianyang, Sichuan 621900, P. R. China;*

[3] *HEDPS, Center for Applied Physics and Technology, and College of Engineering, Peking University, Beijing 100871, P.R. China;*

[4] *Fracture and Reliability Research Institute, School of Engineering, Tohoku University, Sendai 980-8579, Japan.*

**Abstract:** Ordered intermetallics are long believed to be the final products of the aging of U-Nb solid solutions at low temperatures, a crucial property for the practical applications of this alloy in engineering and industry. However, such conjectured ordered compounds have not been experimentally or theoretically established. Herein, numerical evidence for ordered intermetallic U-Nb compounds is presented using thorough first-principles structure predictions up to 500 GPa. Two stable $U_2Nb$ compounds and one metastable $U_2Nb$ and one metastable $U_3Nb$ were discovered. A unique hybridized transition driven by pressure was observed in $U_2Nb$, which is a superposition of one first-order transition and another second-order transition, leading to striking features near the transition pressure of 21.6 GPa. The decomposition limit of these compounds at high temperature was also investigated. The strong stability of $U_2Nb$ in the region of low pressure and high temperature was revealed. This discovery of ordered $U_2Nb$ and its strong stability over a wide pressure range completely changed the phase diagram of U-Nb alloys and shed new light on the dynamic response and aging mechanism of U-Nb alloys.

## I. Introduction

The uranium–niobium (U-Nb) alloy is a complex system with many metastable phases. Depending on the Nb concentration, U-Nb alloys can be quenched to different metastable phases under ambient conditions from the high-temperature Nb-supersaturated BCC solid solution, which is called $\gamma$ phase and becomes unstable below 923 K [1, 2]. They turn to (1) orthorhombic phase $\alpha'$ for 0–4.2 wt. % Nb, (2) monoclinic phase $\alpha''$ for 4.2 –6.9 wt. % Nb, and (3) tetragonal $\gamma^o$ phase for 6.9–8.9 wt. % Nb [3-6]. Among them, the stress-strain curves of $\alpha''$ and $\gamma^o$ phases exhibit apparent double yield behavior and are significant shape-memory alloys [7-11].

It is believed that, at temperatures above 600 K, all metastable phases of the U-Nb alloys would eventually evolve toward or decompose into the $\alpha$ + $\gamma_{Nb}$ final equilibrium state [12], where $\gamma_{Nb}$ is an Nb-rich BCC phase, commonly containing 75 at. % Nb or more at equilibrium [13-17]. For low-temperature (< 600 K) aging of U-Nb alloys, some investigations have attempted to explain the mechanism from different aspects, such as spinodal decomposition [18, 19], chemical redistribution [20] and phase transformation [21], but the jury is still out. In addition to ambient pressure, Zhang et al. experimentally studied U-6 wt. % Nb and U-7.7 wt. % Nb at temperatures and pressures up to 1300 K and 5 GPa, respectively [22, 23]. They found that metastable α" exhibits complex paths upon heating at both ambient and high pressures, leading to diffusional phase decomposition, and is sensitive to the applied pressure.

* *Corresponding authors. E-mail: s102genghy@caep.cn; xrchen@scu.edu.cn*

To date, a few explorations of the possible ordered phase have been conducted experimentally and theoretically. The ordering effect is speculated to be related to the aging mechanism of metastable phases, as suggested by Hsiung et al. [18]. They experimentally studied the aging of metastable U-6 wt. % Nb at 200℃ and concluded that the metastable α" phase will ultimately decompose into Nb-depleted *α* phase and an ordered phase. However, they only provided illustrative sketches for the ordered structure without further details or evidence. Zhang et al. attempted to develop possible candidate-ordered phases with a U:Nb ratio of 3:1 by redistributing atoms in the *Cmcm* lattice of α-U [19]. This work enhanced the belief that ordered phases as the final products of aged U-Nb alloys may exist in the low-temperature region and that the metastable phase α" should decompose into these ordered phases.

To merge this fundamental knowledge gap, also inspired by the complex behavior of uranium compounds under high pressures [24-26], we conducted a thorough structural search of the U-Nb system up to 500 GPa. Several previously unknown ordered intermetallic phases have been discovered that exhibit strong stability, even at ambient pressure. In addition, the ordered phase proposed by Zhang et al. is proven to be unstable, spontaneously relaxing into a new structure with *Pmm2* symmetry that has higher energy than our newly discovered structures and is metastable. The discovery of these novel intermetallics as the final phases of aged U-Nb alloys significantly modifies the current phase diagram and lays a new foundation for further exploration of the aging mechanism of metastable U-Nb alloys. In the next section, the computational method is formulated. The results are presented in Section III. In Section IV, the results are discussed. Finally, Section V summarizes the main findings and conclusions of the study.

## II. Computational Method

First-principles calculations were performed using density functional theory (DFT) [27, 28] as implemented in the Vienna Ab initio Simulation Package (VASP) [29, 30]. The description of the exchange-correlation functional adopted the Perdew-Burke-Ernzerhof (PBE) [31] generalized gradient approximation (GGA). The atomic coordinates and lattice vectors were fully optimized until the Hellmann–Feynman forces acting on each atom less than $10^{-3}$ eV/Å and the convergence of the total energy was better than $10^{-7}$ eV between two successive ionic steps. The cut-off energy for the plane-wave basis was set to 550 eV, and Monkhorst-Pack [32] k-meshes with a grid spacing of $2\pi \times 0.015$ Å$^{-1}$ were used to sample the first irreducible Brillouin zone for all calculations, ensuring that the energy for each structure converges to 1 meV per atom or better. Phonon frequencies were calculated by using the supercell approach (with small displacements) as implemented in the PHONOPY code [33].

The lowest enthalpy candidate structures for U-Nb system up to 500 GPa were predicted by using particle swarm optimization method as implemented in CALYPSO [34], along with VASP. Elaborate structural searches were performed for five different composition stoichiometries (U:Nb = 3:1, 2:1, 1:1, 1:2, 1:3). Every composition has 30 structures in each generation of total 30 generations, and each structure contains up to 4 formula units per simulation cell. The 50 candidate structures with the lowest enthalpy in each composition were then carefully re-optimized with higher accuracy to pin down the most stable structure. The formation enthalpy per atom ΔH was calculated by following expression:

$$\Delta\mathrm{H}=(\mathrm{H}_{total}^{U_x\mathrm{Nb}_y}-y\mathrm{H}^{Nb}-x\mathrm{H}^{U})/(x+y)\,, \tag{1}$$

$$\text{and}\ \ \mathrm{H}=\mathrm{E}+PV\,, \tag{2}$$

with E, $P$, and $V$ stand for the total internal energy, pressure and volume, respectively. $\mathrm{H}_{total}^{U_x\mathrm{Nb}_y}$ is the total enthalpy per formula of the candidate ordered intermetallic phase in U-Nb system, with $x$ and $y$ represent the number of U and Nb atoms in that formula, respectively. $\mathrm{H}^{Nb}$ and $\mathrm{H}^{U}$ are the total enthalpy per atom of BCC-Nb [35] and $\alpha$-U [36], respectively. For pressure higher than 270 GPa where γ'-U is more stable than $\alpha$-U, $\mathrm{H}^{U}$ stands for the total enthalpy per atom of γ'-U [36].

Considering the strong local interactions between the 5*f* electrons of uranium, GGA+*U* is used to verify the results of structure prediction. In addition, the local density approximation (LDA) and LDA+*U* were also used to check the results of structural prediction, and the results were compared with those of GGA. Specifically, we used rotationally invariant DFT+*U* approach from Dudarev et al. [37], which only employs a single effective interaction. The results of these different methods further confirm the reliability of the new structures predicted using GGA.

Though the stoichiometries searched by CALYPSO already cover the commonly encountered actinide intermetallic compounds, in order to include all other possible stoichiometry and exhaust the structure search as complete as possible, the variable-chemical-composition genetic method for whole-composition-space structure search as implemented in USPEX [38] was also employed to generate another independent set of candidate structures. These structures were then compared with those given by CALYPSO. In this structure search, maximal number of atoms in the primitive cell is 20. The first generation includes 60 structures, and each subsequent generation contains 30 structures. A total of 30 generations were generated.

# III. Results

## A. Structural stability at zero Kelvin

To determine the thermodynamic stability of U-Nb system at low temperatures, the enthalpy convex hull of the predicted candidate structures was constructed and shown in Fig. 1. For all these structures, both $U_3Nb$ (approximately 100–200 GPa) and $U_2Nb$ (approximately 0–200 GPa) have a negative formation enthalpy, suggesting the existence of ordered intermetallics at the U-rich end. In contrast, all formation enthalpies at the Nb-rich end are positive, indicating that intermetallics in this composition range are unstable, and phase separation at low temperatures or solid solutions at high temperatures might be formed. Additionally, the variation trend of formation enthalpy with pressure is different for Nb-depleted or Nb-rich regimes. Particularly, in the Nb-rich regime, it first increases when the pressure goes from 0 to 100 GPa, decreases when the pressure is approximately 100–300 GPa, and increases again when the pressure is larger than 300 GPa. In the Nb-depleted regime, it decreases first (up to 200 GPa) and then increases rapidly to a high value with higher pressure. From the convex hull, it is evident that $U_2Nb$ is the most stable compound within the whole chemical composition. In contrast, $U_3Nb$ is sufficiently close to the convex hull and can be a metastable intermetallic.

The predicted $U_2Nb$ compound underwent several structural transitions when the pressure ranged from 0 to 500 GPa. Figure 2 presents the crystalline lattices of $U_2Nb$ and $U_3Nb$. The low-pressure $U_2Nb$ is ordered in a hexagonal $CaIn_2$-type [39] structure (hP6 in the Pearson's symbol) with a space group *P6₃/mmc* between -10 GPa (the lowest pressure we explored) and 21.6 GPa. It transforms into a hexagonal $AlB_2$-type [40] structure (hP3) with a space group *P6/mmm* when beyond 21.6 GPa. At pressures higher than 368 GPa, $U_2Nb$ became an orthogonal structure with *Fmmm* symmetry. In contrast, $U_3Nb$ has an orthogonal structure with *Cmcm* symmetry, and no other competitive structure was discovered between 0 and 500 GPa. It should be emphasized that the ordered $U_3Nb$ structure proposed by Zhang et al [19] is in fact unstable. It spontaneously changes into a new structure with a space group of *Pmm2*, which is a distorted $Cu_3Au$-type ($L1_2$) structure. This relaxed structure, however, has a formation enthalpy per atom approximately 0.2 eV higher than that of our newly discovered *Cmcm*-$U_3Nb$ and is thermodynamically unstable. The phonon dispersions of $U_2Nb$ and *Cmcm*-$U_3Nb$ were calculated to examine their dynamic stability, and no imaginary modes were found, confirming that they are all dynamically stable. This result also implies that *Cmcm*-$U_3Nb$ is metastable, although it is not on the convex hull. Note that these structures have distinct atomic environments and can be easily detected through X-ray diffraction experiments [41].

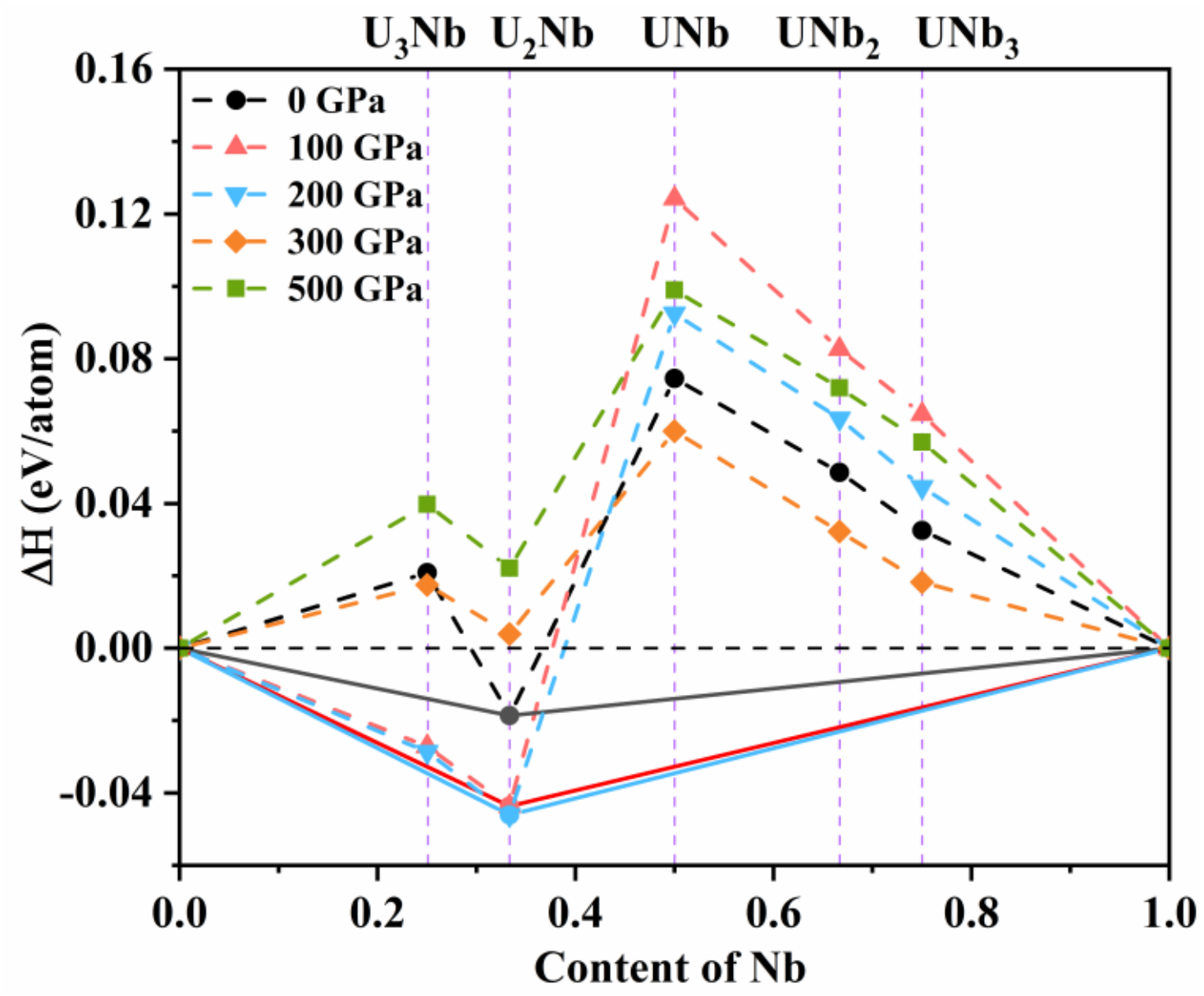

**Fig. 1** (color online) Convex hulls of predicted structures at different pressures. Solid lines connect thermodynamically stable phases at the given pressure, and dashed-line connections represent instable or metastable phases.

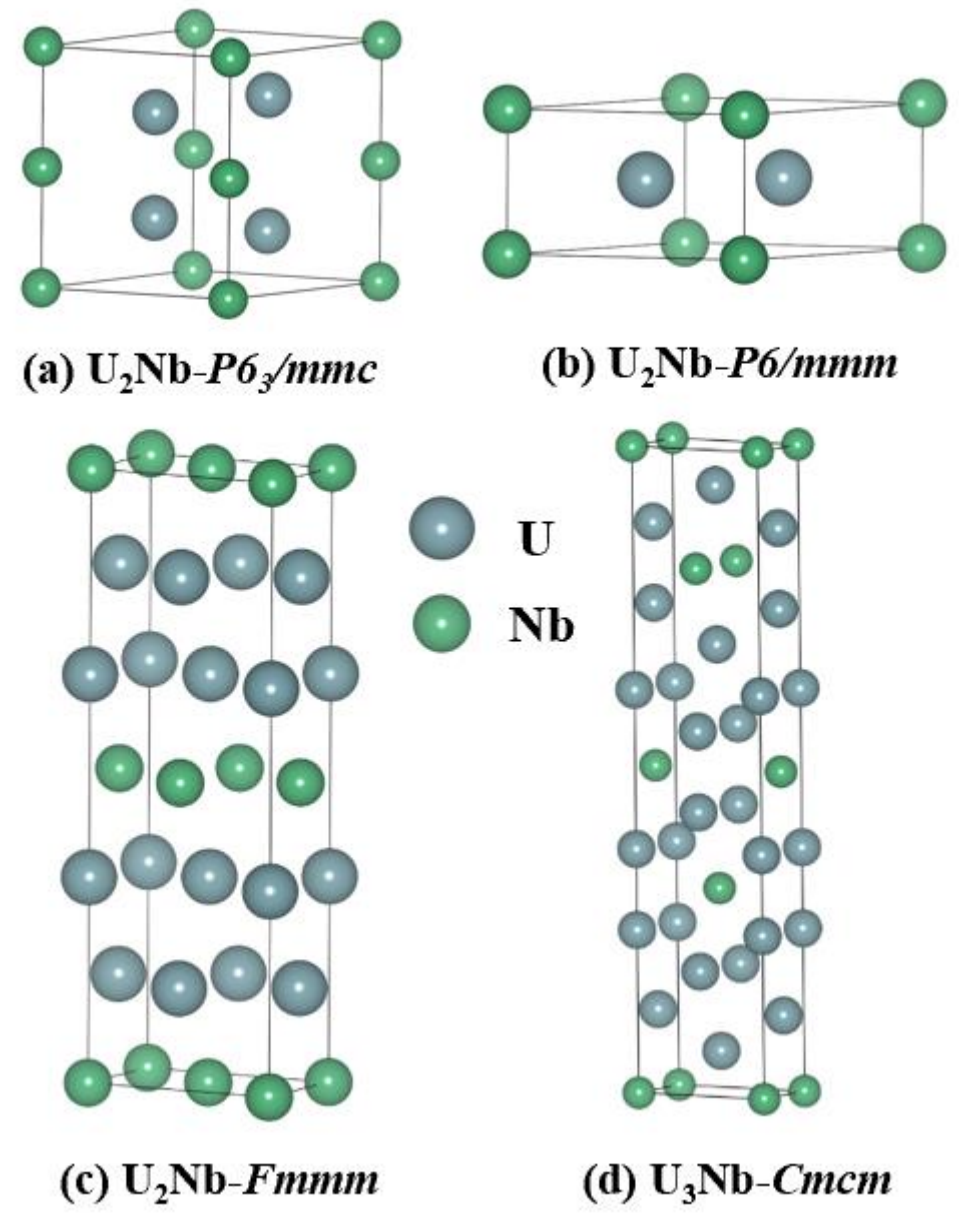

**Fig. 2** (color online) Crystal structures of $U_2Nb$ in the phase of (a) *P*$6_3$*/mmc* (being stable up to approximately 21.6 GPa), (b) *P6/mmm* (stable within approximately 21.6–368 GPa) and (c) *Fmmm* (beyond 368 GPa). (d) The predicted *Cmcm* structure of $U_3Nb$.

Considering the strong local interactions between the 5*f* electrons of uranium, we also use GGA+*U* to conduct structural search of the U-Nb system at 0 and 100 GPa to verify the reliability of the GGA results. The corresponding results are shown in Fig. 3(a), along with those of GGA as comparison. When GGA+*U* is considered, the convex hull of formation enthalpy has an obvious downward trend under the same pressure, especially in the niobium-rich region. Notably, $U_2Nb$ is still located on the convex hull, and $U_3Nb$ is close to the convex hull at 100 GPa. Moreover, the new phases predicted using GGA+*U* are *P6₃/mmc*-$U_2Nb$ at zero pressure, *P6/mmm*-$U_2Nb$ at 100 GPa, and metastable *Cmcm*-$U_3Nb$ at 100 GPa, which is consistent with the results obtained using GGA. In addition to the GGA exchange correlation function, we also used the LDA and LDA+*U* to check the results of structure prediction, as shown in Fig. 3(b). The results of both LDA and LDA+*U* indicate that $U_2Nb$ lies on the convex hull at 0 and 100 GPa, and $U_3Nb$ is close to the convex hull at 100 GPa, which is consistent with the results of GGA and GGA+*U*. Furthermore, convex hulls of different *U* values (*U*= 0, 0.5, 1.0, 1.5, 2.0 eV) at 0 and 100 GPa in Fig. 4 further confirm the reliability of our structural prediction. As the *U* value increases, the convex hulls at both 0 and 100 GPa show an obvious downward trend. For the convex hulls of all *U* values at 0 and 100 GPa, $U_2Nb$ is always the most stable phase. $U_3Nb$ is close to the convex hulls of all *U* values at 100 GPa. In general, the calculated results of DFT+*U* further confirm the reliability of the structure prediction using GGA.

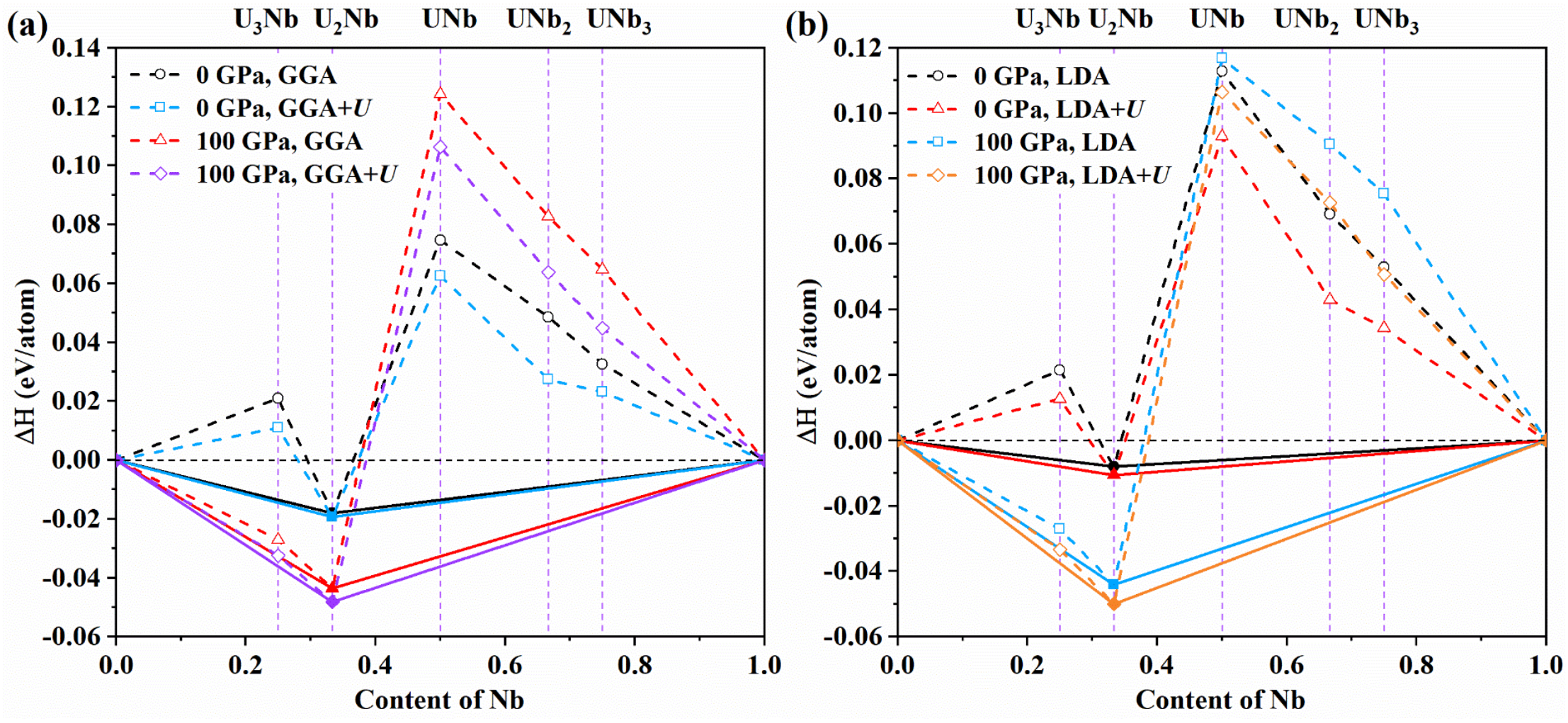

**Fig. 3** (color online) Convex hulls of predicted structures at 0 and 100 GPa using (a) GGA and GGA+*U* and (b) LDA and LDA+*U* (*U*=1.0 eV). Solid lines connect thermodynamically stable phases at the given pressure, and dashed-line connections represent instable or metastable phases.

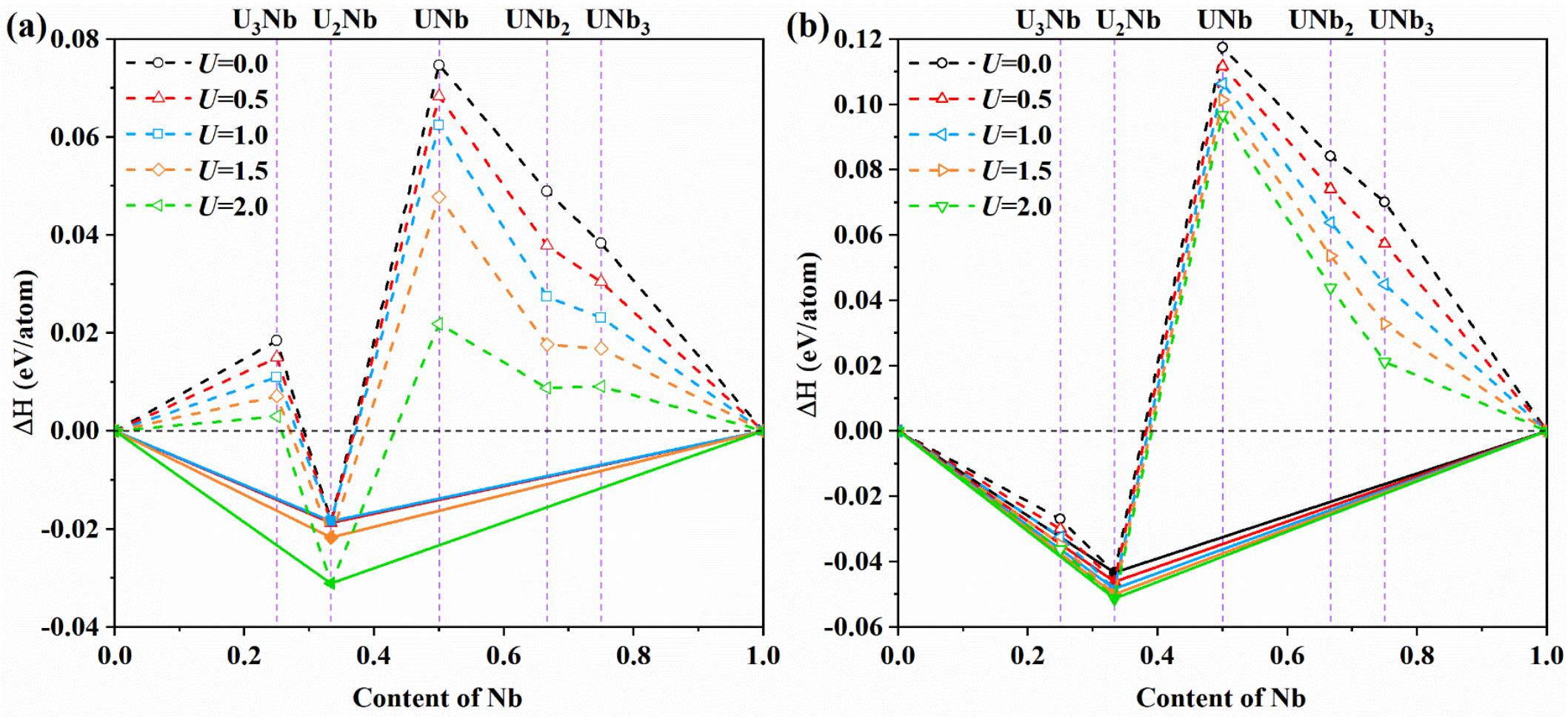

**Fig. 4** (color online) Convex hulls of different *U* values calculated by GGA+*U* (*U*= 0,

0.5, 1.0, 1.5, 2.0 eV) at (a) 0 GPa and (b) 100 GPa. Solid lines connect thermodynamically stable phases at the given pressure, and dashed-line connections represent instable or metastable phases.

In order to confirm the reliability of the ordered phases of the U-Nb system predicted by CALYPSO, we also performed another independent structural search at 0 and 200 GPa using variable-chemical-composition genetic method as implemented in USPEX. Results of the variable-composition structure prediction are completely consistent with those of CALYPSO, as displayed in Fig. 5. The structure with the lowest enthalpy is *P6₃/mmc*-$U_2Nb$ at zero pressure and *P6/mmm*-$U_2Nb$ at 200 GPa. Metastable *Cmcm*-$U_3Nb$ are also predicted at 200 GPa. This cross-check guarantees the completeness of our structure search.

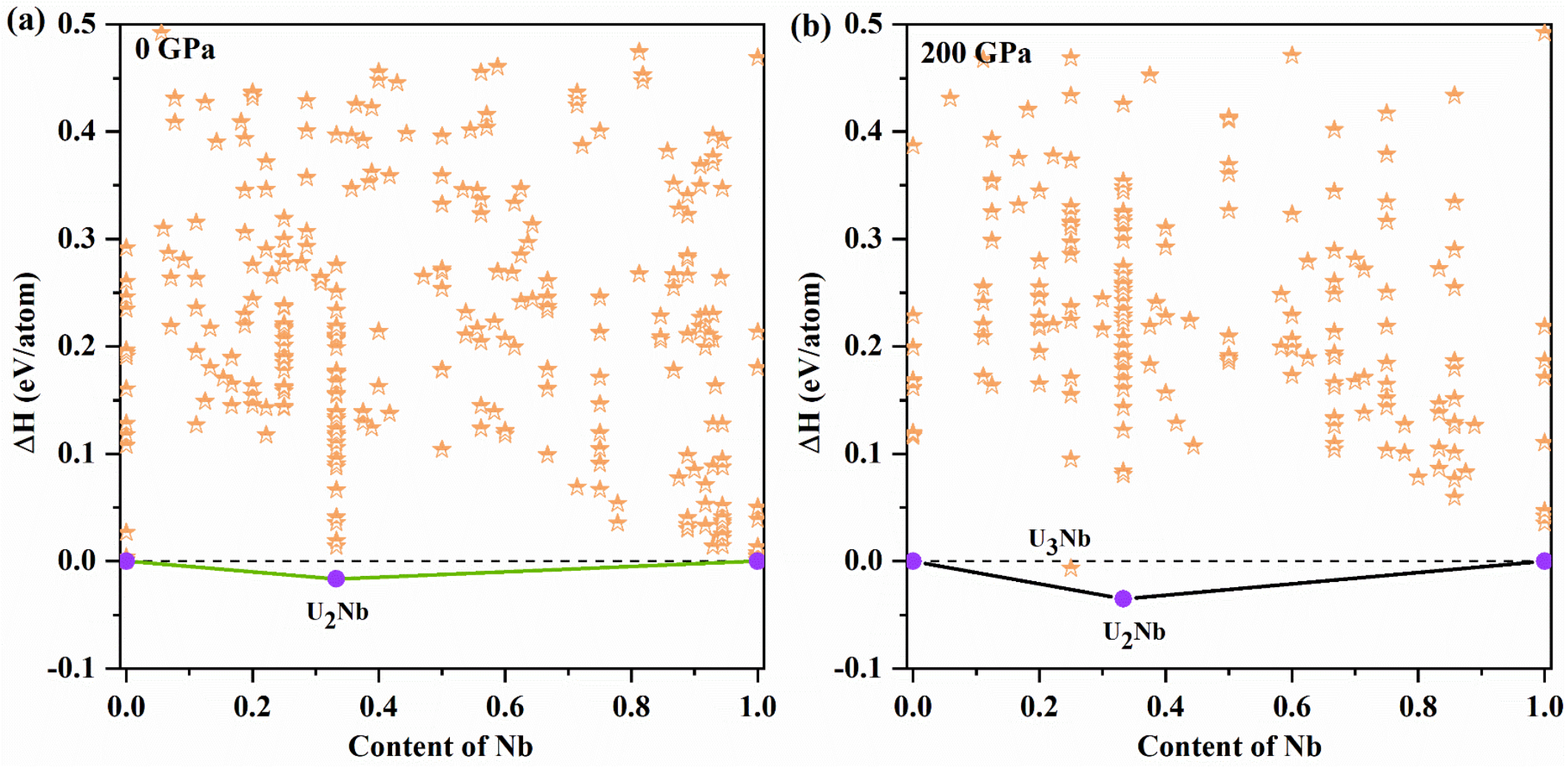

**Fig. 5** Results of variable-composition structure search using USPEX at (a) 0 GPa and (b) 200 GPa.

Relationships between the newly predicted structures can be described by their interatomic separation histograms, as shown in Fig. 6. The shortest U-U separation in *P6₃/mmc*-$U_2Nb$ phase is 2.58 Å, which is slightly shorter than the 2.71 Å of α-U, suggesting a stronger interaction between U atoms in *P6₃/mmc*-$U_2Nb$. If we do not distinguish U and Nb, the averaged local atomic environment of U atoms in *P6₃/mmc*-$U_2Nb$ is similar to that of α-U, in which both have four atoms with a separation distance less than 3 Å as the nearest neighboring shell; just above 3 Å, α-U has eight atoms in the second shell, while *P6₃/mmc*-$U_2Nb$ has seven atoms; at longer distances from 5 to 6 Å, α-U has 26 atoms, and *P6₃/mmc*-$U_2Nb$ has 29 atoms. This structural resemblance may be one of the reasons for the stabilization of the intermetallic $U_2Nb$ phase. On the other hand, compared to *P6₃/mmc*-$U_2Nb$, the high-pressure *P6/mmm*-$U_2Nb$ phase has a higher symmetry. The former is derived from the latter by spontaneous symmetry breaking, which is reflected in the shell splitting in the U-Nb separation from the U atom just above 3, 5, and 5.75 Å, whereas the Nb-Nb separations only slightly change. Notably, the U-U separation and its neighboring count also change. In addition, both histograms of the interatomic separation from the Nb atom for *P6₃/mmc*-$U_2Nb$ and *P6/mmm*-$U_2Nb$ exhibit a big shell gap between 3.2 and 4.8 Å, which is similar to BCC-Nb [41].

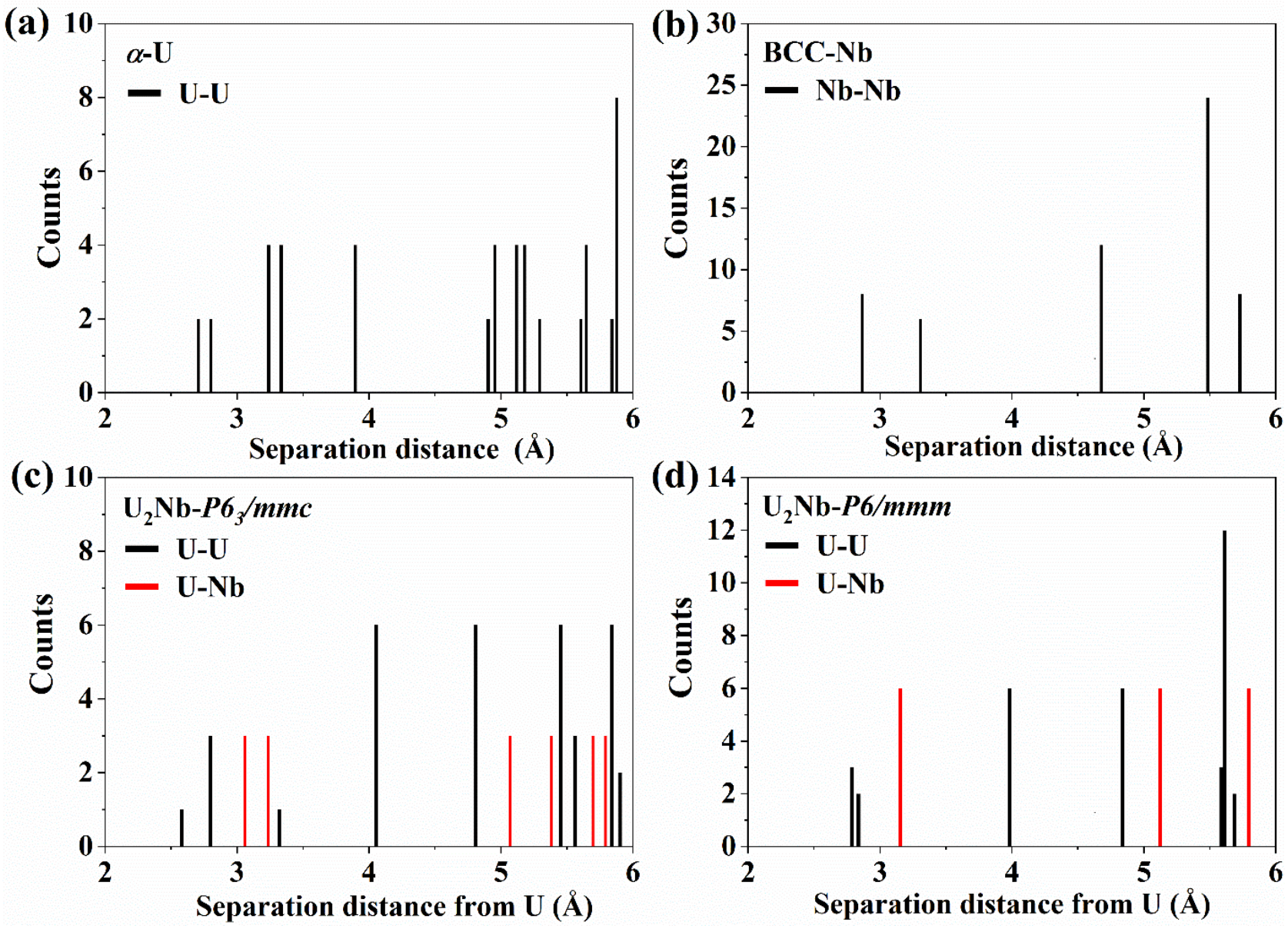

**Fig. 6** (color online) Histograms of interatomic separations for (a) *α*-U, (b) BCC-Nb, (c) *P6₃/mmc*-$U_2Nb$, and (d) *P6/mmm*-$U_2Nb$ at zero pressure.

## B. Electronic structure

All stable and metastable intermetallics obtained are metallic within the explored pressure range, with U-5*f* electrons dominating the electronic structure near the Fermi level, as shown by the band structure and density of state (DOS) in Fig. 7(a). For all of them, only the high-pressure *Fmmm*-$U_2Nb$ showed a strong presence of U-6*d* electrons around the Fermi level [41]. In all of these structures, the contribution of Nb-4*d* to the total DOS is very small, suggesting a prominent charge transfer from Nb-4*d* to U-5*f* orbitals. The residual Nb-4*d* electrons slightly hybridize with the U-6*d* or U-5*f* states. The magnitude of charge transfer between the U and Nb atoms was characterized by Bader charge analysis [42]. In Fig. 7(b), the average valence state of U (Nb) is increased from -0.11 (+0.22) at 0 GPa to -0.33 (+0.64) at 368 GPa. The average valence state of U (Nb) is -0.18 (+0.35) when the pressure is higher than 368 GPa, which slightly increases with pressure within the *Fmmm* phase. Evidently, these intermetallic ordered phases are weakly ionic, and their vibrational modes can be infrared-active.

The differential charge density shown in Figs. 7(c) and (d) reveals the electron redistribution between U and Nb atoms, of which particular interest is that in the high symmetric *P6/mmm*-$U_2Nb$ phase, the charge is transferred between neighboring U atoms and formed weak σ bonds that resemble the hexatomic ring in graphene. Weak bonds between the adjacent U-layers were also observed. Similar σ bonds between metallic atoms with 4*f* electrons have also been previously reported in compressed $CeO_2$ [43]. It should be noted that *Cmcm*-$U_3Nb$ also possesses such kind of intralayer U-ring and interlayer U-U bonding [41]. However, electron localization function analysis shows that these covalent bonds should be very weak. These features show that the high-pressure U-Nb ordered phases are metallic compounds governed by delocalized 5*f* electrons, exhibiting weak ionic and covalent interactions simultaneously.

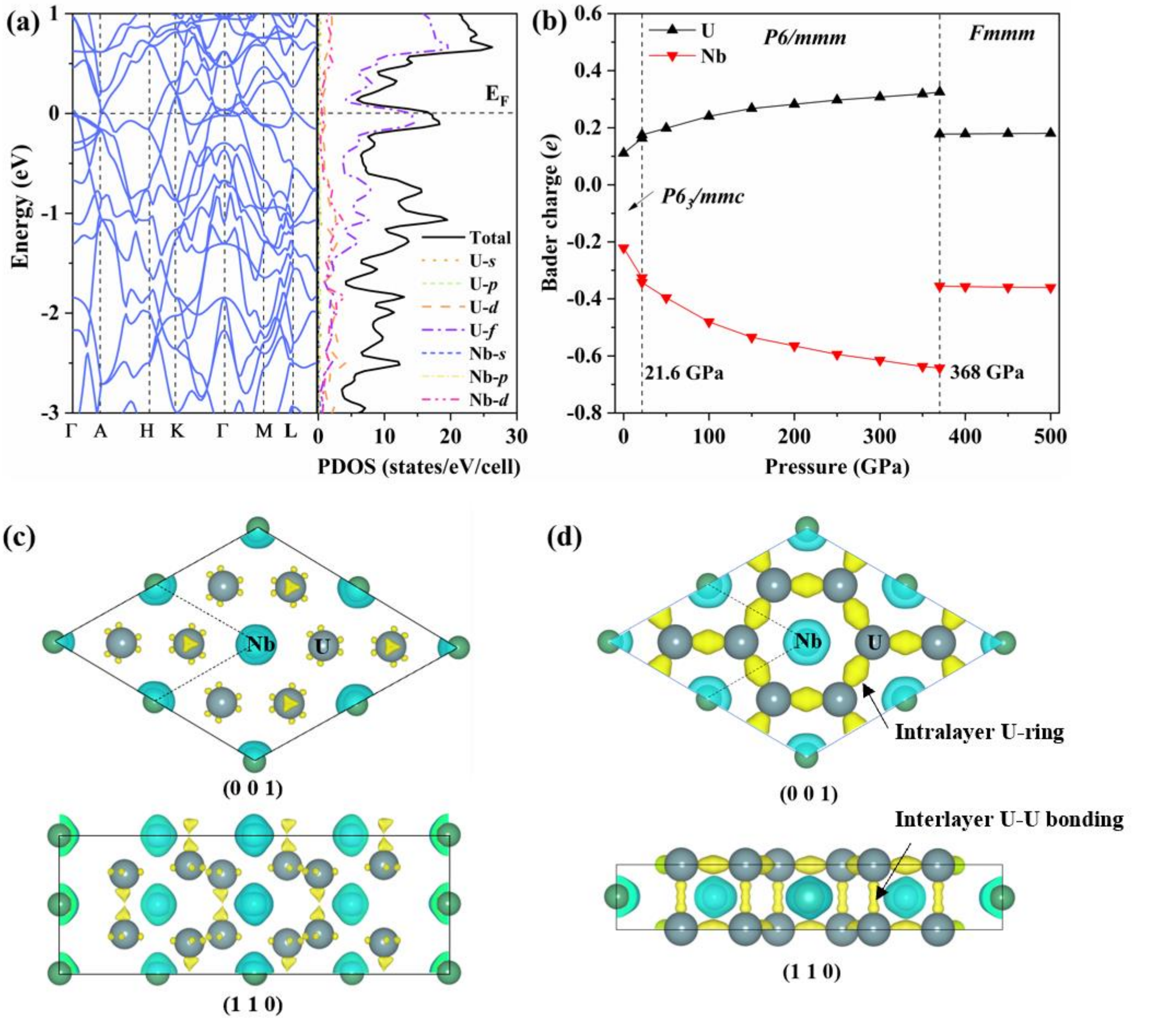

**Fig. 7** (color online) (a) Electronic band structure and projected DOS of *P6₃/mmc*-$U_2Nb$ at zero pressure. (b) The averaged Bader charge of U and Nb atoms in each $U_2Nb$ phase. (c) Differential charge density (isosurface = 0.02 *e*/bohr$^3$) of *P6₃/mmc*-$U_2Nb$ at 0 GPa, and (d) *P6/mmm*-$U_2Nb$ at 200 GPa. The blue color indicates the loss of electrons from Nb atoms and the yellow color represents the grain of electrons between U atoms.

## C. Equation of state

The equation of state (EOS) of $U_2Nb$ at 0 K was calculated and compared with those of $\alpha$-U and BCC-Nb up to 100 GPa, as shown in Fig. 8. The DFT-calculated EOS for $U_2Nb$ exhibits a sharp volume collapse of 0.96% at a transition pressure $P_t$ of 21.6 GPa, a signature of a first-order structure transition. Similarly, there was also a volume collapse of 1.59% at 368 GPa when *P6/mmm*-$U_2Nb$ transformed into *Fmmm*-$U_2Nb$.

To quantify the effect of ordering on the EOS of $U_2Nb$, the compression curve of U-Nb solid solution is also estimated by using the ideal mixing model [44] and the EOS of $\alpha$-U and BCC-Nb. Both the EOS curve of this model and the direct DFT-calculated EOS curve are plotted in Fig. 8 for comparison. At the composition of U:Nb = 2:1, the EOS of the solid solution is roughly in line with the ordered phase. The deviation was approximately 1.42% at 0 GPa. The maximal deviation was approximately 1.58% within a wide pressure range of 0–500 GPa [41]. The EOS data below 21.6 GPa (for the low-pressure *P6₃/mmc*-$U_2Nb$ phase) can be fitted to a Vinet EOS model, which gives an equilibrium volume ($V_0$) of 19.69 Å$^3$ per atom, bulk modulus ($B_0$) of 141 GPa, and pressure derivative of B (B') of 3.08 GPa at zero pressure. For the ideal mixing model of the solid solution with a chemical composition of $U_2Nb$, the corresponding EOS parameters are $V_0$ (19.43 Å$^3$ per atom), $B_0$ (148 GPa), and B' (4.80 GPa) at zero pressure).

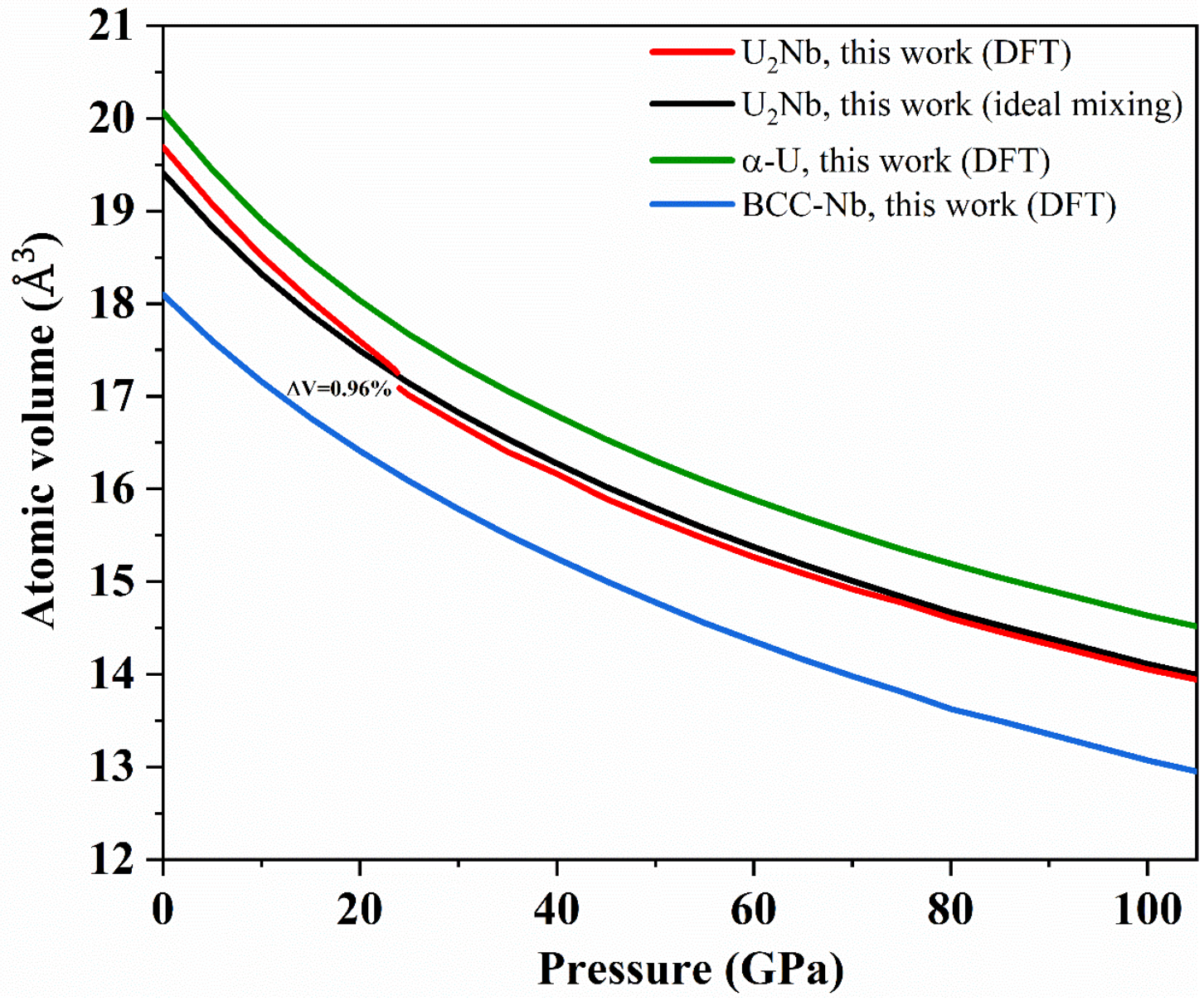

**Fig. 8** (color online) Calculated equation of state of $U_2Nb$, α-U, and BCC-Nb at 0 K. The hypothetical U-Nb solid solution with a composition of U:Nb=2:1 as estimated by using the ideal mixing model is also shown for comparison.

# IV. Discussion

## A. Unique phase transition in $U_2Nb$

The intermetallic ordered phase $U_2Nb$ is thermodynamically stable up to 294 GPa [41] and could be metastable up to (or beyond) 500 GPa, within which it undergoes several structure transitions. Figure 9 shows the enthalpy differences of the *$P6_3/mmc$* and *Fmmm* phases with respect to the intermediate-pressure phase *P6/mmm* as a function of pressure. Evidently, the transition from the *$P6_3/mmc$* to the *P6/mmm* phase occurs at 21.6 GPa. The most interesting feature of this transition is that the low-pressure phase *$P6_3/mmc$* has very little hysteresis (i.e., over-pressuring) beyond the transition pressure $P_t$. In contrast, the high-pressure phase *P6/mmm* can survive and extend to very low pressure metastably. The enthalpy curves in Fig. 9 show a remarkable half-cross, as indicated by the dashed-line of the *$P6_3/mmc$* branch beyond $P_t$, where it becomes completely unstable and spontaneously collapses to the *P6/mmm* phase. The curved arrow in Fig. 9 illustrates this kind of "*semi-continuous*" change in the structure, which is a typical feature of a continuous transition. On the other hand, the reverse transition from *P6/mmm* back to *$P6_3/mmc$* exhibits a typical hysteresis of a first order transition. In this sense, this transition is an ideal hybridization of the first-order and second-order transitions and has distinct features of both. To the best of our knowledge, this should be the first report of this kind of unique phenomenon, and we would like to nominally term it as the "$1\frac{1}{2}$-order" transition to differentiate it from other mixing transitions of first-order and second-order, where the first-order feature is significantly weakened, such as that occuring in compressed vanadium [45]. Further compression of *P6/mmm* will transform into the *Fmmm* phase at approximately 368 GPa, for which the enthalpy curve exhibits a clear intersection and is a typical first-order transition.

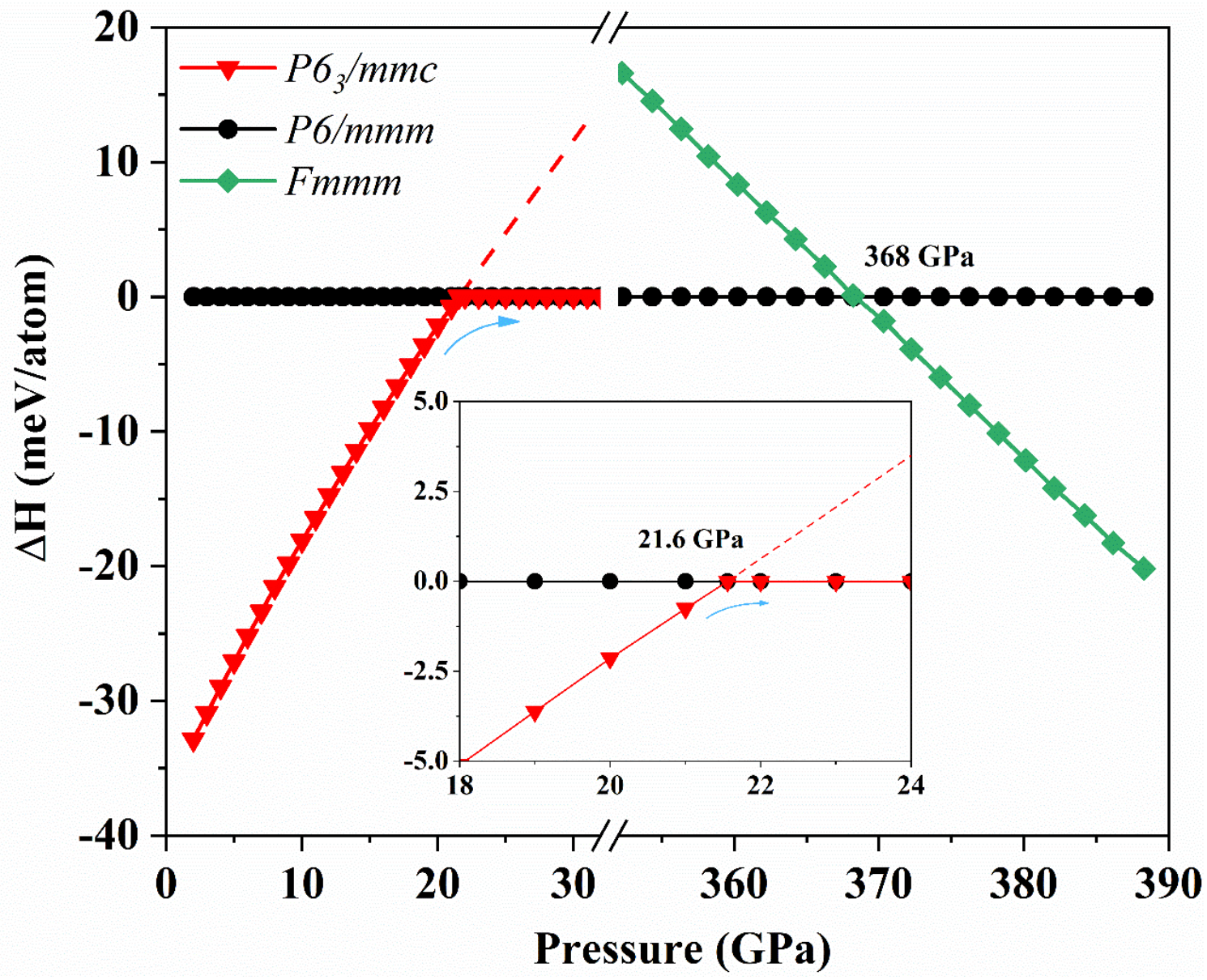

**Fig. 9** (color online) Enthalpy difference of $U_2Nb$ in *P6₃/mmc* and *Fmmm* phases with respect to the *P6/mmm* phase. The dashed-line indicates the region where *P6₃/mmc* becomes spontaneously unstable.

The variation in the lattice constant of $U_2Nb$ with pressure is shown in Fig. 10 to elucidate the transition mechanism. We find that the lattice vector lengths *a* and *c* take distinct values when $P < P_t$ for *P6₃/mmc* and *P6/mmm*. Moreover, they collapse into that of *P6/mmm* when beyond $P_t$. The *c/a* ratio clearly illustrates this collapse behavior near the transition point. It first gradually decreases and then approaches that of *P6/mmm* at a faster rate when it is closer to $P_t$. However, this variation in *c/a* is very similar to the order parameter of a second-order transition, except for the sudden jump at $P_t$, which is a characteristic of a first-order transition. Therefore, the hybridization nature of this structural change is unequivocal.

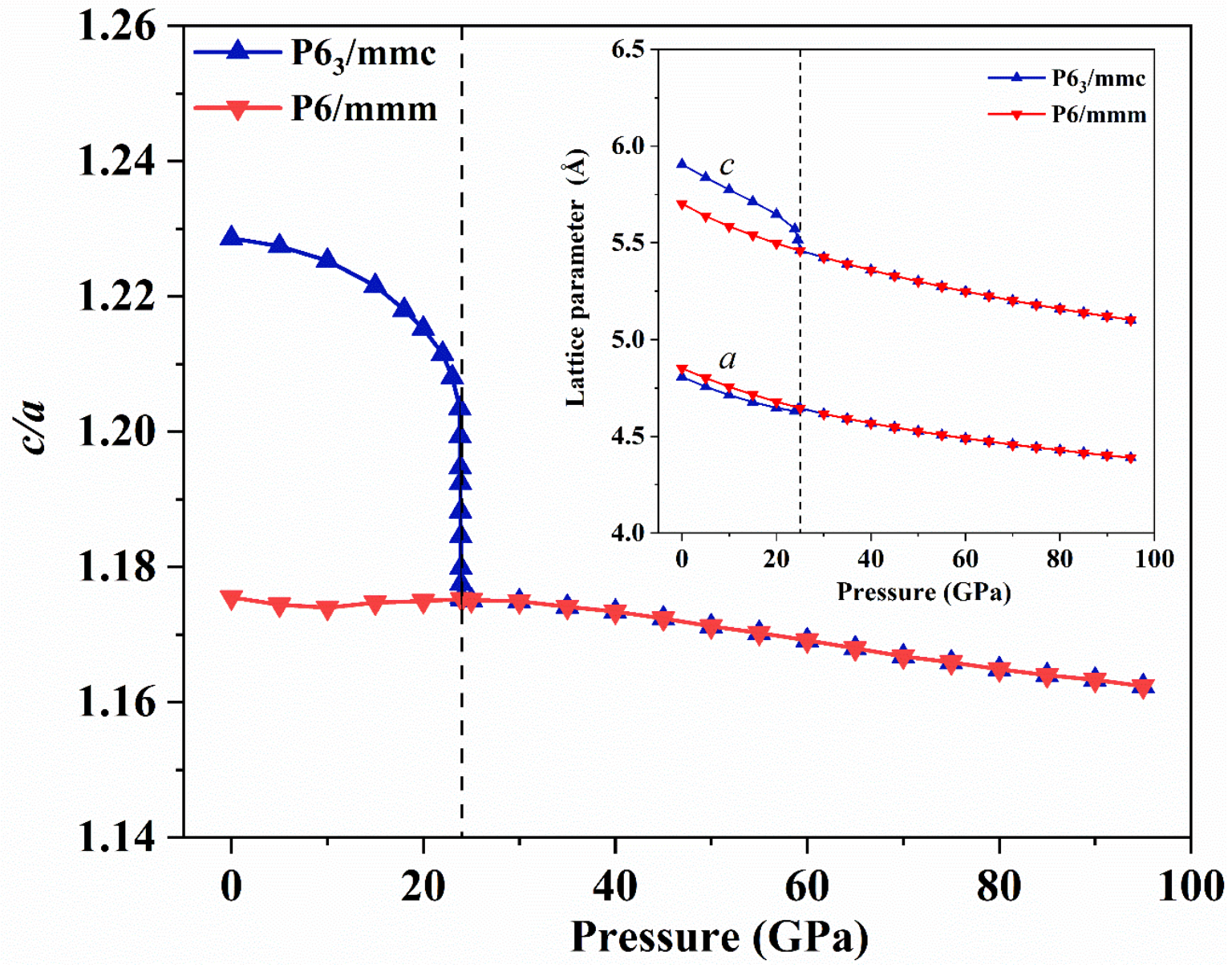

**Fig. 10** (color online) Variation of lattice vectors *a* and *c*, and their ratio *c/a* with pressure in *P6₃/mmc*-$U_2Nb$ and *P6/mmm*-$U_2Nb$ phases, respectively. The dashed-vertical line indicates the transition pressure.

According to the relationship between the group and subgroup, *P6₃/mmc* is a Klassengleiche subgroup of *P6/mmm* with a doubled basis vector *c* [46, 47]. They belong to the same crystallographic class, with the U atoms of adjacent layers in the *P6₃/mmc* phase moving in the direction of [0 0 1] to increase symmetry and form the *P6/mmm* phase. This also determines the transition path, and *P6₃/mmc* loses translational symmetry. The Klassengleiche group-subgroup transition is usually associated with a second-order continuous phase transition. This is in line with the observations shown in Figs. 9 and 10, demonstrating that there is a significant portion of the second-order transition in the *P6₃/mmc-P6/mmm* structural change. Furthermore, in a typical second-order transition, there is a soft mode associated with the breaking of symmetry [48]. Indeed, a soft mode in the *P6/mmm* phase can be observed at the high symmetry point *A* in the irreducible Brillouin zone when the pressure is decreased to below $P_t$ with the lattice parameter *c/a* along the transition path in Fig. 10, as shown in Fig. 11. This vibrational soft mode describes the internal rearrangement of the atoms during the pressure-driven transition.

Since second order transition is usually accompanied with divergent higher-order derivatives of the free energy, we also examined the variation of the elastic constants and sound velocity in $U_2Nb$. All elastic constants show a clear jump at $P_t$ = 21.6 GPa [41]. Particularly, the low-pressure branch of $C_{33}$ exhibits a divergent signature when approaching $P_t$. We can extend this curve up to 24 GPa, which is the metastable limit of *P6₃/mmc*, where $C_{33}$ becomes completely divergent and *P6₃/mmc* spontaneously collapses into a highly symmetric *P6/mmm* structure. The divergence in $C_{33}$ indicates that this transition also has a feature of λ transition (a type of continuous transition), which is terminated prematurely by the first-order change at 21.6 GPa and is buried in the domain of *P6/mmm*. This type of divergence is also reflected in the sound velocity, as evidenced by the variation in the shear ($C_s$) and longitudinal ($C_l$) sound velocities with pressure, as shown in Fig. 12. The sharp volume collapse, soft vibrational mode, and discontinuous jump and divergence in elastic constants and sound velocity confirm the hybridization nature of the first-order and second-order transitions for this unique *P6₃/mmc* to *P6/mmm* phase transition.

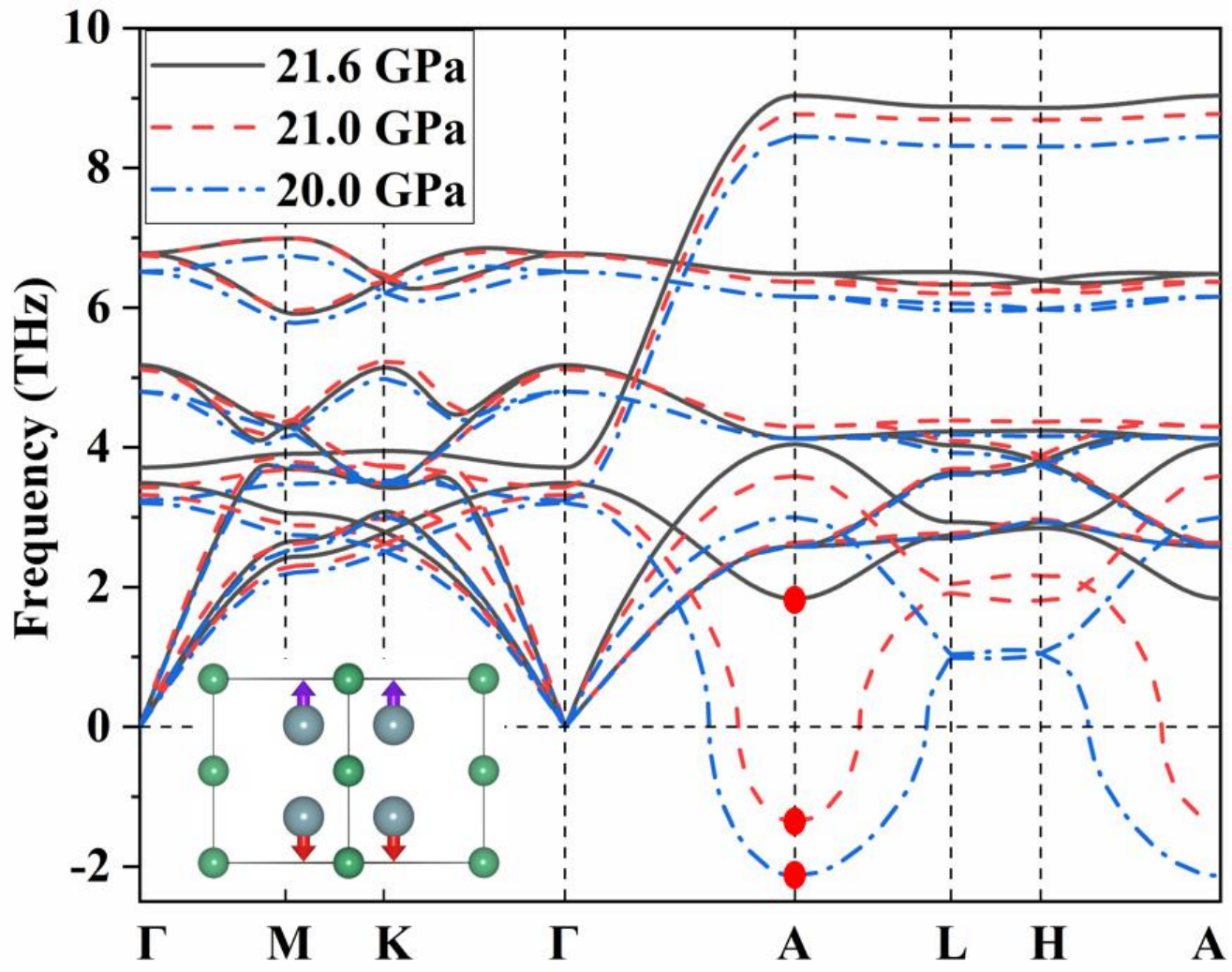

**Fig. 11** (color online) Calculated phonon spectra of *P6/mmm*-$U_2Nb$ at 20.0, 21.0 and 21.6 GPa. The vibrational soft mode at high symmetry point *A* (0, 0, 1/2) is marked by the solid circles and shown in the inset.

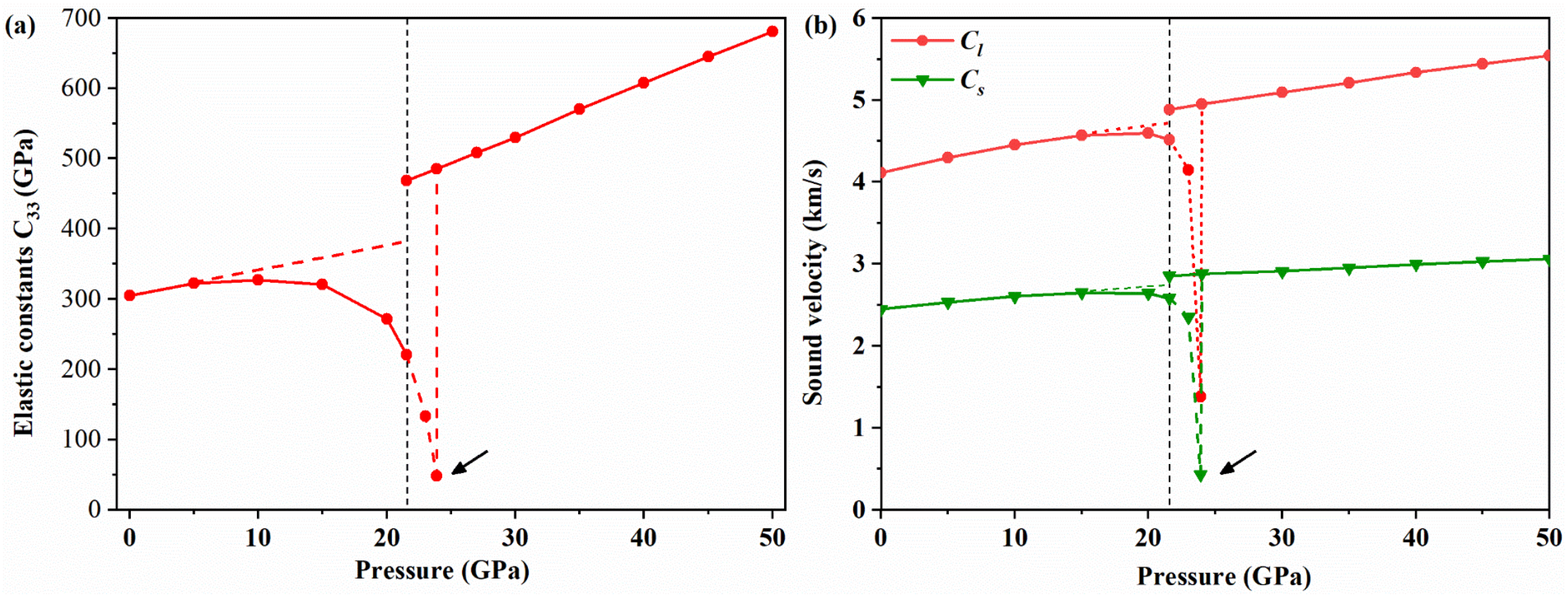

**Fig. 12** (color online) Variation of (a) elastic constant $C_{33}$ and (b) shear ($C_s$) and longitudinal ($C_l$) sound velocity with pressure. The vertical dotted-line indicates the first-order transition pressure, whereas the arrows mark the underlying λ transition buried by the former and exhibits divergency in the elastic constant $C_{33}$ and sound velocities.

## B. Landau model for phase transition in $U_2Nb$

Subsequently, we show that this hybridized transition can be described using Landau theory. As mentioned above, the lattice parameter *c/a* is strongly coupled with the internal coordinates of the U atoms. In principle, any of these can be taken as the order parameter. However, since the internal coordinates are directly related to the symmetry change, we chose them as the order parameter to describe the hybridized transition process. The enthalpy differences as a function of the order parameter, at 21.0, 21.6 and 22.0 GPa, were calculated and are shown in Fig. 13. Here, the order parameter *d* is defined as the deviation of the Z coordinate of the U atoms in the *P6₃/mmc* phase from that of the *P6/mmm* phase, as shown in Fig. 14.

Compared to a typical double well potential energy surface (PES), a triple well PES is observed, and the potential energy barrier is approximately 7.8 meV at a $P_t$ of 21.6 GPa. This PES can be expanded around the transition point using pressure as the driven thermodynamic variable. Thus, the obtained Landau enthalpy model for the phase transition is [49]:

$$\mathrm{H}(P,\mathrm{d})-\mathrm{H}(P,0)=a_0(P-P_0)d^2-b_0d^4+\mathrm{c}_0d^6, \tag{3}$$

where H(*P*, 0) is the enthalpy of the ground state (with $d$=0) at a given pressure *P*, and $a_0$, $b_0$, $c_0$ and $P_0$ are positive constants. Notably, $P_0$ is the absolute low-pressure instability limit for the high-pressure phase, rather than the transition pressure. The sixth-order term in Eq. (3) is necessary to stabilize the system. For a thermodynamically stable system, the transition occurs when

$$\left(\frac{\partial \mathrm{H}}{\partial d}\right)_P=2a_0(P-P_0)d-4b_0d^3+6\mathrm{c}_0d^5=0, \tag{4}$$

and H(*P*, *d*)=H(*P*, 0) is simultaneously satisfied. The stable domain of the phases is defined by

$$\left(\frac{\partial^2 H}{\partial d^2}\right)_P=2a_0(P-P_0)-12b_0d^2+30\mathrm{c}_0d^4=0. \tag{5}$$

According to this model, the low-symmetric phase has an order parameter of

$$d^2=(b_0/3c_0)[1+\sqrt{1-3a_0(P-P_0)c_0/b_0^{\,2}}], \tag{6}$$

which decays to $\frac{|b_0|}{2c_0}$ at a rate of $(P-P_0)^{1/4}$. This rate of 1/4 is half of a second-order transition in mean field theory, and the critical order parameter at the transition pressure is $\pm\left|\frac{\mathrm{b}_0}{2c_0}\right|^{1/2}$.

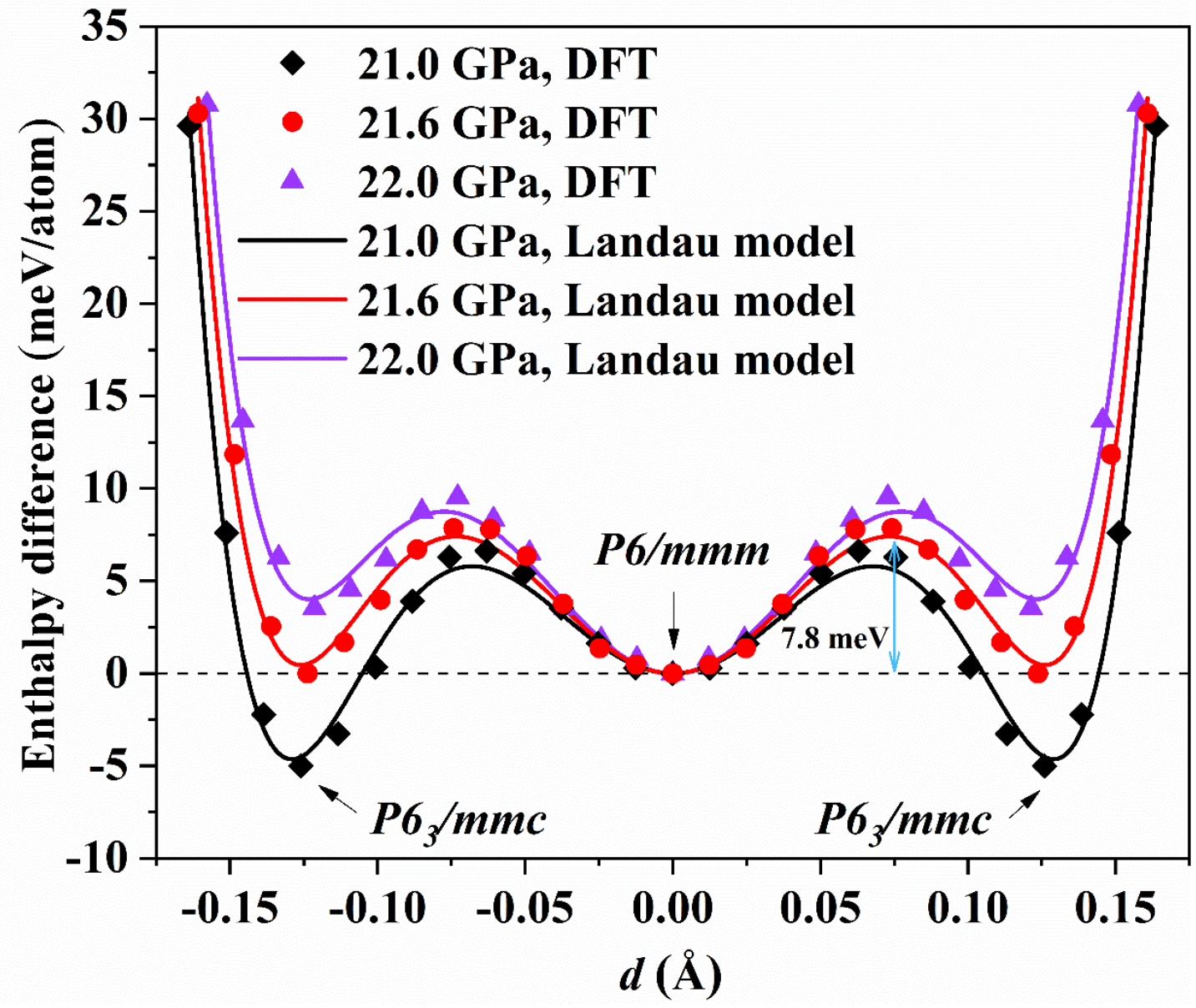

**Fig. 13** (color online) Performance of the Landau model against the direct DFT data at a pressure of 21.0, 21.6, and 22.0 GPa.

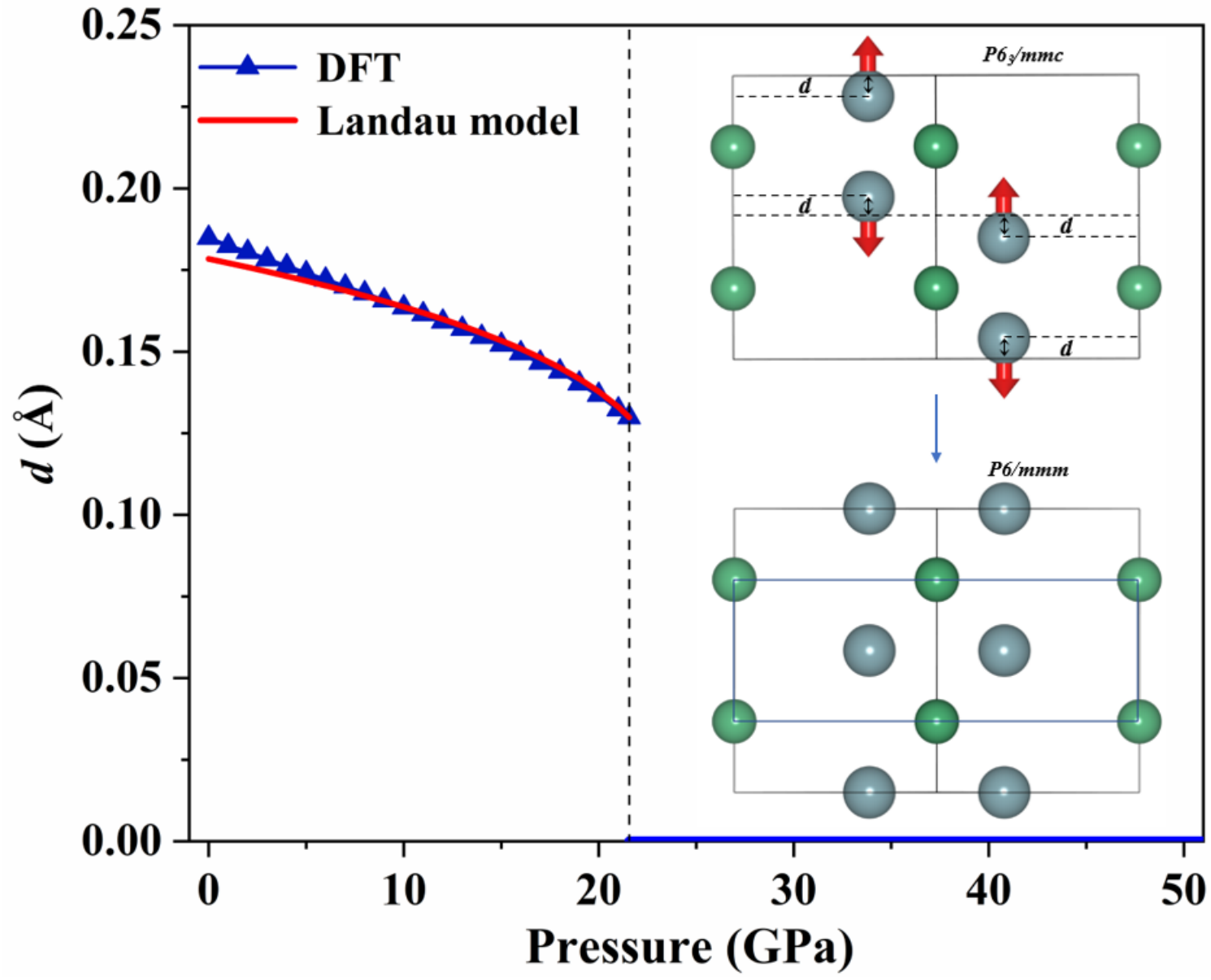

**Fig. 14** (color online) Evolution of the order parameter *d* from structure *P6₃/mmc*-$U_2Nb$ to *P6/mmm*-$U_2Nb$ as a function of pressure. The discontinuous jump at the transition pressure is evident. Inset schematics illustrate the atomic rearrangement during the structural transition, in which bold-line box represents the primitive cell of *P6/mmm*-$U_2Nb$.

The parameters of this model can be determined by fitting the enthalpy model to the direct DFT data. The obtained PES as a function of the order parameter is shown as a curve in Fig. 13. The variation in the order parameter with pressure is expressed in Eq. (6) and is compared to the direct DFT data in Fig. 14. The order parameter first decreased slowly and then jumped to zero at $P_t$ when approaching the transition pressure $P_t$ from the low-pressure end, revealing the termination of a continuous transition by a sharp first-order change. The good agreement between Eqs. (3) and (6) and the DFT data confirms the validity of the Landau model for the pressure-driven *P6₃/mmc* to *P6/mmm* transition in $U_2Nb$. This model predicts an enthalpy barrier of $E_b$ = 7.4 meV, the position of the barrier $d_b$ = 0.076 Å, and the critical order parameter $d_c$ = 0.132 Å, which are compared to the direct DFT results of $E_b$ = 7.8 meV, $d_b$ = 0.075 Å, and $d_c$ = 0.130 Å. It also predicts an over-compression limit of 24.05 GPa, and an under-compression limit of 13.99 GPa. Furthermore, based on this mathematical model, there will be a tricritical point when $b_0$=0, and it will become a second-order transition when $a_0$ reverses its sign.

## C. Thermal stability and decomposition limit of $U_2Nb$

The previous sections discuss phase stability and structural transition at absolute zero. Now, we turn to chemical stability at finite temperature. The contribution of lattice vibrations to the Gibbs free energy was considered using the quasi-harmonic approximation method. The calculated formation Gibbs free energy (ΔG) of $U_2Nb$ under pressure is shown in Fig. 15 at temperatures of 0, 300 and 600 K. The thermal stability of ordered intermetallic $U_2Nb$ decreases as the temperature increases. At zero pressure, $U_2Nb$ decomposes into α-U and BCC-Nb or their solid solution at approximately 520 K, which is consistent with previous experiments, showing that the miscibility zone in U-Nb alloy is above 600 K [50]. At the lowest pressure (-10 GPa) we investigated, the decomposition temperature increases to 890 K. Here, the decomposition temperature is defined as the temperature at which the formation Gibbs free energy becomes zero at a given pressure. It is the lower bound of the stability of the ordered phase, since it might still exist at higher temperature metastably. Notably, the ordered $U_2Nb$ is not thermodynamically stable and might segregate into other phases between 14 and 29 GPa and beyond 294 GPa. On the other hand, *P6/mmm* phase of $U_2Nb$ has the strongest stability, and its decomposition temperature is greater than 4410 K at 150 GPa. We also noticed that the stability of $U_3Nb$ was also weakened at high temperatures [41]. It can metastably exist as a mixture of $U_2Nb$, α-U and BCC-Nb at low temperatures (approximately 25 GPa).

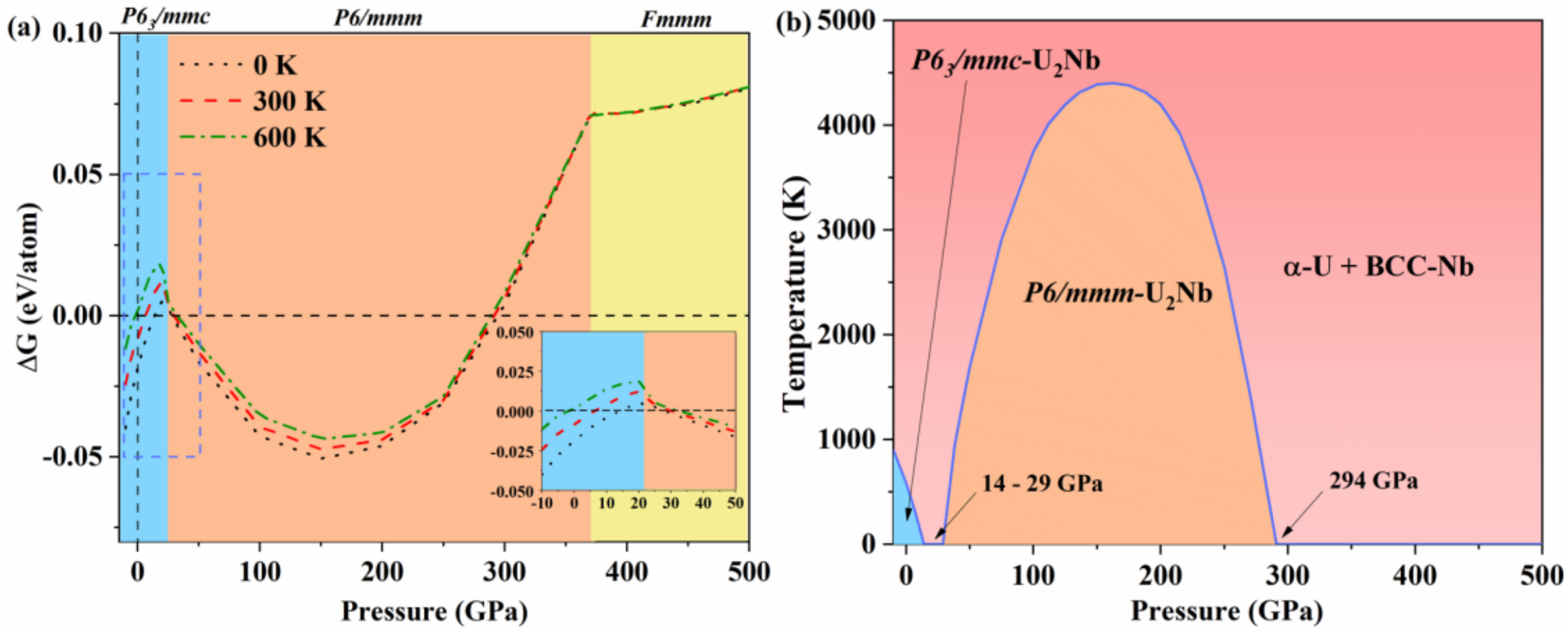

**Fig. 15** (color online) (a) Calculated formation Gibbs free energy (ΔG) of $U_2Nb$ from -10 to 500 GPa with respect to α-U and BCC-Nb at 0, 300 and 600 K. The respective region of each phase is also indicated. (b) Calculated finite temperature phase diagram of $U_2Nb$, which highlights the stable domain of $U_2Nb$ with respect to decomposition.

Based on the finite temperature stability of $U_2Nb$ at ambient pressure, we proposed a decomposition route for metastable U-Nb alloys (α' or α") aged at room temperature, as shown in Fig. 16. In the figure, the U-rich metastable phase (U-2.3 wt. % Nb or U-6.5 wt.% Nb) decomposes first into α-U and $U_2Nb$, and then $U_2Nb$ undergoes a phase separation into α-U and BCC-Nb when it is beyond its stable zone. Furthermore, for the dynamic loading of U-rich U-Nb alloys, it transforms into ordered *P6/mmm*-$U_2Nb$ beyond 29 GPa based on our newly established phase diagram. The strong stability of this phase implies that it exists over a broad pressure-temperature range. In contrast, when released from the dynamically compressed state, the strong stability of *P6₃/mmc*-$U_2Nb$ in the dilated region (with negative pressure) suggests that the tensile sample might transition into this ordered phase and induce premature fracture because of the dynamic coupling of phase transition and plasticity.

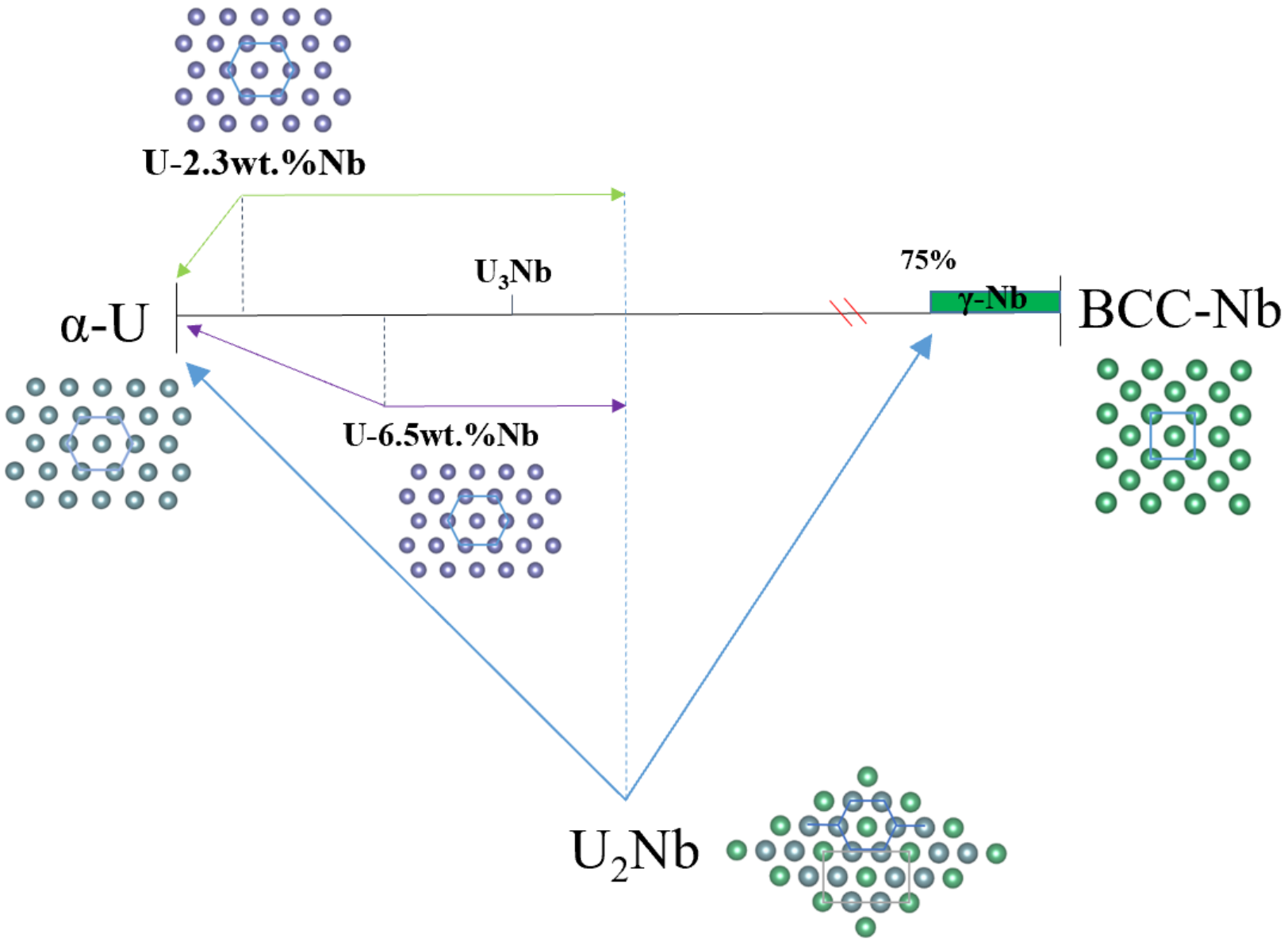

**Fig. 16** (color online) Schematic decomposition route of aged metastable $\alpha'$ (U-2.3 wt. % Nb) or $\alpha''$ (U-6.5 wt. % Nb) phase in U-Nb system via the intermediate $U_2Nb$ ordered phase.

## V. Conclusions

Ordered intermetallic $U_2Nb$ was predicted to exist over a wide pressure range based on comprehensive DFT calculations and structural searches. The previously proposed $U_3Nb$ was proven to be unstable, and a new metastable *Cmcm*-$U_3Nb$ was predicted, which might coexist (as a mixture) with ordered $U_2Nb$, α-U, and BCC-Nb at approximately 25 GPa. For the ordered $U_2Nb$, its first structure change from *P6₃/mmc* to *P6/mmm* occurs at 21.6 GPa and has both features of first-order (sharp volume collapse and finite energy barrier) and continuous transitions (soft mode and discontinuous jump and "λ" shape divergency in elastic constants and sound velocities).

This hybridization nature with striking features of both transitions is rare, and we nominally termed it as the "$1\frac{1}{2}$-order" transition to differentiate it from other mixing transitions, where the first-order feature is sufficiently weak. The proposed Landau model properly describes this symmetry-breaking transition. Both $U_2Nb$ and $U_3Nb$ exhibit prominent charge transfer from Nb to U atoms and are 5*f* metals with features of weak ionic and covalent bonding.

The thermal stability of $U_2Nb$ at finite temperature reveals that *$P6_3/mmc$* has a decomposition temperature of 520 K at zero pressure, which increases to a higher temperature in the dilated region, whereas *P6/mmm*-$U_2Nb$ possesses very strong stability over a wide pressure-temperature domain. This also suggests that the low-temperature metastable α' and α" phases could be shocked and segregated into α-U and *P6/mmm*-$U_2Nb$ phases, whereas *$P6_3/mmc$*-$U_2Nb$ could be formed and induce premature fractures by the phase transition in the released and tensile regime. These newly discovered $U_2Nb$ phases as the final products of aging not only completely change the low-temperature phase diagram of U-Nb alloys but also shed new light on their aging mechanism and peculiar mechanical properties. Furthermore, it would stimulate further theoretical and experimental investigations on this important subject.

## Acknowledgments

This work was supported by the NSAF under Grant Nos. U1730248 and U1830101, the National Natural Science Foundation of China under Grant Nos.11672247, 11872056, 11904282, 12074274. Part of the simulation was performed on resources provided by the center for Comput. Mater. Sci. (CCMS) at Tohoku University, Japan.

## Author Contribution

Xiao L. Pan: Calculation, Analysis, Writing, Original draft preparation. Hao Wang: Analysis, Validation. Lei L. Zhang: Analysis, Validation. Yu F. Wang: Analysis, Validation. Xiang R. Chen: Writing, Reviewing and Editing, Fund. Hua Y. Geng: Conceive the idea, Design the project, Writing, Reviewing and Editing, Fund. Ying Chen: Analysis.

## References

[1] K. H. Eckelmeyer, A. D. Romig, L. J. Weirick, The effect of quench rate on the microstructure, mechanical properties, and corrosion behavior of U-6 wt pct Nb, Philos. Mag., **15** (1984) 1319-1330.

[2] R. A. Vandermeer, Phase transformations in a uranium-14 at. % niobium alloy, Acta Metall., **28** (1980) 383-393.

[3] A. Landa, P. Söderlind, A. Wu, Phase stability in U-6Nb alloy doped with Ti from the first principles theory, Appl. Sci., **10** (2020) 3417.

[4] C. D. Amato, F. S. Saraceno, T. B. Wilson, Phase transformations and equilibrium structures in uranium-rich niobium alloys, J. Nucl. Mater., **12** (1964) 291-304.

[5] K. Tangri, D. K. Chaudhuri, Metastable phases in uranium alloys with high solute

solubility in the BCC gamma phase. Part I — the system U-Nb, J. Nucl. Mater., **15** (1965) 278-287.
[6] M. Anagnostidis, M. Colombié, H. Monti, Phases metastables dans les alliages uranium-niobium, J. Nucl. Mater., **11** (1964) 67-76.
[7] R. A. Vandermeer, J. C. Ogle, W. B. Snyder, Shape memory effects in a uranium-14 at. % niobium alloy, Scripta Mater., **12** (1978) 243-248.
[8] R. A. Vandermeer, J. C. Ogle, W. G. Northcutt, A phenomenological study of the shape memory effect in polycrystalline uranium-niobium alloys, Met. Trans. A, **12** (1981) 733-741.
[9] R. D. Field, D. W. Brown, D. J. Thoma, Texture development and deformation mechanisms during uniaxial straining of U–Nb shape-memory alloys, Philos. Mag., **85** (2005) 2593-2609.
[10] D. W. Brown, M. A. Bourke, R. D. Field, W. L. Hults, D. F. Teter, D. J. Thoma, S. C. Vogel, Neutron diffraction study of the deformation mechanisms of the uranium–7 wt. % niobium shape memory alloy, Mat. Sci. Eng. A-Struct., **421** (2006) 15-21.
[11] Z. Changsheng, W. Hong, L. Jian, P. Beibei, X. Yuanhua, L. Yaoguang, S. Guangai, Z. Xinjian, F. Tao, W. Xiaolin, The aging-effect-modulated mechanical behavior in U-Nb shape memory alloys through the modified twinning-detwinning process of the α″ phase, Mater. Design, **162** (2019) 94-105.
[12] H. M. Volz, R. E. Hackenberg, A. M. Kelly, W. L. Hults, A. C. Lawson, R. D. Field, D. F. Teter, D. J. Thoma, X-ray diffraction analyses of aged U–Nb alloys, J. Alloy. Compd., **444-445** (2007) 217-225.
[13] B. Djurić, Decomposition of gamma phase in a uranium-9.5 wt % niobium alloy, J. Nucl. Mater., **44** (1972) 207-214.
[14] R. E. Hackenberg, H. M. Volz, P. A. Papin, A. M. Kelly, R. T. Forsyth, T. J. Tucker, K. D. Clarke, Kinetics of Lamellar Decomposition Reactions in U-Nb Alloys, Solid State Phenom., **172-174** (2011) 555-560.
[15] J. Zhang, D. W. Brown, B. Clausen, S. C. Vogel, R. E. Hackenberg, In Situ Time-Resolved Phase Evolution and Phase Transformations in U-6 wt. pct. Nb, Met. Trans. A, **50** (2019) 2619-2628.
[16] T. C. Duong, R. E. Hackenberg, V. Attari, A. Landa, P. E. A. Turchi, R. Arróyave, Investigation of the discontinuous precipitation of U-Nb alloys via thermodynamic analysis and phase-field modeling, Comp. Mater. Sci., **175** (2020) 109573.
[17] R. E. Hackenberg, M. G. Emigh, P. A. Papin, A. M. Kelly, R. T. Forsyth, T. J. Tucker, K. D. Clarke, Influence of Initiation Site and Lamellar Divergency on the Overall Kinetics of Cellular Growth and Coarsening in Aged U-Nb Alloys, Mater. Sci. Forum, **941** (2018) 863-868.
[18] L. Hsiung, J. Zhou, Spinodal decomposition and order-disorder transformation in a Water-Quenched U-6 wt. % Nb Alloy, LLNL Report, UCRL-TR-224432, 2006.
[19] C. Zhang, L. Xie, Z. Fan, H. Wang, X. Chen, J. Li, G. Sun, Straightforward understanding of the structures of metastable α″ and possible ordered phases in uranium–niobium alloys from crystallographic simulation, J. Alloy. Compd., **648** (2015) 389-396.
[20] A. J. Clarke, R. D. Field, R. E. Hackenberg, D. J. Thoma, D. W. Brown, D. F. Teter,

M. K. Miller, K. F. Russell, D. V. Edmonds, G. Beverini, Low temperature age hardening in U–13 at. % Nb: An assessment of chemical redistribution mechanisms, J. Nucl. Mater., **393** (2009) 282-291.
[21] Y. Zhang, X. Wang, Q. Xu, Y. Li, X-ray diffraction study of low temperature aging in U–5.8 wt. % Nb, J. Nucl. Mater., **456** (2015) 41-45.
[22] J. Zhang, S. Vogel, D. Brown, B. Clausen, R. Hackenberg, Equation of state, phase stability, and phase transformations of uranium-6 wt. % niobium under high pressure and temperature, J. Appl. Phys., **123** (2018) 175103.
[23] J. Zhang, R. E. Hackenberg, S. C. Vogel, D. W. Brown, Equation of state and phase evolution of U-7.7Nb with implications for the understanding of dynamic behavior of U-Nb alloys, Appl. Phys. Lett., **114** (2019) 221901.
[24] H. Y. Geng, Y. Chen, Y. Kaneta, M. Kinoshita, Structural behavior of uranium dioxide under pressure by LSDA+U calculations, Phys. Rev. B, **75** (2007) 054111.
[25] H. X. Song, H. Y. Geng, Q. Wu, Pressure-induced group-subgroup phase transitions and post-cotunnite phases in actinide dioxides, Phys. Rev. B, **85** (2012) 064110.
[26] H. Y. Geng, H. X. Song, Q. Wu, Anomalies in nonstoichiometric uranium dioxide induced by a pseudo phase transition of point defects, Phys. Rev. B, **85** (2012) 144111.
[27] P. Hohenberg, W. Kohn, Inhomogeneous Electron Gas, Phys. Rev., **136** (1964) B864-B871.
[28] W. Kohn, L. J. Sham, Self-Consistent Equations Including Exchange and Correlation Effects, Phys. Rev., **140** (1965) A1133-A1138.
[29] G. Kresse, J. Hafner, Ab initio molecular dynamics for liquid metals, Phys. Rev. B, **47** (1993) 558-561.
[30] G. Kresse, J. Furthmuller, Efficiency of ab-initio total energy calculations for metals and semiconductors using a plane-wave basis set, Comput. Mater. Sci., **6** (1996) 15-50.
[31] J. P. Perdew, K. Burke, M. Ernzerhof, Generalized gradient approximation made simple, Phys. Rev. Lett., **77** (1996) 3865.
[32] H. J. Monkhorst, J. D. Pack, Special points for Brillouin-zone integrations, Phys. Rev. B, **13** (1976) 5188-5192.
[33] A. Togo, I. Tanaka, First principles phonon calculations in materials science, Scripta Mater., **108** (2015) 1-5.
[34] Y. Wang, J. Lv, L. Zhu, Y. Ma, CALYPSO: A method for crystal structure prediction, Comput. Phys. Commun., **183** (2012) 2063-2070.
[35] S. P. Kramynin, E. N. Akhmedov, Equation of state and properties of Nb at high temperature and pressure, J. Phys. Chem. Solids, **135** (2019) 109108.
[36] I. A. Kruglov, A. Yanilkin, A. R. Oganov, P. Korotaev, Phase diagram of uranium from ab initio calculations and machine learning, Phys. Rev. B, **100** (2019) 174104.
[37] S. L. Dudarev, G. A. Botton, S. Y. Savrasov, C. J. Humphreys, A. P. Sutton, Electron-energy-loss spectra and the structural stability of nickel oxide: An LSDA+ U study, Phys. Rev. B, **57** (1998) 1505.
[38] A. R. Oganov, C. W. Glass, Crystal structure prediction using ab initio evolutionary techniques: Principles and applications, J. Chem. Phys., **124** (2006) 244704.

[39] S. Qin, S. Liu, C. Zhang, J. Xin, Y. Wang, Y. Du, Thermodynamic modeling of the Ca–In and Ca–Sb systems supported with first-principles calculations, Calphad, **48** (2015) 35-42.
[40] F. Ling, K. Luo, L. Hao, Y. Gao, Z. Yuan, Q. Gao, Y. Zhang, Z. Zhao, J. He, D. Yu, Universal Phase Transitions of AlB2-Type Transition-Metal Diborides, ACS Omega, **5** (2020) 4620-4625.
[41] See Supplementary Material (xxx) for other detailed information.
[42] G. Henkelman, A. Arnaldsson, H. Jónsson, A fast and robust algorithm for Bader decomposition of charge density, Comput. Mater. Sci., **36** (2006) 354-360.
[43] H. X. Song, L. Liu, H. Y. Geng, Q. Wu, First-principle study on structural and electronic properties of CeO2 and ThO2 under high pressures, Phys. Rev. B, **87** (2013) 184103.
[44] H. Y. Geng, N. X. Chen, M. H. F. Sluiter, First-principles equation of state and phase stability for the Ni−Al system under high pressures, Phys. Rev. B, **70** (2004) 094203.
[45] Y. X. Wang, Q. Wu, X. R. Chen, H. Y. Geng, Stability of rhombohedral phases in vanadium at high-pressure and high-temperature: first-principles investigations, Sci. Rep., **6** (2016) 32419.
[46] T. Hahn, U. Shmueli, J. W. Arthur, International tables for crystallography, Reidel Dordrecht, 1983.
[47] U. Muller, Symmetry Relations between Crystal Structures, New York: Oxford University Press, 2013.
[48] J. Lazewski, P. Piekarz, K. Parlinski, Mechanism of the phase transitions in MnAs, Phys. Rev. B, **83** (2011) 054108.
[49] E. M. Lifshitz, L. P. Pitaevskii, Statistical physics: theory of the condensed state, Oxford: Pergamon, 1980.
[50] J. Koike, M. E. Kassner, R. E. Tate, R. S. Rosen, The Nb-U (niobium-uranium) system, J. Phase Equilib. Diff., **19** (1998) 253-260.

# Supplementary Material for "Prediction of novel final phases in aged uranium-niobium alloys"

Xiao L. Pan[1, 2], Hao Wang[1, 2], Lei L. Zhang[2], Yu F. Wang[2], Xiang R. Chen[1*], Hua Y. Geng[2, 3]*, Ying Chen[4]

[1] *College of Physics, Sichuan University, Chengdu 610065, P. R. China;*

[2] *National Key Laboratory of Shock Wave and Detonation Physics, Institute of Fluid Physics, China Academy of Engineering Physics, Mianyang, Sichuan 621900, P. R. China;*

[3] *HEDPS, Center for Applied Physics and Technology, and College of Engineering, Peking University, Beijing 100871, P.R. China;*

[4] *Fracture and Reliability Research Institute, School of Engineering, Tohoku University, Sendai 980-8579, Japan.*

## SI. Supplementary Methods

Vibrational Helmholtz free energy was calculated at the given volume $V$ by integration over the phonon occupation states, as follows:

$$F_{Ph}(T,V) = \int_0^{\infty} g(\omega)\left[\frac{\hbar\omega}{2} + k_B T\, \ln\left(1 - \exp\left(-\frac{\hbar\omega}{k_B T}\right)\right)\right] d\omega, \qquad \textbf{(7)}$$

in which g(ω) is the density of states of phonons. The dispersion of phonons (from which g(ω) is obtained) was computed using the frozen phonon and small-displacement method as implemented in PHONOPY code [1]. A 2 × 2 × 2 supercell for $U_2Nb$ (48 atoms), 2 × 1 × 2 supercell for $U_3Nb$ (64 atoms), 3 × 3 × 3 supercell for BCC-Nb (54 atoms) and 3 × 2 × 2 supercell for α-U (48 atoms) were employed in the calculations. The Gibbs free energy at given temperature T and pressure P of lattice vibrational contribution is calculated by using the vibrational Helmholtz free energy and the quasi-harmonic approximation (QHA), where the volume dependence of phonon frequencies $\omega(V)$ is introduced to partially account for the anharmonic effect. As the temperature increases, the volume dependence of phonon Helmholtz free energy changes, and the equilibrium volume at the given temperature changes accordingly. In QHA, the Gibbs free energy is defined at a given pressure by the following minimization formula:

$$G(T,P) = \min_V [\mathrm{E}(V) + F_{ph}(T, V) + PV] \qquad \textbf{(8)}$$

where E($V$) is the cold part of the internal energy. The QHA Gibbs free energies of $U_2Nb$, $U_3Nb$, α-U, γ'-U and BCC-Nb under pressure are then calculated. Further, the formation Gibbs free energy per atom for $U_2Nb$ and $U_3Nb$ is calculated by following expression:

$$\Delta G = (G^{U_x\mathrm{Nb}_y} - xG^{U} - yG^{\mathrm{Nb}}) / (x+y) \qquad \textbf{(9)}$$

where $G^{U_x\mathrm{Nb}_y}$ is the Gibbs free energy per formula of a compound, in which $x$ and $y$

* *Corresponding authors. E-mail: s102genghy@caep.cn; xrchen@scu.edu.cn*

represent the number of uranium and niobium atoms in that formula, respectively; $G^U$ and $G^{Nb}$ represent the Gibbs free energy per atom of the ground state structure of uranium and niobium at the given pressure, respectively.

The elastic constants of single crystal were calculated by using the energy-strain method, where nine strains around the equilibrium lattice constant were employed. There are six independent elastic constants for the hexagonal lattice. Using the single crystal elastic constants, the bulk modulus (*B*) and shear modulus (*G*) of polycrystal are evaluated by the Voigt-Reuss-Hill (VRH) method [2]:

$$B = \frac{B_V + B_R}{2},\ G = \frac{G_V + G_R}{2}, \tag{10}$$

where

$$B_V = \frac{2(C_{11} + C_{12}) + 4C_{13} + C_{33}}{9}, \tag{11}$$

$$B_R = \frac{(C_{11} + C_{12})C_{33} - 2C_{13}^2}{C_{11} + C_{12} + 2C_{33} - 4C_{13}}, \tag{12}$$

$$G_v = \frac{C_{11} + C_{12} + 2C_{33} - 4C_{13} + 12C_{44} + 12C_{66}}{30}, \tag{13}$$

$$G_R = \frac{5[(C_{11} + C_{12})C_{33} - 2C_{13}^2]C_{44}C_{66}}{2B_v C_{44} C_{66} + 2[(C_{11} + C_{12})C_{33} - 2C_{13}^2](C_{44} + C_{66})}. \tag{14}$$

Furthermore, the longitudinal $C_l$ and transverse $C_s$ sound velocities of the polycrystalline sample are evaluated by using the shear modulus *G*, bulk modulus *B*, and mass density *ρ* from Navier's equation as follows [3]:

$$C_l = \left( \frac{B + \frac{4G}{3}}{\rho} \right)^{1/2},\ C_S = \left( \frac{G}{\rho} \right)^{1/2}. \tag{15}$$

For the proposed Landau model of Eq. (3) in the main text, it can be derived that at the transition pressure $P_t$, the critical order parameter $d_c$ is $d_c = \pm \left| \frac{b_0}{2c_0} \right|^{1/2}$, the energy barrier for the transition is $E_b = \frac{b_0^3}{54c_0^2}$, and its corresponding order parameter (i.e., the position of the barrier) is $d_b = \frac{d_c}{\sqrt{3}}$. In addition, the transition pressure ($P_t$) and over-compression limit ($P^*$) can be easily derived, and are listed as follows:

$$a_0(P_t - P_0) = \frac{b_0^2}{4c_0}; a_0(P^* - P_0) = \frac{b_0^2}{3c_0}. \quad \textbf{(16)}$$

It should be noted that there is a tricritical point when $b_0 = 0$, for which along the first-order line, there is coexistence of three phases: one disordered ($d = 0$) and two ordered phases (with $d = \pm |d_c|$).

## SII. Structure stability

To illustrate the thermodynamic stability of $U_2Nb$, its formation enthalpy (ΔH) at absolute zero under pressures was calculated and shown in Fig. S1(a). *P6₃/mmc*-$U_2Nb$ is the stable phase in the negative and low-pressure region (from -10 to 14 GPa). There is a region (14–29 GPa) where $U_2Nb$ has a slightly positive formation enthalpy with a maximum value of 5.14 meV/atom at 20 GPa. It should be emphasized that *P6/mmm*-$U_2Nb$ has a large negative ΔH over a wide range of pressures (29–294 GPa), whereas *Fmmm* phase has a positive ΔH and could be metastable. Overall, the *P6₃/mmc*-$U_2Nb$ is stable between -10 and 14 GPa, and *P6/mmm*-$U_2Nb$ is stable between 29 and 294 GPa. At other pressures, they could be metastable. The formation enthalpy of *Cmcm*-$U_3Nb$ under pressure is plotted in Fig. S1(b). It is positive over the whole pressure range, suggesting that it is not thermodynamically stable. But a minimum ΔH of 2.17 meV/atom is achieved at 29 GPa against the decomposition into *α*-U and BCC-Nb, revealing that it could be metastable, or form a mixture with *α*-U and BCC-Nb.

The enthalpy of the *Cmcm*-$U_3Nb$ predicted in this work is compared with the $U_3Nb$ (*Pmm2*) structure relaxed from the unstable configuration proposed by Zhang et al. [4], as shown in Fig. S2. Zhang et al. proposed the hypothetical ordered phase of $U_3Nb$ based on the lattice of *α*-U with *Cmcm* symmetry. As can be seen in Fig. S2, the enthalpy of the relaxed *Pmm2*-$U_3Nb$ is significantly higher than that of *Cmcm*-$U_3Nb$, suggesting that *Cmcm* structure is more stable than the *Pmm2* structure for $U_3Nb$.

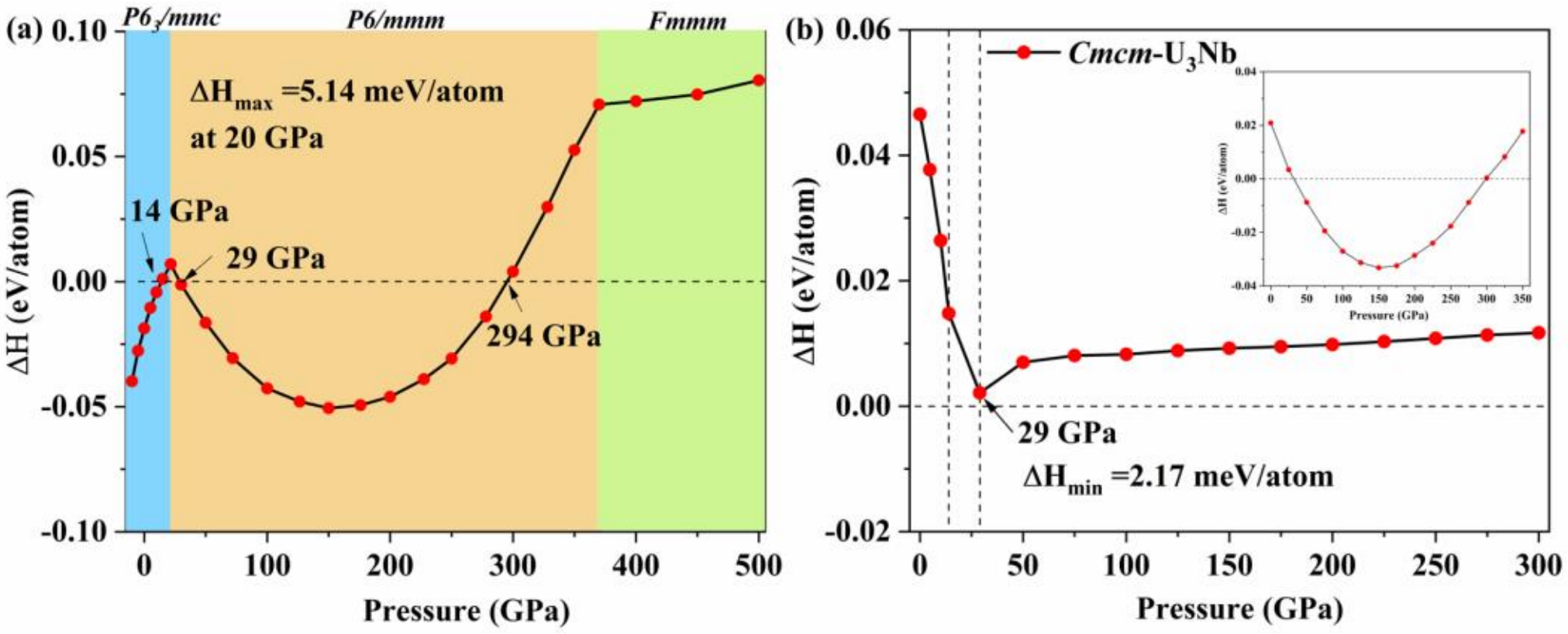

**Fig. S1.** (a) Calculated formation enthalpy of $U_2Nb$ with respect to elemental U and Nb states under pressure; (b) Calculated formation enthalpy of $U_3Nb$ with respect to *α*-U and *P6₃/mmc* -$U_2Nb$ at 0–14 GPa, w.r.t. *α*-U and BCC-Nb at 14–29 GPa, and w.r.t. *α*-U and *P6/mmm* -$U_2Nb$ at 29–300 GPa. Inset of (b) displays the ΔH of *Cmcm*-$U_3Nb$ with respect to $\alpha$-U and BCC-Nb.

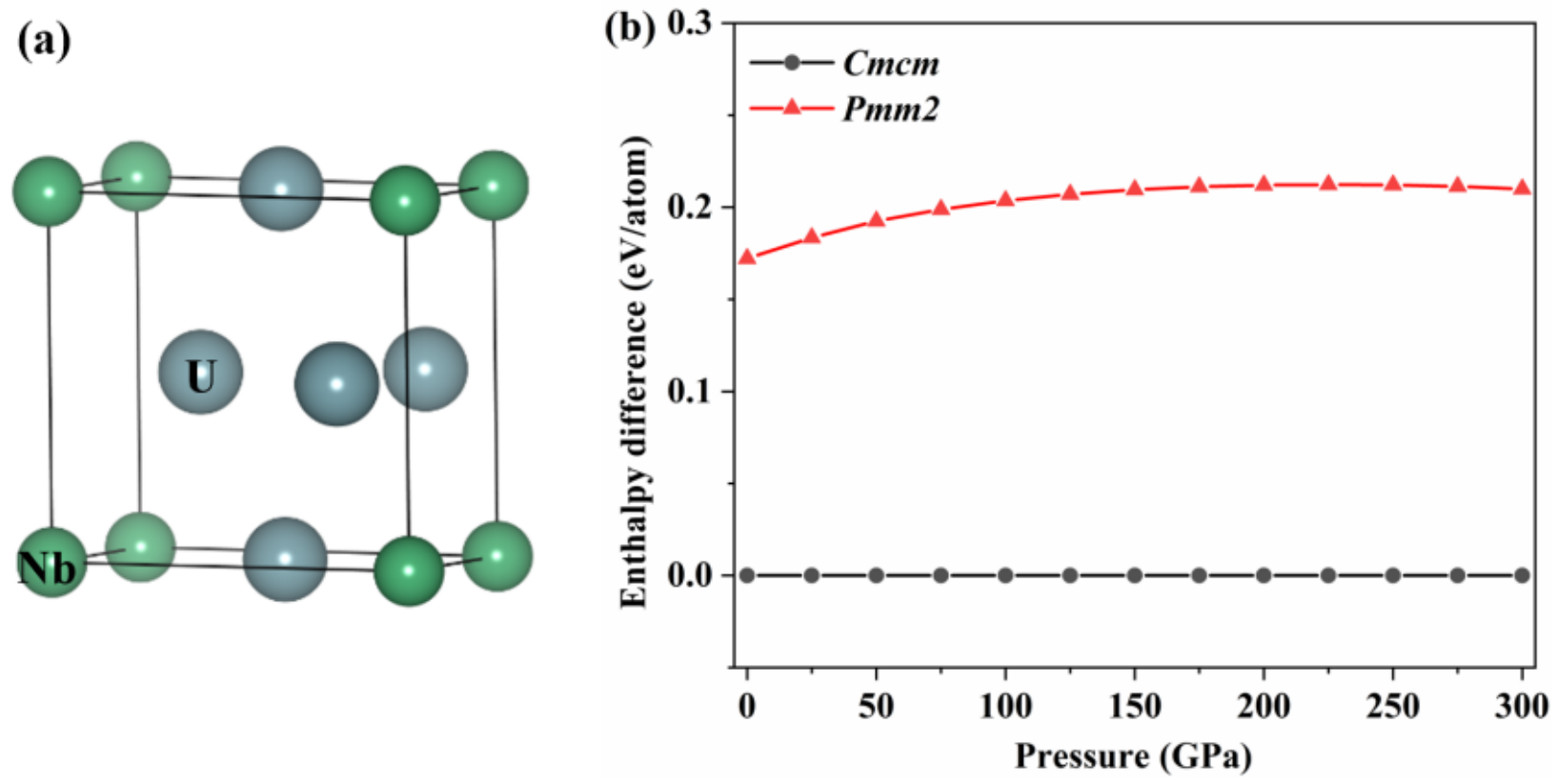

**Fig. S2.** (a) The structure of $U_3Nb$ (*Pmm2*, which is a distorted $Cu_3Au$-type ($L1_2$) structure) relaxed from the candidate structure proposed by Zhang et al. [4] which is unstable. (b) Calculated enthalpy difference between *Pmm2*-$U_3Nb$ and our newly predicted *Cmcm*-$U_3Nb$ as a function of pressure.

The calculated phonon dispersion curves of $U_2Nb$ and $U_3Nb$ are shown in Figs. S3-S6. There are no imaginary modes, suggesting that they are all dynamically stable. With the increase of pressure, the interatomic distance decreases and the interaction between atoms enhances, leading to the increasement of the phonon frequency. The vibrational modes of the optical branches with the highest frequency at gamma point for the *P6₃/mmc* and *P6/mmm* phases of $U_2Nb$ are shown in Fig. S7. In these modes, the Nb atoms of the adjacent layers vibrate along the Z direction with the same amplitude but opposite direction, while U atoms keep stationary.

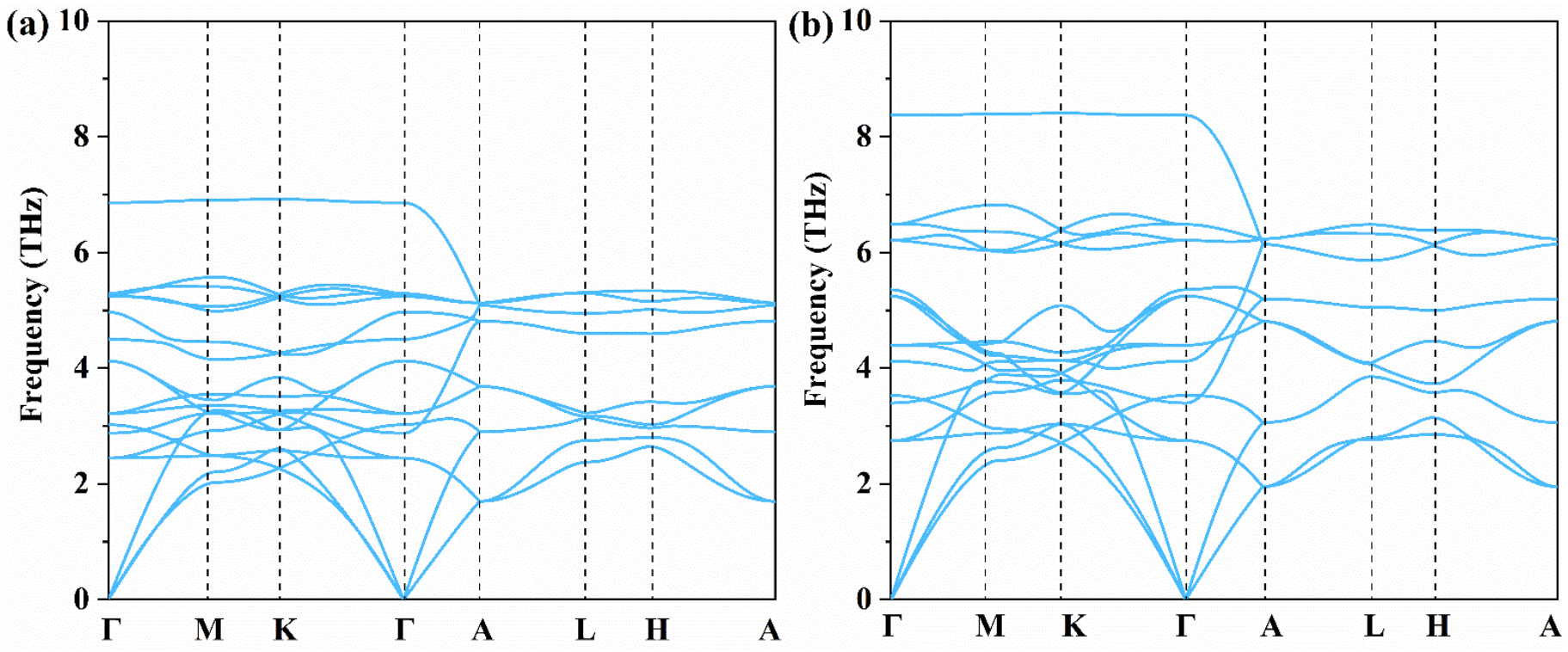

**Fig. S3.** Calculated phonon spectra of *$P6_3/mmc$*-$U_2Nb$ at (a) 0 GPa and (b) 20 GPa.

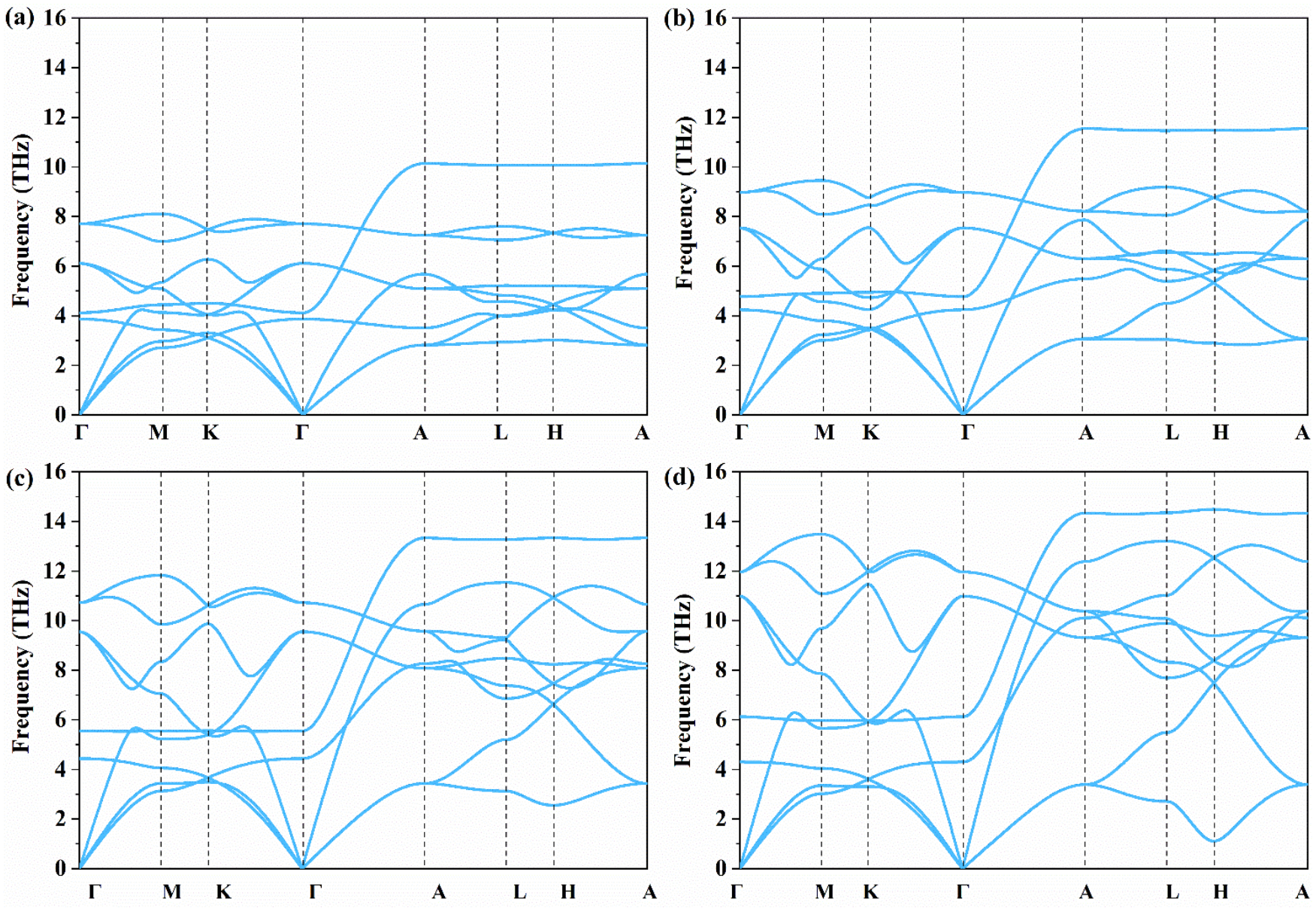

**Fig. S4.** Calculated phonon spectra of *P6/mmm*-$U_2Nb$ at (a) 50 GPa, (b) 100 GPa, (c) 200 GPa and (d) 300 GPa.

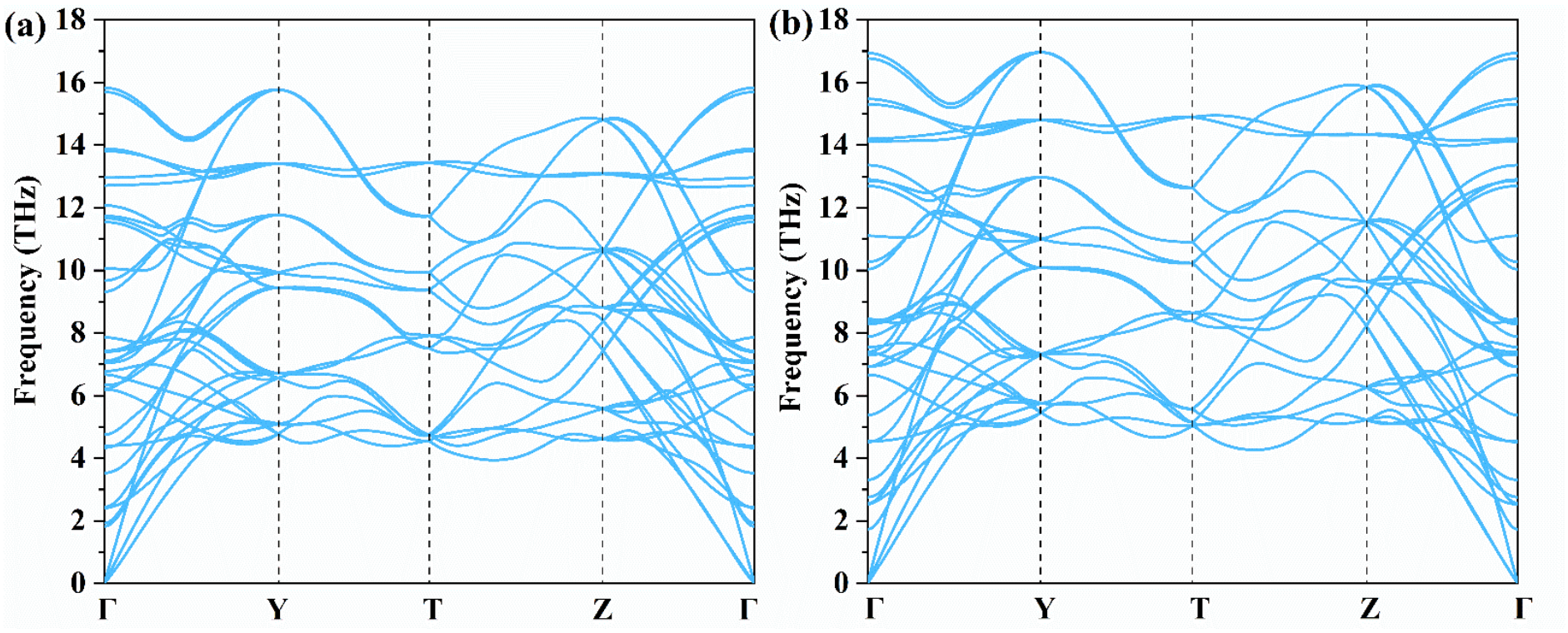

**Fig. S5.** Calculated phonon spectra of *Fmmm*-$U_2Nb$ at (a) 400 GPa and (b) 500 GPa.

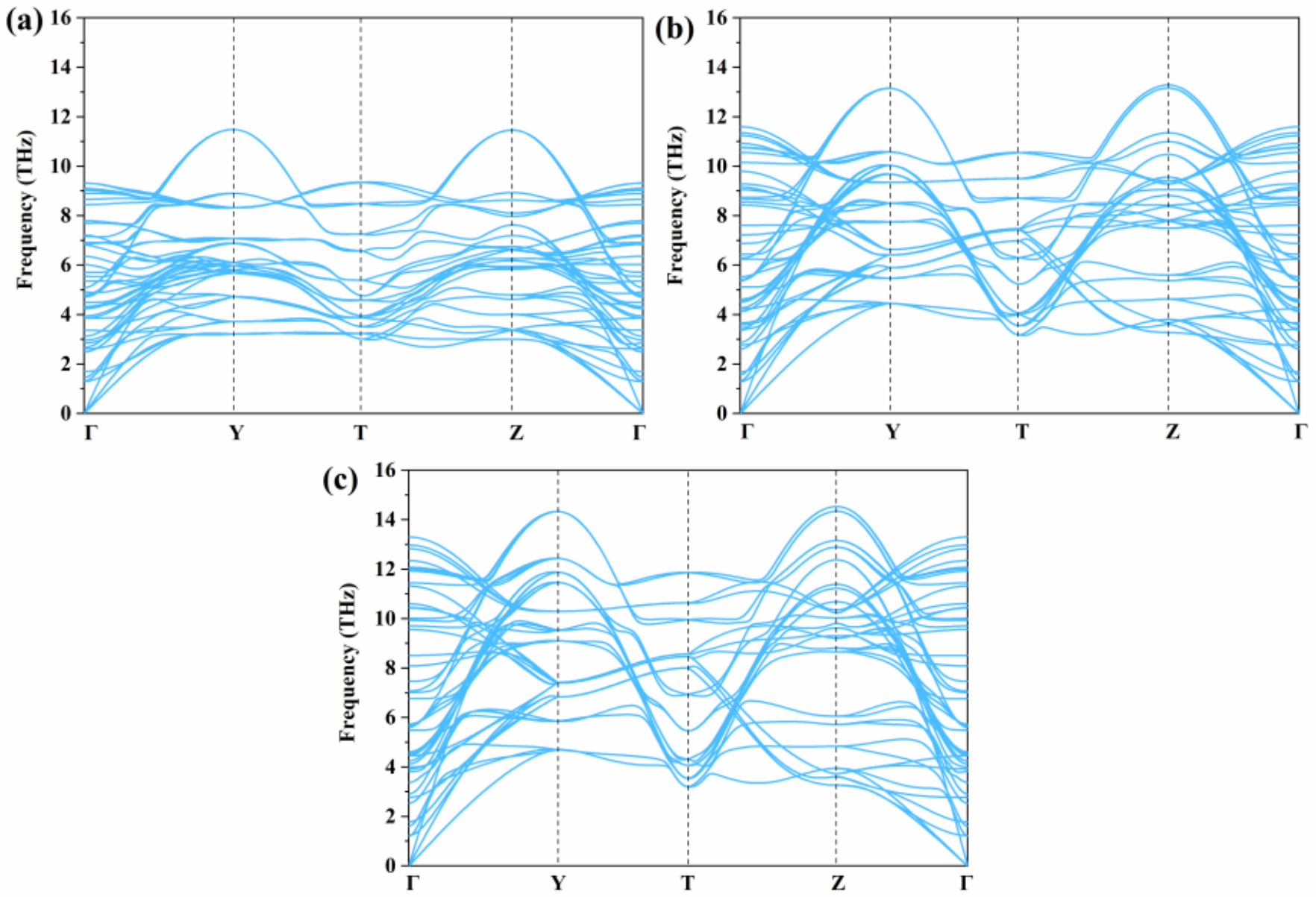

**Fig. S6.** Calculated phonon spectra of *Cmcm*-$U_3Nb$ at (a) 100 GPa, (b) 200 GPa and (c)

300 GPa.

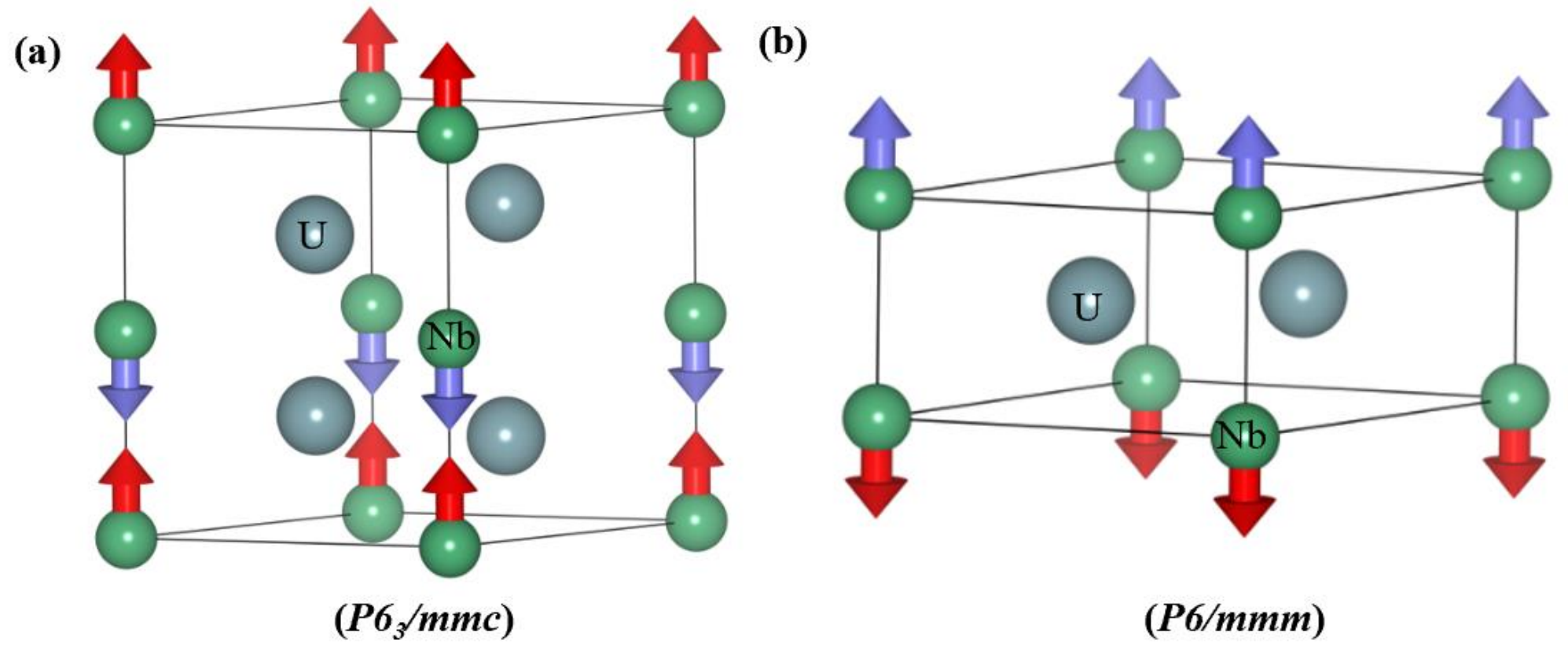

**Fig. S7.** The vibrational modes of the optical branches with the highest frequency at the gamma point for the (a) *P6₃/mmc* and (b) *P6/mmm* phases of $U_2Nb$.

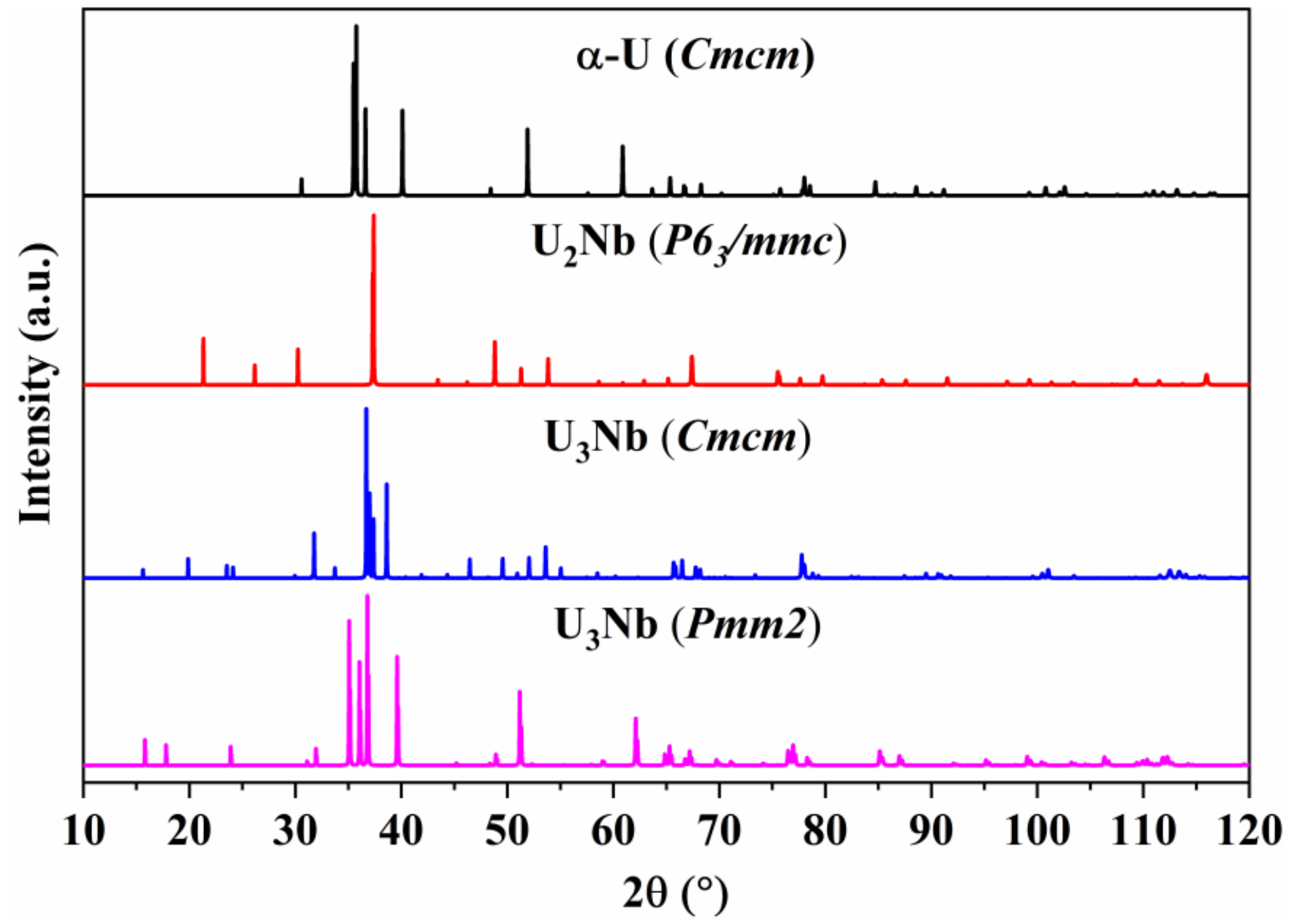

**Fig. S8.** Simulated X-ray diffraction pattern for *α*-U, *P6₃/mmc*-$U_2Nb$, *Cmcm*-$U_3Nb$, and *Pmm2*-$U_3Nb$ by using a wave-length of λ= 1.54 Å at 0 GPa.

Interatomic separation histograms from Nb to other atoms for *P6₃/mmc*-$U_2Nb$ and *P6/mmm*-$U_2Nb$ at zero pressure are shown in Fig. S9. For *P6₃/mmc*-$U_2Nb$, the total number of U atoms in the first neighborhood shell of Nb atoms are 12, and the number of Nb atoms in the first neighborhood are 2. There is a clear gap between 3.25 Å and 4.75 Å for both *P6₃/mmc*-$U_2Nb$ and *P6/mmm*-$U_2Nb$, which is similar to the BCC-Nb. In addition, the distribution of U atoms in *P6₃/mmc*-$U_2Nb$ phase is narrower than in *α*-U. Compared with *P6₃/mmc*-$U_2Nb$, the atomic distribution in *P6/mmm*-$U_2Nb$ is also narrower, implying that the latter has higher stability under high pressure due to the smaller *PV* term in enthalpy.

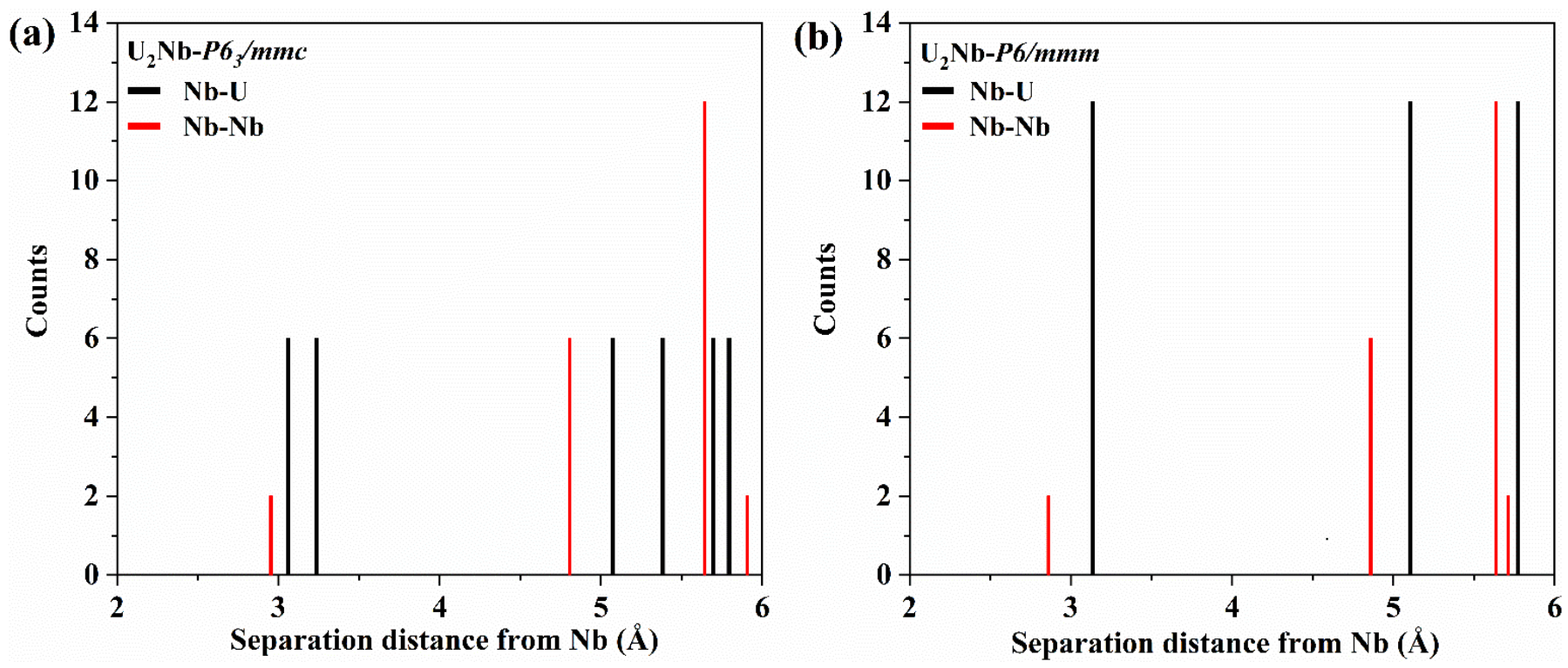

**Fig. S9.** Interatomic separation histograms for (a) *$P6_3/mmc$*-$U_2Nb$ and (b) *P6/mmm*-$U_2Nb$ at zero pressure.

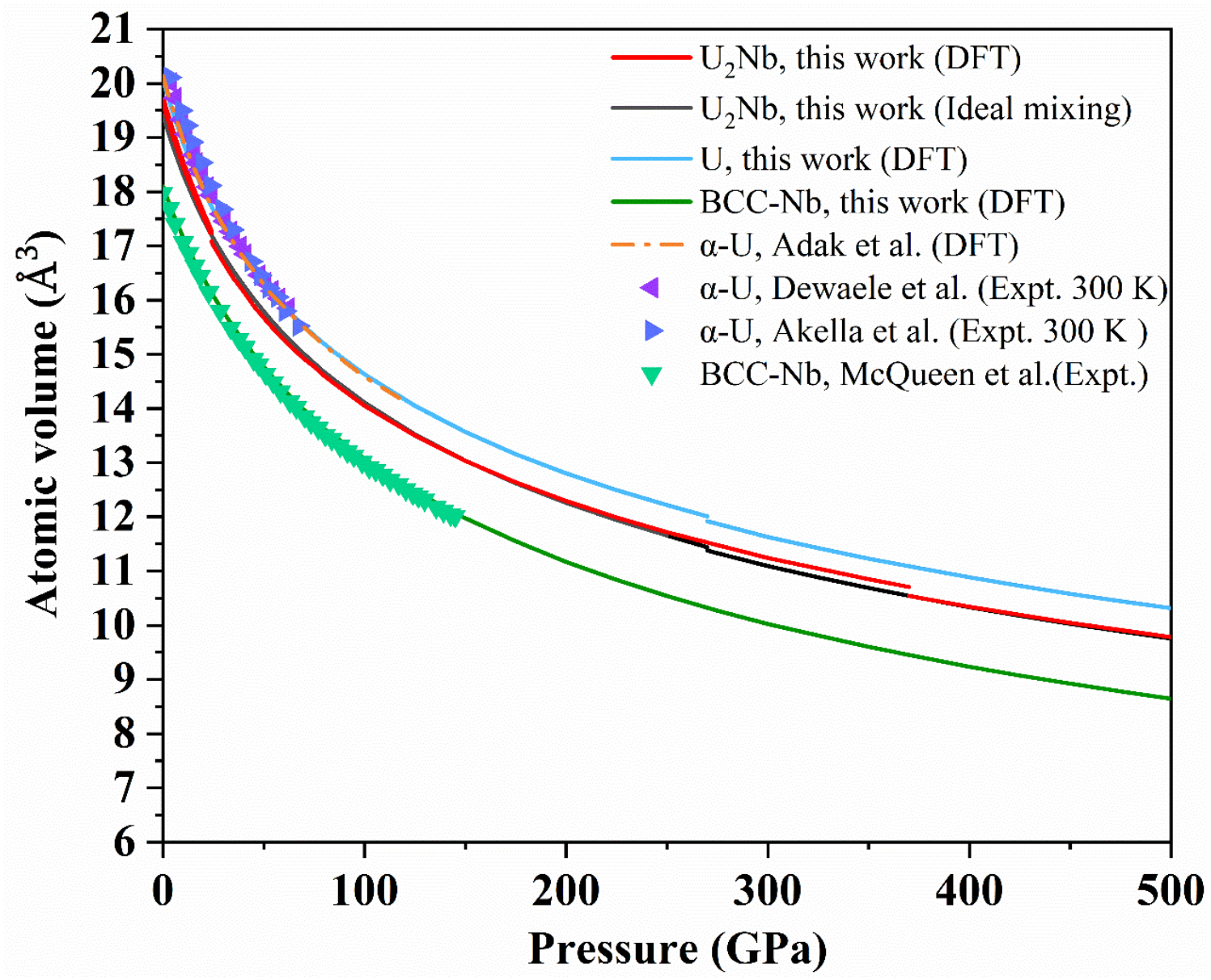

**Fig. S10.** Calculated equation of state of $U_2Nb$, α-U (0–270 GPa), γ'-U (270–500 GPa), and BCC-Nb at zero Kelvin as compared to previous DFT and experimental data [5-8]. The hypothetical U-Nb solid solution with a composition of U:Nb=2:1 as estimated by the ideal mixing model is also shown for comparison.

## SIII. Electronic bonding properties

The calculated electronic band structures and projected density of states (PDOS) of *P6/mmm*-$U_2Nb$ at 100 GPa, *Fmmm*-$U_2Nb$ at 400 GPa, and *Cmcm*-$U_3Nb$ at 100 GPa are displayed in Fig. S11. It is obvious that both $U_2Nb$ and $U_3Nb$ are metallic. Their band structure near the Fermi level is mainly contributed by 5*f* electrons of the U atoms. Besides, it can be seen from the PDOS that there is a hybrid interaction between the *f* orbitals and *d* orbitals of U atoms, suggesting that there is a week covalent bonding between U atoms. To investigate the electron transfer between atoms in $U_2Nb$ and $U_3Nb$, Bader charges of each phase were calculated and listed in Table S2. There is a clear transfer of electrons between U and Nb atoms, with U gaining electrons and Nb losing electrons. This indicates that there is also an ionic interaction between them.

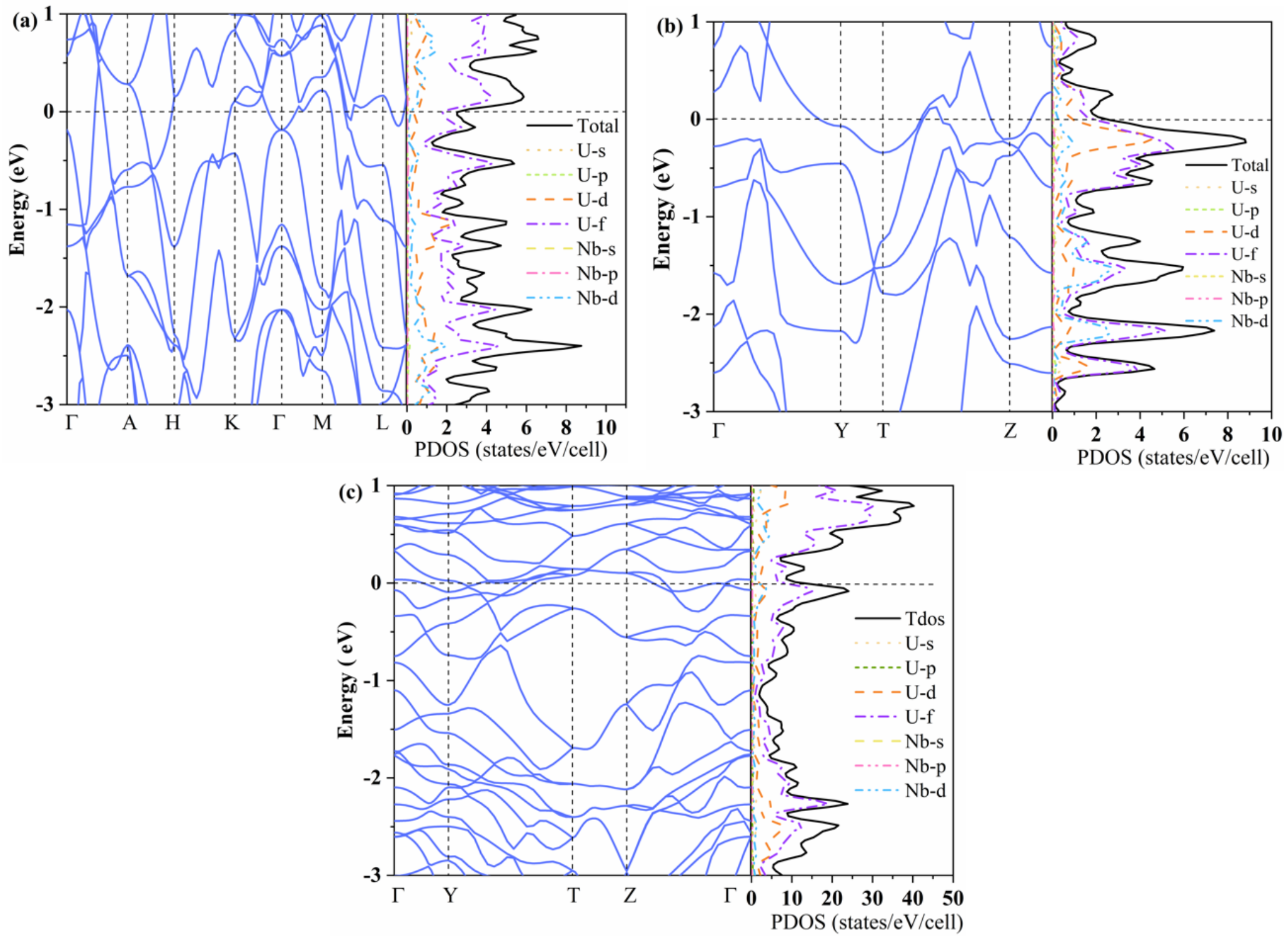

**Fig. S11.** Electronic band structures (left panel) and projected density of states (right panel) of (a) *P6/mmm*-$U_2Nb$ at 100 GPa, (b) *Fmmm*-$U_2Nb$ at 400 GPa and (c) *Cmcm*-$U_3Nb$ at 100 GPa.

Furthermore, the electron localization function (ELF) and differential charge density for $U_2Nb$ and $U_3Nb$ were calculated. The results are shown in Figs. S12-S14. It can be seen in ELF that the electrons of *P6₃/mmc*-$U_2Nb$ spread over the lattice at zero pressure, and there is no obvious localization between U-Nb or U-U atoms. Differential charge density illustrates how the electrons redistribute when forming the compound. Figure S13 reveals that some electrons move from Nb atoms to the bridge between adjacent U atoms, confirming that very weak covalent bonding may exist in the *$P6_3/mmc$*-$U_2Nb$ phase. In *P6/mmm*-$U_2Nb$ phase, due to the decrease of interatomic distance at high pressures, the interaction between U and Nb atoms is enhanced, and there is still no clear electron localization between atoms. However, obvious charge aggregation between U atoms in the same layer is observed, confirming that there might be very weak covalent bonding in *P6/mmm*-$U_2Nb$ phase at 200 GPa. Similar features are also observed in *Fmmm*-$U_2Nb$ and *Cmcm*-$U_3Nb$.

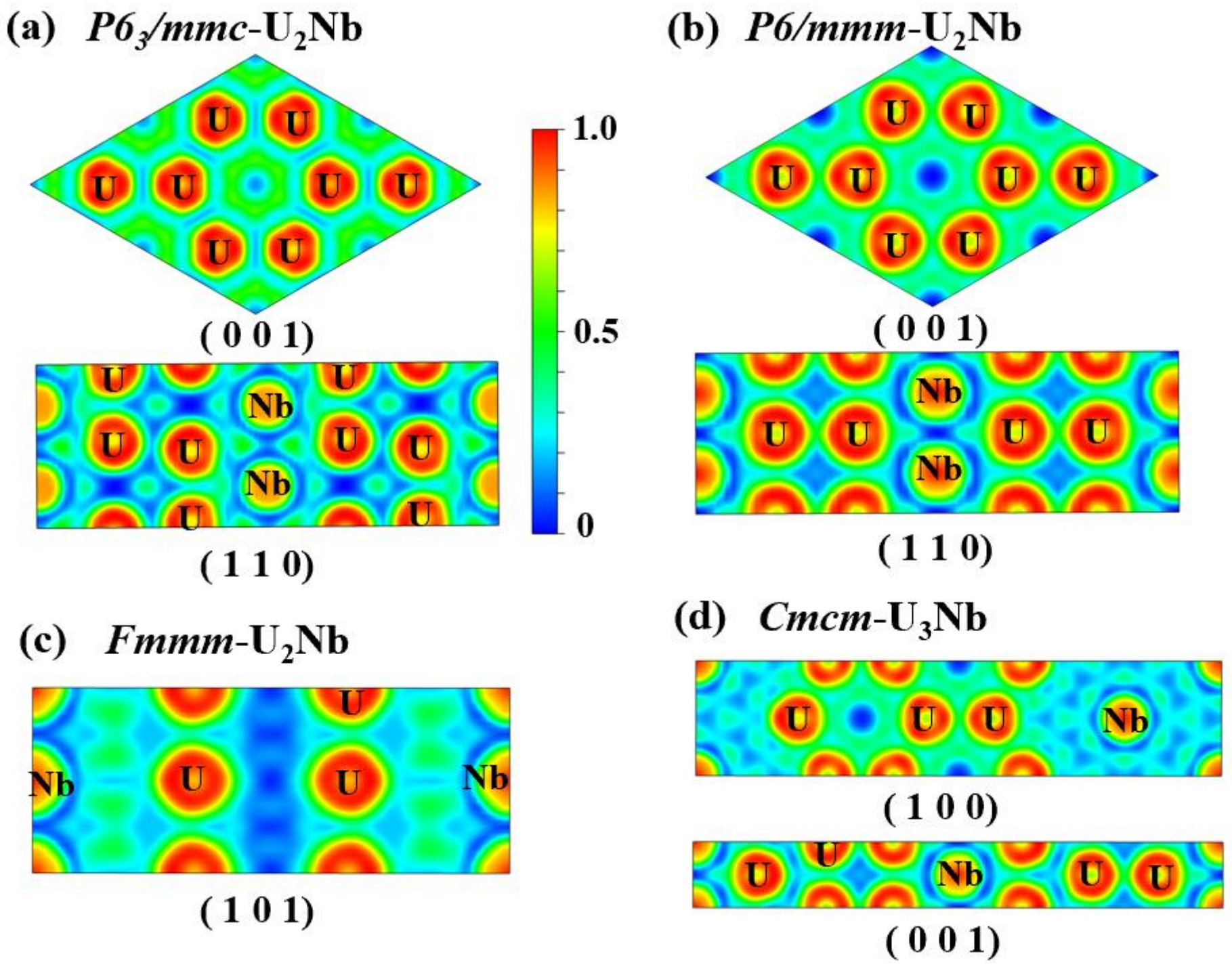

**Fig. S12.** Electron localization function (ELF) of (a) *P6₃/mmc*-$U_2Nb$ at 0 GPa, (b) *P6/mmm*-$U_2Nb$ at 200 GPa on the plane of (0 0 1) and (1 1 0), (c) *Fmmm*-$U_2Nb$ at 400 GPa on the plane of (1 0 1) and (d) *Cmcm*-$U_3Nb$ at 200 GPa on the plane of (1 0 0) and (0 0 1). ELF isosurfaces are drawn along with the value from 0 (blue) to 1.0 (red).

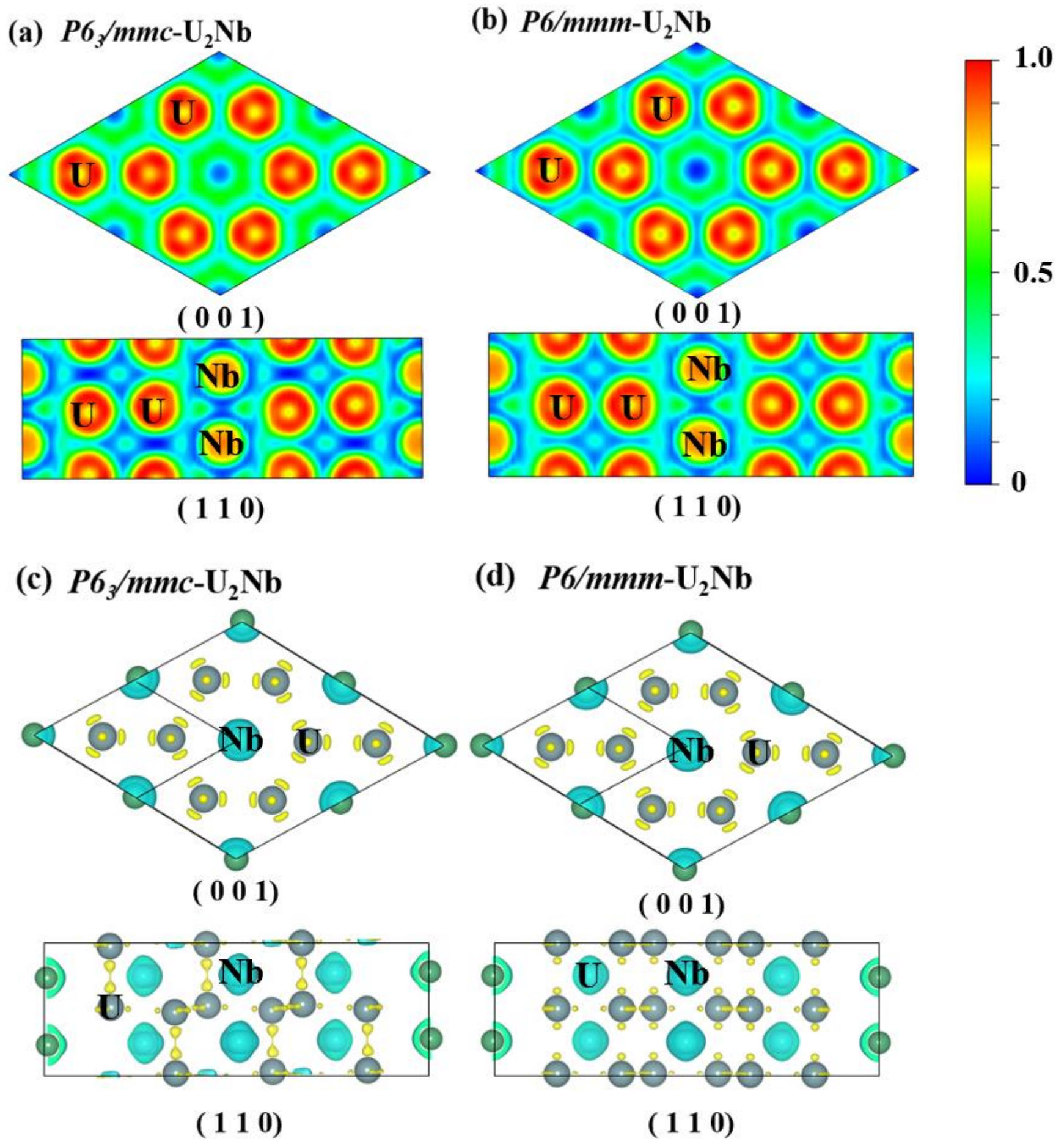

**Fig. S13.** Electron localization function (ELF) of (a) *P*$6_3$*/mmc*-$U_2Nb$ and (b) *P6/mmm*-$U_2Nb$ at 21.6 GPa. The differential charge density (isosurface = 0.02 *e*/bohr$^3$) of (c) *P*$6_3$*/mmc*-$U_2Nb$ and (d) *P6/mmm*-$U_2Nb$ at 21.6 GPa. The blue color indicates the loss of electrons and yellow color represents the gain of charge.

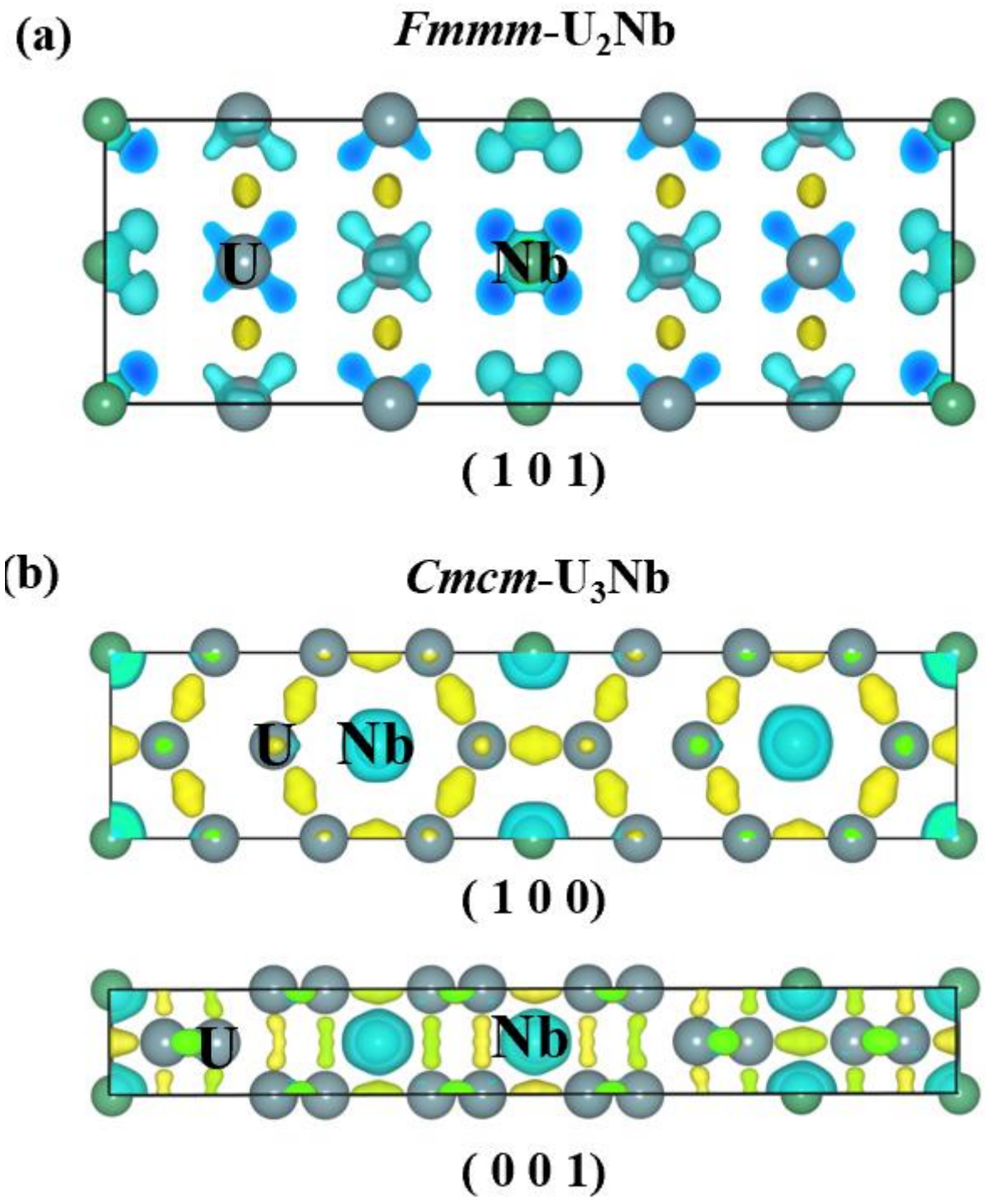

**Fig. S14.** Differential charge density (isosurface = 0.02 *e*/bohr$^3$) of (a) *Fmmm*-$U_2Nb$ at 400 GPa and (b) *Cmcm*-$U_3Nb$ at 200 GPa. The blue color indicates the loss of electrons and yellow color represents the gain of charge between U atoms.

## SIV. Influence of temperature on stability of ordered phase

With QHA, the formation Gibbs free energy (ΔG) of *Cmcm*-$U_3Nb$ in the range of 0 to 300 GPa at temperatures of 0, 300 and 600 K was calculated and shown in Fig. S15. When the reference terminal states are α-U and BCC-Nb, the pressure range of $U_3Nb$ with negative formation enthalpy decreases with the increase of temperature, while its stability slightly increases with the increase of temperature at 16–25 GPa, as displayed in Fig. S15(a). When the reference terminal states are α-U and $U_2Nb$, on the other hand, the formation Gibbs free energy of *Cmcm*-$U_3Nb$ is negative only in a narrow pressure range of approximately 16–25 GPa. It implies that although *Cmcm*-$U_3Nb$ is not thermodynamically stable, it could be metastable in this pressure range.

Furthermore, the influence of vibrational free energy on the convex hull for each candidate structure at zero pressure was investigated. The results are presented in Fig. S16. It is evident that the stability of each candidate phase decreases with increasing temperature. $U_2Nb$ is always at the bottom of the convex hull, suggesting that it is the most stable ordered phase in U-Nb system.

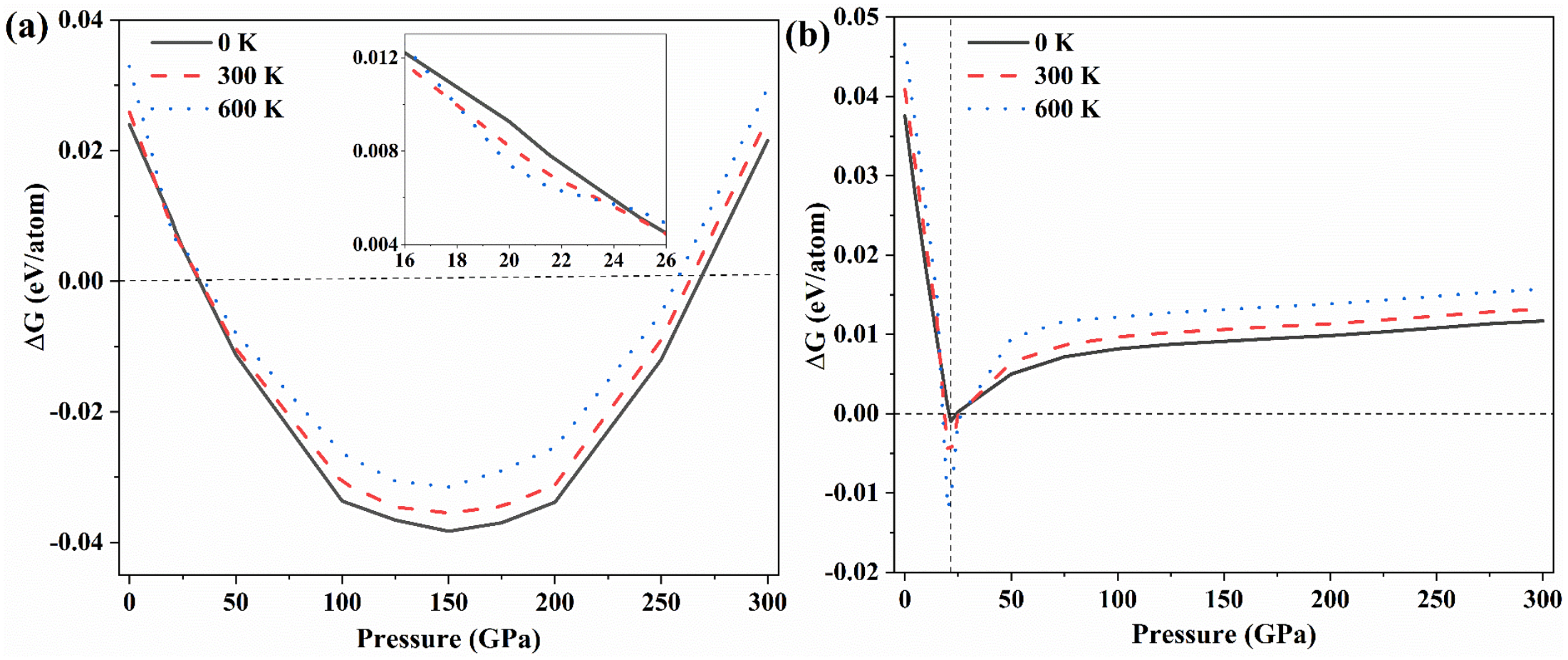

**Fig. S15.** Calculated formation Gibbs free energy ΔG as a function of pressure at different temperatures of 0, 300 and 600 K for *Cmcm*-$U_3Nb$. The vertical dotted-line in (b) corresponds to a pressure of 21.6 GPa. The reference terminal states are $\alpha$-U and BCC-Nb for (a), and $\alpha$-U and $U_2Nb$ for (b).

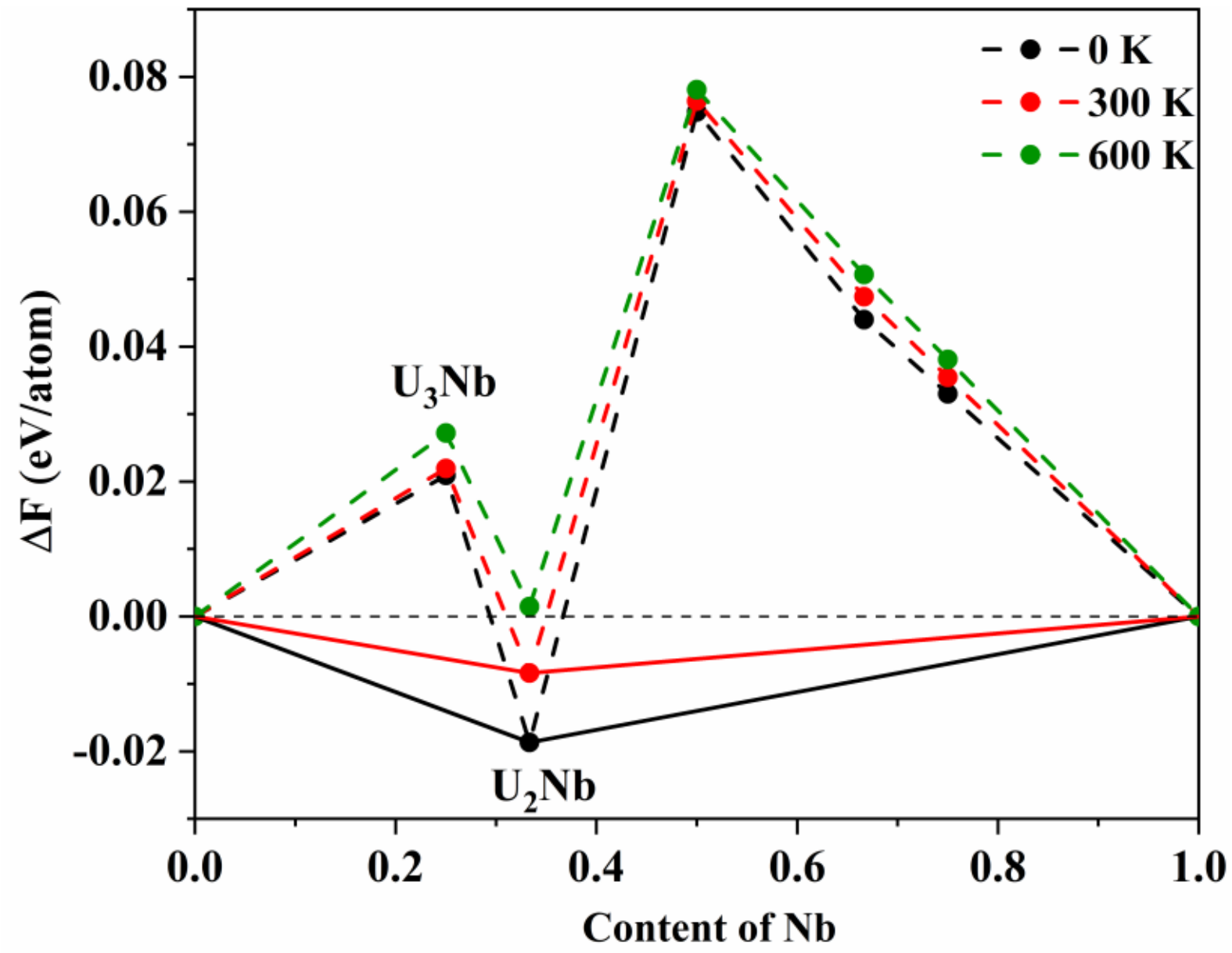

**Fig. S16.** Calculated formation free energy ΔF at zero pressure at temperatures of 0, 300, and 600 K.

## SV. Effect of core-core overlap on pseudopotential

The "Nb_sv" pseudopotential of Nb with a core radius of 2.4 bohr (1.27 Å) and "U" pseudopotential of U with a core radius of 2.8 bohr (1.48 Å) in VASP library were used in our calculations, which will produce slight core-core overlap in $U_2Nb$ at pressures near 500 GPa. Therefore, in order to quantify the influence of this core-core overlap on the calculated results under high pressure, we also employed other pseudopotential with a smaller core radius for comparison. They include the ultra-soft pseudopotentials (USPPs) of CASTEP software with a core radius of 1.4 bohr (0.74 Å) and 2.1 bohr (1.11 Å) for Nb and U, respectively. In addition, the USPPs of Quantum

Espresso (QE) software with a core radius of 1.5 bohr (0.79 Å) for Nb and 2.1 bohr (1.11 Å) for U were also used for comparison.

The nearest-neighboring U-U, U-Nb and Nb-Nb bonds in $U_2Nb$ at 500 GPa in our calculation are 2.35, 2.35 and 2.31Å, respectively. Thus, neither USPPs of CASTEP nor that of QE have a core-core overlap. We used these pseudopotentials to calculate the equation of state (EOS) for U, Nb and $U_2Nb$, respectively. The results are presented in Fig. S17. It can be seen that the EOS calculated by VASP is very consistent with those of CASTEP and QE, with the maximum difference in atomic volume less than 1%, which guarantees the reliability of the results calculated by VASP under high pressure. In general, for the PAW potential implemented in the VASP code, a slight core-core overlap has negligible impact on the calculated results [9].

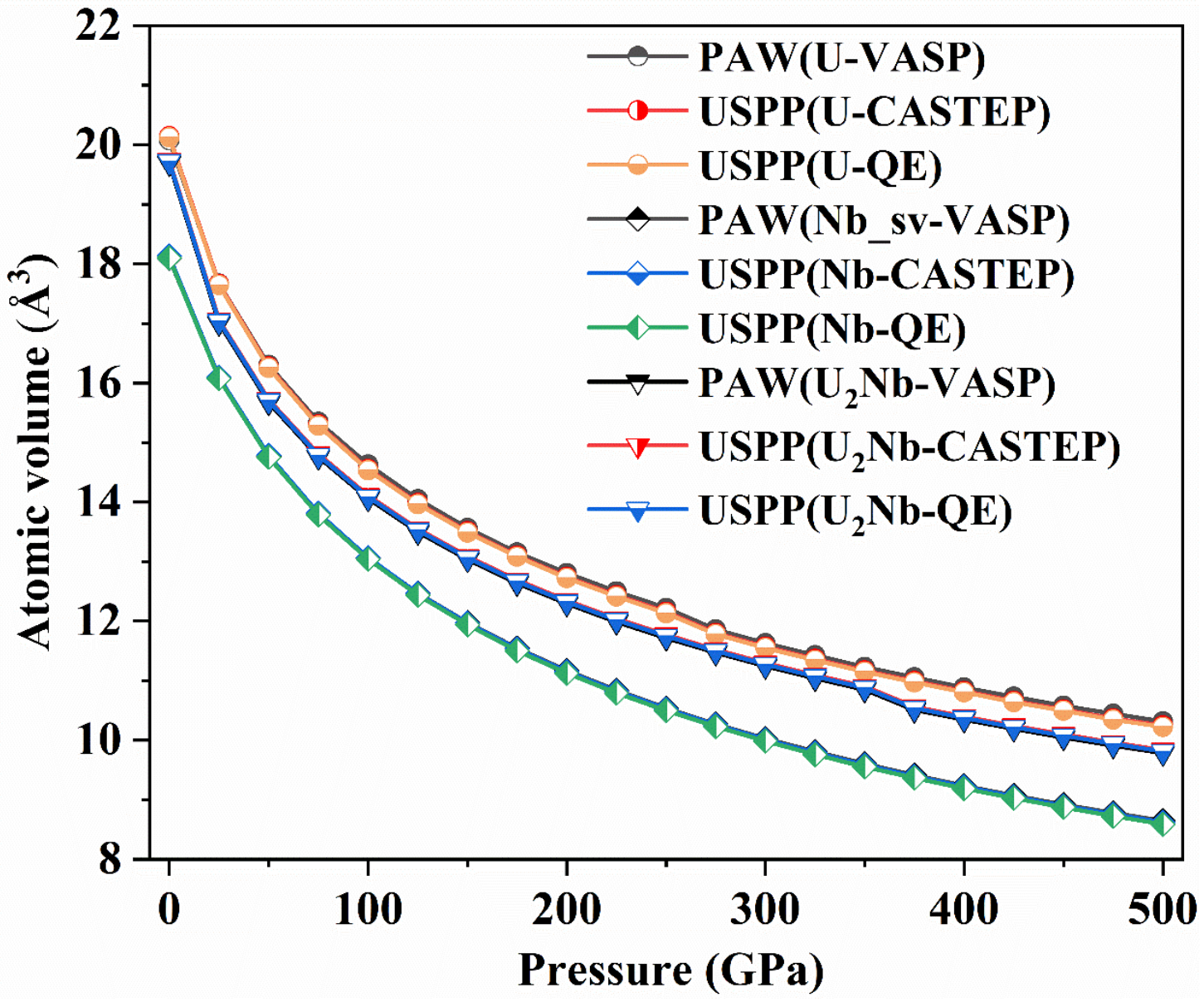

**Fig. S17.** Calculated equation of state of $U_2Nb$, U and Nb at zero Kelvin with different pseudopotentials.

## SVI. Results of GGA+*U* and LDA+*U*

Dynamic stability of the new predicted structures is also demonstrated by the phonon dispersion curves without virtual frequency using LDA (Fig. S18). In addition, we also calculated the electronic projected density of states (PDOS) of $P6_3/mmc$-$U_2Nb$, *P6/mmm*-$U_2Nb$ and *Cmcm*-$U_3Nb$ using LDA and LDA+*U*, and the results are presented in Fig. S19. These newly predicted structures are all metallic with the *f* electrons of the U atoms dominating the Fermi level. In general, the calculated results of LDA and LDA+*U* further confirm the reliability of the GGA results.

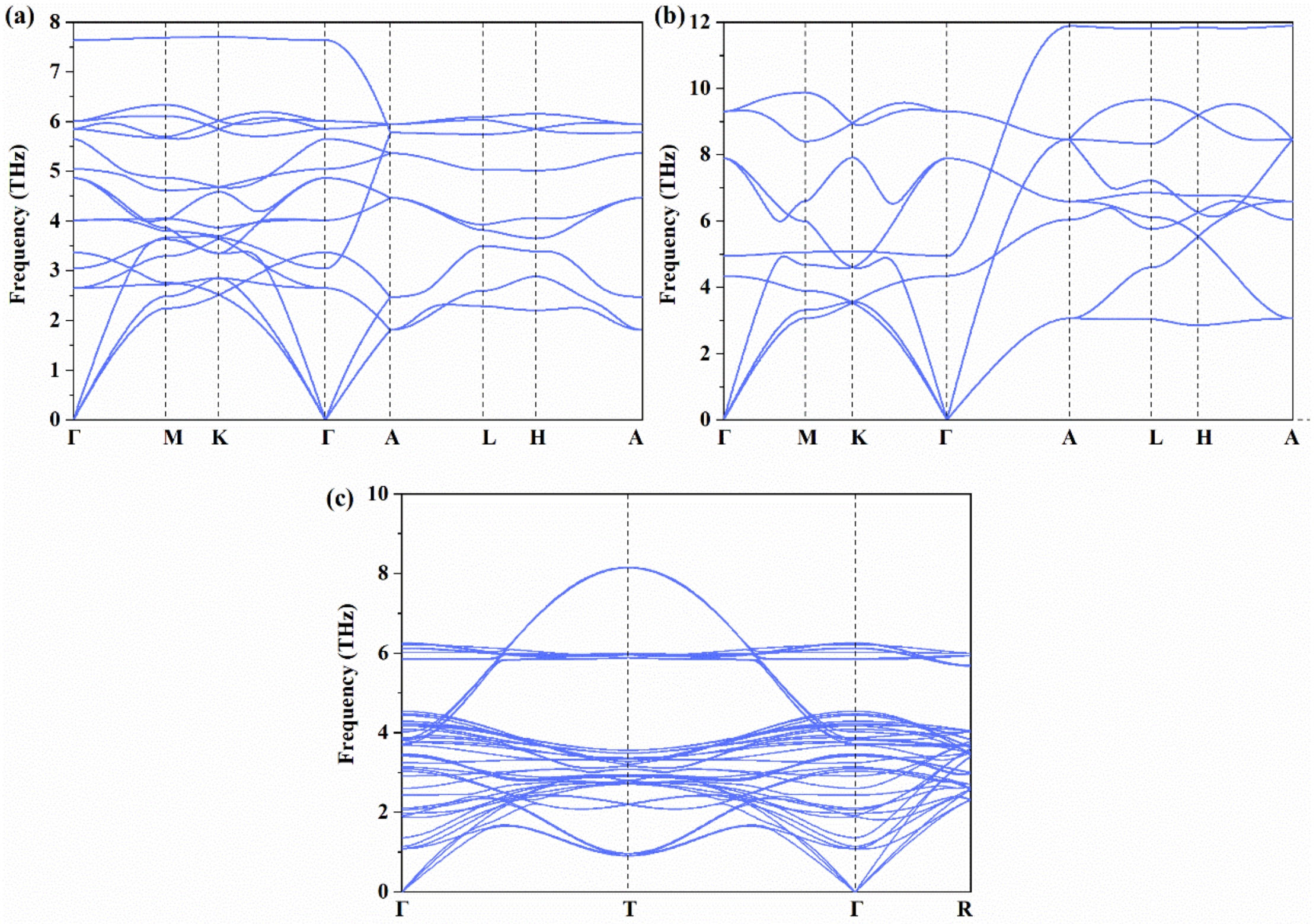

**Fig. S18.** Calculated phonon dispersion curves of (a) *P6₃/mmc*-$U_2$Nb at 0 GPa, (b) *P6/mmm*-$U_2$Nb at 100 GPa, and (c) *Cmcm*-$U_3$Nb at 100 GPa using LDA.

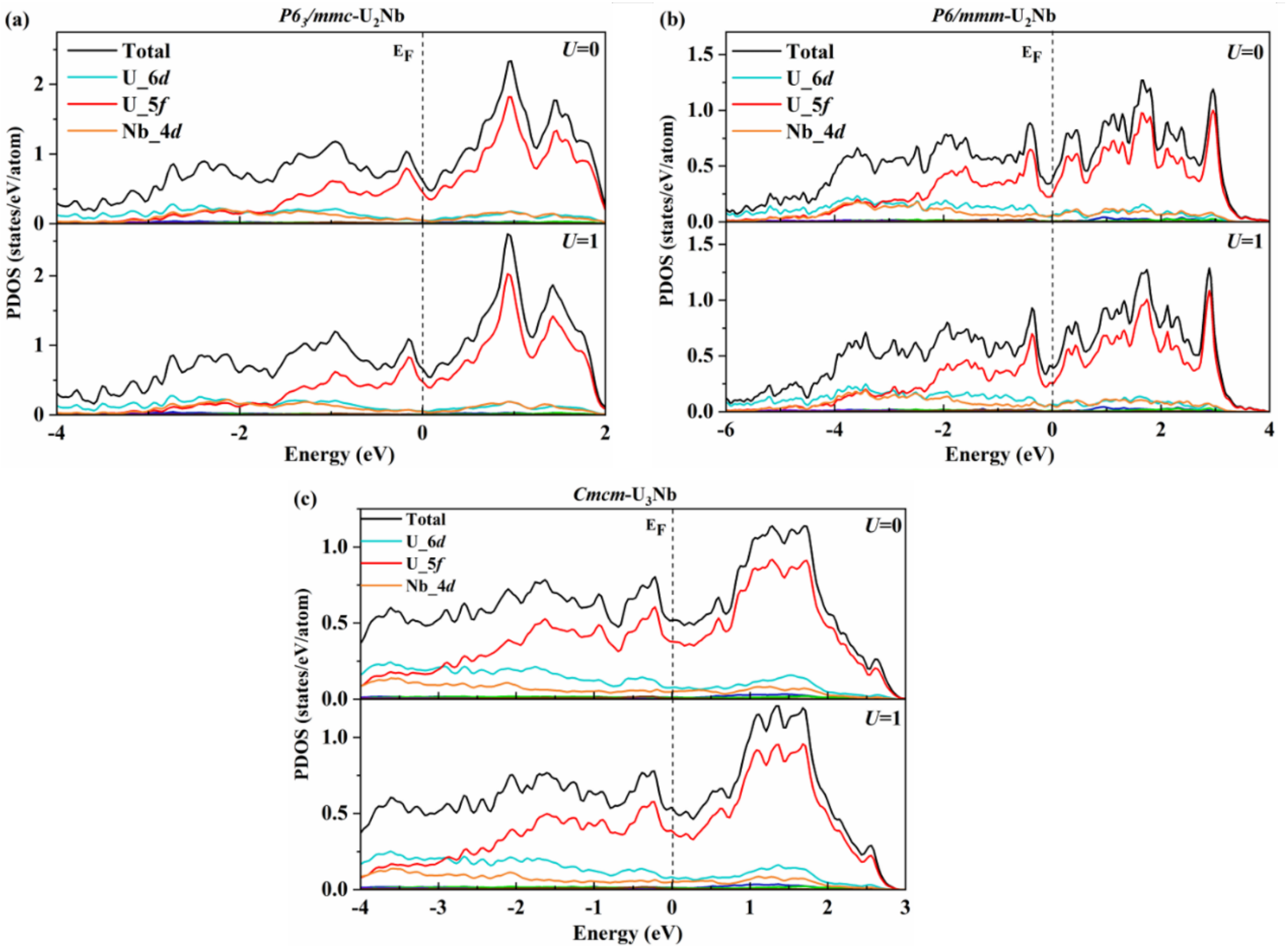

**Fig. S19.** Calculated projected density of states (PDOS) using LDA (*U*=0) and LDA+*U* (*U*=1) for (a) *P6₃/mmc*-$U_2$Nb at 0 GPa, (b) *P6/mmm*-$U_2$Nb at 100 GPa, and (c) *Cmcm*-$U_3$Nb at 100 GPa.

In order to verify whether the predicted phases are all metallic, the electronic projected density of states (PDOS) of *P6₃/mmc*-$U_2Nb$, *P6/mmm*-$U_2Nb$ and *Cmcm*-$U_3Nb$ at different *U* values were calculated by GGA+*U* and shown in Figs. S20-22. These results confirm that the newly predicted structures are all metallic with the 5*f* electrons of the U atoms dominating the electronic density of states near the Fermi level.

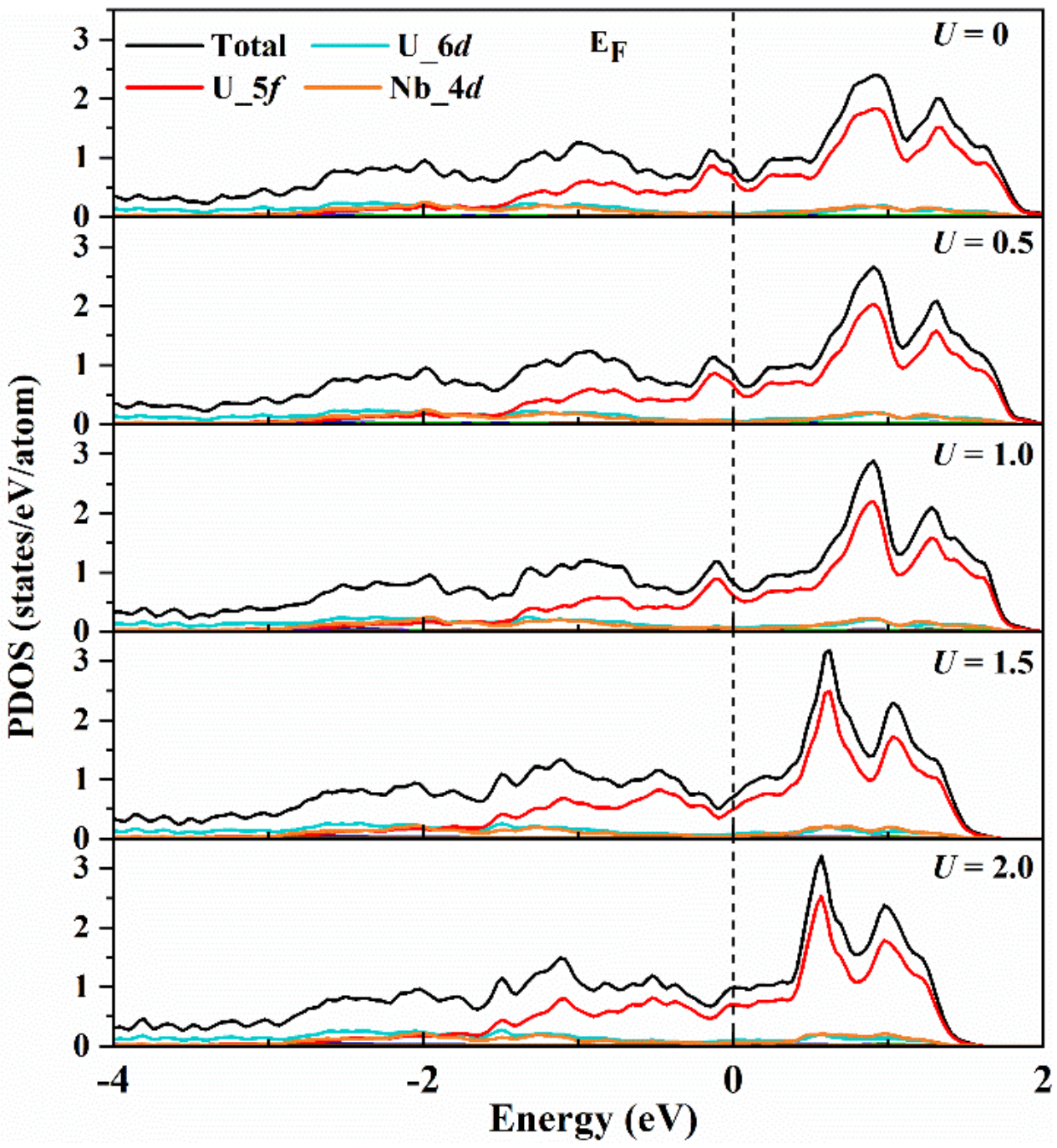

**Fig. S20.** Calculated projected density of states (PDOS) with different *U* values for *P6₃/mmc*-$U_2Nb$ at 0 GPa.

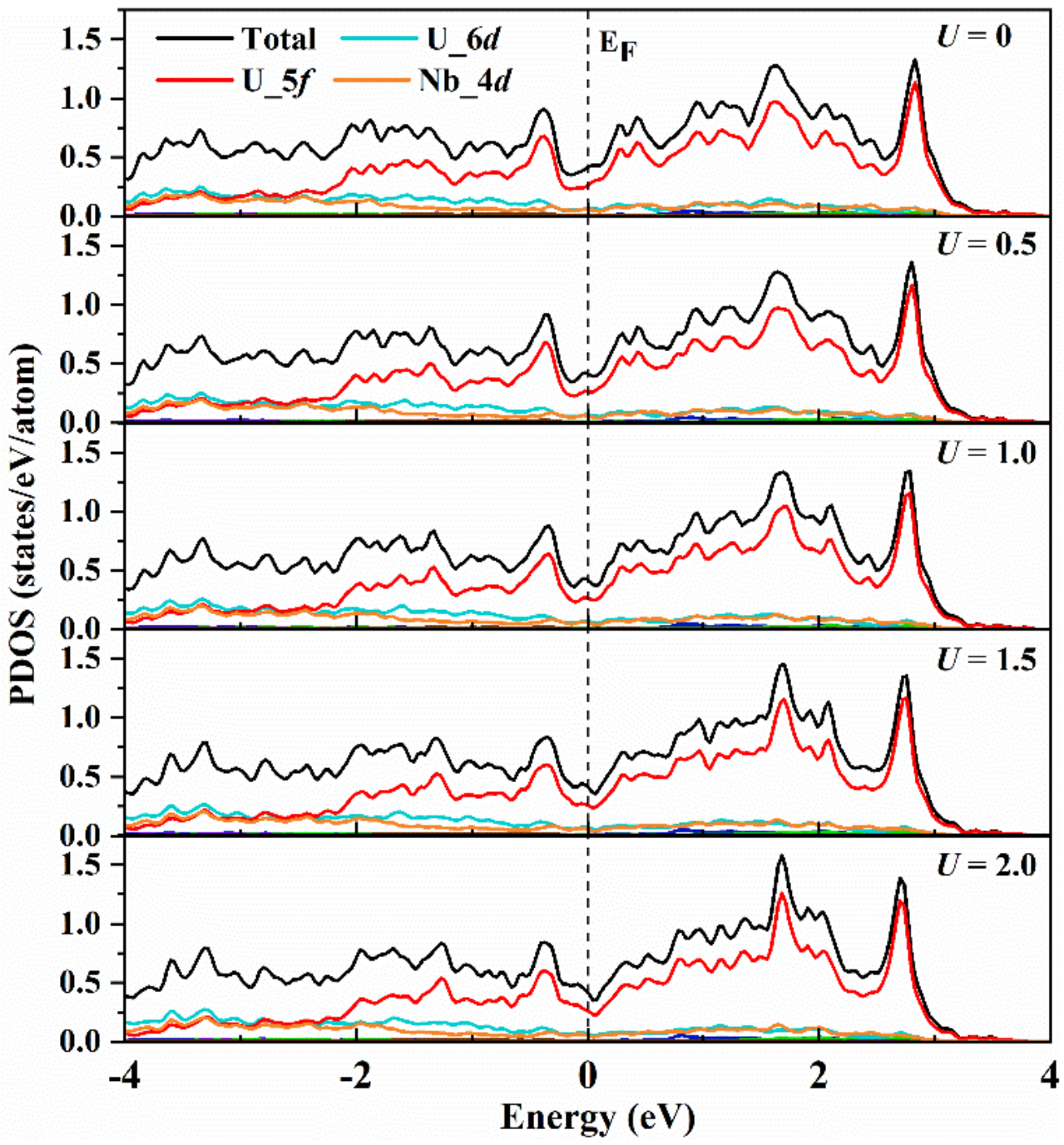

**Fig. S21.** Calculated projected density of states (PDOS) with different *U* values for

*P6/mmm*-$U_2Nb$ at 100 GPa.

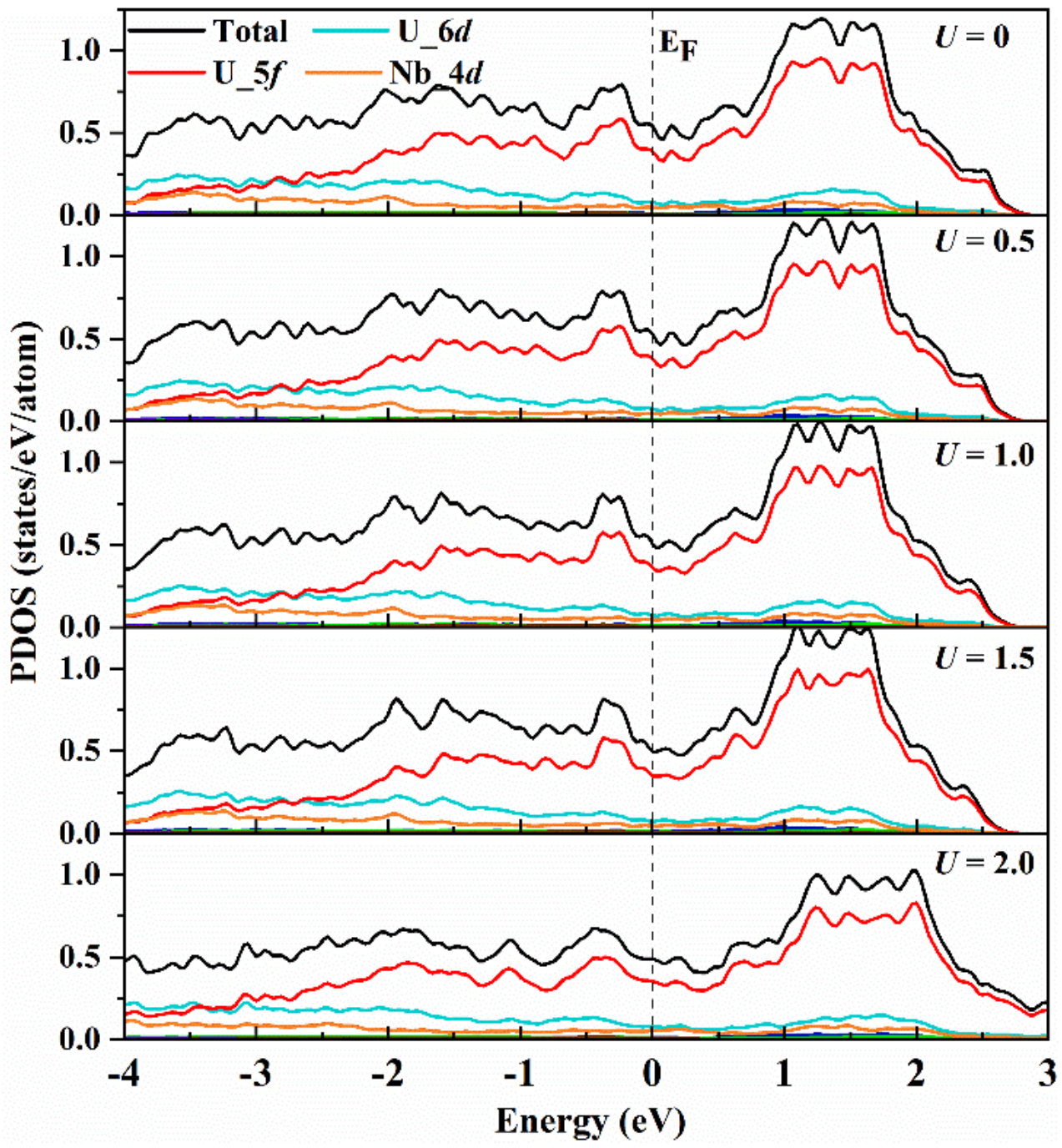

**Fig. S22.** Calculated projected density of states (PDOS) with different *U* values for *Cmcm*-$U_3Nb$ at 100 GPa.

# SVII. Supplementary Tables

**Table S1.** Lattice parameters, atomic coordinates and Wyckoff site occupation of *P6₃/mmc*-$U_2Nb$, *P6/mmm*-$U_2Nb$, *Fmmm*-$U_2Nb$, *Cmcm*-$U_3Nb$, and *Pmm2*-$U_3Nb$ at the given pressure.

| phase | lattice parameters(Å) | atom | site | atomic coordinates |
|---|---|---|---|---|
| ***$P6_3/mmc$*-$U_2Nb$ (0GPa)** | a=b=4.807<br>c=5.906<br>α=β=90°<br>γ=120° | Nb | *2b* | (0.0, 0.0, 1/4)<br>(0.0, 0.0, 1/3) |
| | | U | *4f* | (1/3, 2/3, 1/2+d)<br>(2/3, 1/3, d)<br>(1/3, 2/3, -d)<br>(2/3, 1/3, 1/2-d)<br>d=0.0313 |
| ***P6/mmm*-$U_2Nb$ (100GPa)** | a=b=4.377<br>c=2.542<br>α=β=90°<br>γ=120° | Nb | *1a* | (0.0, 0.0, 0.0) |
| | | U | *2d* | (1/3, 2/3, 1/2)<br>(2/3, 1/3, 1/2) |
| ***Fmmm*-$U_2Nb$ (400GPa)** | a=12.199<br>b=4.069<br>c=2.500 | Nb | *4a* | (0.0, 0.0, 0.0) (1/2, 0.0, 1/2)<br>(1/2, 1/2, 0.0) (0.0, 1/2, 1/2) |
| | | U | *8h* | (0.0, 1/2-y, 0.0) (0.5, 1/2-y, 0.5) |
| | α=β=γ=90° | | | (1/2, y, 0) (0, y, 1/2)<br>(0, -y, 1/2) (1/2, -y, 0)<br>(1/2, 1/2+y, 1/2) (0, 1/2+y, 0)<br>y=0.164 |
| ***Cmcm*-$U_3$Nb**<br>**(200GPa)** | a=2.372<br>b=19.013<br>c=4.201<br>α=β=γ=90° | Nb | *4c* | (1/2, 1/2-y, z+1/2) (1/2, 1/2+y, z)<br>(0, -y, z+1/2) (0, y, z)<br>y=0.092, z=0.25 |
| | | U | *4c* | (0, 1/2-y, z+1/2) (0, 1/2+y, z), y=0.215<br>(1/2, -y, z+1/2) (1/2, y, z), y=0.215<br>(0, 1/2+y, z+1/2) (0, 1/2-y, z), y=0.156<br>(1/2, y, z+1/2) (1/2, -y, z), y=0.156<br>(0,1/2+y, z+1/2) (0, 1/2-y, z), y=0.03<br>(1/2, y, z+1/2) (1/2, -y, z), y=0.03<br>z=1/4 |
| ***Pmm2*-$U_3$Nb** | a=2.873 | Nb | *1a* | (0, 0, -y), y=0.016 |
| | b=4.979 | U | *1b* | (0, 1/2, y), y=0.018 |
| | c=5.600 | U | *1c* | (1/2, 0, 1/2-y), y=0.02 |
| | α=β=γ=90° | U | *1d* | (1/2, 1/2, 1/2+y), y=0.185 |

**Table S2**. Bader charge analysis for $U_2$Nb and $U_3$Nb.

| Phase | Fractional Coordinates | | | Charge state (\|*e*\|) | Atom |
|---|---|---|---|---|---|
| | *x* | *y* | *z* | | |
| *P6₃/mmc*-$U_2$Nb (0GPa) | 0.3333 | 0.6667 | 0.5313 | -0.1893 | U |
| | 0.6667 | 0.3333 | 0.0313 | -0.0630 | U |
| | 0.6667 | 0.3333 | 0.4687 | -0.1478 | U |
| | 0.3333 | 0.6667 | 0.9687 | -0.0428 | U |
| | 0.0000 | 0.0000 | 0.2500 | 0.1877 | Nb |
| | 0.0000 | 0.0000 | 0.7500 | 0.2552 | Nb |
| *P6/mmm*-$U_2$Nb (100GPa) | 0.6667 | 0.3333 | 0.5000 | -0.2207 | U |
| | 0.3333 | 0.6667 | 0.5000 | -0.2599 | U |
| | 0.0000 | 0.0000 | 0.0000 | 0.4806 | Nb |
| *Fmmm*-$U_2$Nb (400GPa) | 0.0000 | 0.3360 | 0.0000 | -0.1737 | U |
| | 0.0000 | 0.6639 | 0.0000 | -0.1793 | U |
| | 0.0000 | 0.8360 | 0.5000 | -0.1737 | U |
| | 0.0000 | 0.1639 | 0.5000 | -0.1793 | U |
| | 0.5000 | 0.3360 | 0.5000 | -0.1737 | U |
| | 0.5000 | 0.6639 | 0.5000 | -0.1793 | U |
| | 0.5000 | 0.8360 | 0.0000 | -0.1737 | U |
| | 0.5000 | 0.1639 | 0.0000 | -0.1793 | U |
| | 0.5000 | 0.0000 | 0.5000 | 0.3529 | Nb |
| | 0.5000 | 0.5000 | 0.0000 | 0.3529 | Nb |
| | 0.0000 | 0.0000 | 0.0000 | 0.3529 | Nb |
| | 0.0000 | 0.5000 | 0.5000 | 0.3529 | Nb |
| *Cmcm*-$U_3Nb$ (100GPa) | 0.0000 | 0.7856 | 0.7500 | -0.1460 | U |
| | 0.0000 | 0.1564 | 0.7500 | -0.0625 | U |
| | 0.0000 | 0.8435 | 0.2500 | -0.0625 | U |
| | 0.0000 | 0.2143 | 0.2500 | -0.1460 | U |
| | 0.0000 | 0.0314 | 0.7500 | -0.3524 | U |
| | 0.0000 | 0.9685 | 0.2500 | -0.3631 | U |
| | 0.5000 | 0.2856 | 0.7500 | -0.1460 | U |
| | 0.5000 | 0.6564 | 0.7500 | -0.0625 | U |
| | 0.5000 | 0.3435 | 0.2500 | -0.0625 | U |
| | 0.5000 | 0.7143 | 0.2500 | -0.1460 | U |
| | 0.5000 | 0.5314 | 0.7500 | -0.3524 | U |
| | 0.5000 | 0.4685 | 0.2500 | -0.3631 | U |
| | 0.5000 | 0.9084 | 0.7500 | 0.5663 | Nb |
| | 0.5000 | 0.0915 | 0.2500 | 0.5663 | Nb |
| | 0.0000 | 0.4084 | 0.7500 | 0.5663 | Nb |
| | 0.0000 | 0.5915 | 0.2500 | 0.5663 | Nb |

**Table S3.** Calculated equilibrium atomic volume $V_0$ (Å$^3$), modulus $B_0$ (GPa), and pressure derivative $B'$ at 0 GPa for $U_2Nb$, $\alpha$-U and BCC-Nb, along with the theoretical and experimental data [5-8].

| | $U_2Nb$ (GGA) | $U_2Nb$ (ideal mixing) | $\alpha$-U (GGA) | BCC-Nb (GGA) | $\alpha$-U (DET) [5] | $\alpha$-U (Expt.) [6] | $\alpha$-U (Expt.) [7] | BCC-Nb (Expt.) [8] |
|---|---|---|---|---|---|---|---|---|
| $V_0$ | 19.69 | 19.43 | 20.08 | 18.11 | 20.19 | 20.66 | 20.38 | 17.98 |
| $B_0$ | 141 | 148 | 135 | 172 | 136 | 136 | 129 | 169 |
| $B'$ | 3.08 | 4.80 | 5.30 | 3.66 | 4.97 | 3.78 | 5.22 | 3.43 |

**Table S4.** The calculated elastic constants (in GPa) and shear ($C_s$) and longitudinal ($C_l$) wave sound velocities (in km/s) for $U_2Nb$ at given pressures (in GPa).

| **Phase** | **Pressure** | $C_{11}$ | $C_{12}$ | $C_{13}$ | $C_{33}$ | $C_{44}$ | $C_{66}$ | $C_s$ | $C_l$ |
|---|---|---|---|---|---|---|---|---|---|
| *P6₃/mmc* | 0 | 0 | 93 | 58 | 305 | 84 | 93 | 4.11 | 2.45 |
| | 5 | 327 | 107 | 71 | 322 | 92 | 110 | 4.30 | 2.53 |
| | 10 | 371 | 123 | 85 | 327 | 99 | 124 | 4.45 | 2.60 |
| | 15 | 415 | 140 | 100 | 321 | 106 | 138 | 4.57 | 2.65 |
| | 20 | 456 | 158 | 120 | 272 | 113 | 149 | 4.60 | 2.64 |
| | 21.6 | 469 | 163 | 132 | 221 | 114 | 153 | 4.51 | 2.58 |
| *P6/mmm* | 21.6 | 456 | 141 | 116 | 476 | 125 | 156 | 4.88 | 2.85 |
| | 25 | 482 | 159 | 124 | 498 | 128 | 162 | 4.95 | 2.88 |
| | 30 | 524 | 184 | 134 | 530 | 133 | 170 | 5.09 | 2.91 |
| | 35 | 564 | 209 | 144 | 570 | 138 | 177 | 5.21 | 2.95 |
| | 40 | 602 | 234 | 155 | 608 | 142 | 184 | 5.33 | 2.99 |
| | 45 | 640 | 258 | 166 | 645 | 146 | 191 | 5.44 | 3.03 |
| | 50 | 675 | 280 | 176 | 681 | 150 | 197 | 5.54 | 3.06 |

**Table S5.** The parameters for the Landau model of Eq. (3) given in the main text.

| $a_0$ (eV/GPa·Å$^{-2}$) | $b_0$ (eV/Å$^4$) | $c_0$ (eV/Å$^6$) | $P_0$ (GPa) | $E_b$(DFT) (meV) | $E_b$(model) (meV) | $d_b$(DFT) (Å) | $d_b$(model) (Å) |
|---|---|---|---|---|---|---|---|
| 0.37 | 3.16×10$^2$ | 8.97×10$^3$ | 13.99 | 7.8 | 7.4 | 0.075 | 0.076 |

| $d_c$(DFT) (Å) | $d_c$(model) (Å) | $P^*$(model) (GPa) |
|---|---|---|
| 0.130 | 0.132 | 24.05 |

## Supplementary references

[1] A. Togo, I. Tanaka, First principles phonon calculations in materials science, Scripta Mater., **108** (2015) 1-5.
[2] R. Hill, The elastic behaviour of a crystalline aggregate, Proc. Phys. Soc. Sect. A, **65** (1952) 349.
[3] O. L. Anderson, A simplified method for calculating the debye temperature from elastic constants., J. Phys. Chem. Solids, **24** (1963) 909-917.
[4] C. Zhang, L. Xie, Z. Fan, H. Wang, X. Chen, J. Li, G. Sun, Straightforward understanding of the structures of metastable α″ and possible ordered phases in uranium–niobium alloys from crystallographic simulation, J. Alloy. Compd., **648** (2015) 389-396.
[5] S. Adak, H. Nakotte, P. F. de Châtel, B. Kiefer, Uranium at high pressure from first principles, Physica B, **406** (2011) 3342-3347.
[6] A. Dewaele, J. Bouchet, F. Occelli, M. Hanfland, G. Garbarino, Refinement of the equation of state of α-uranium, Phys. Rev. B, **88** (2013) 134202.
[7] J. Akella, G. S. Smith, R. Grover, Y. Wu, S. Martin, Static EOS of uranium to 100 GPa pressure, High Pressure Res., **2** (1990) 295-302.
[8] R. G. McQueen, S. P. Marsh, J. W. Taylor, J. N. Fritz, W. J. Carter, in High Velocity Impact Phenomena, edited by R. Kinslow (Academic Press, New York, 1970), p. 293.
[9] H. Y. Geng, H. X. Song, J. F. Li, Q. Wu, High-pressure behavior of dense hydrogen up to 3.5 TPa from density functional theory calculations, J. Appl. Phys., **111** (2012) 063510.